\documentclass[10pt]{amsart}
\usepackage[utf8]{inputenc}
\usepackage[a4paper,width=185mm,top=10mm,bottom=15mm,left=13mm,right=13mm]{geometry}
\usepackage{etex}
\usepackage[all]{xy}
\usepackage{etoolbox}
\usepackage{lmodern}
\usepackage{textcomp}
\usepackage{array}
\usepackage{xstring}
\usepackage{longtable}
\usepackage[listings]{tcolorbox}
\tcbuselibrary{breakable}
\tcbuselibrary{skins}
\usepackage[pdftex]{hyperref}
\usepackage{amsmath}
\usepackage{tikz,tikz-cd,stmaryrd}
\usepackage{fancyhdr}
\usepackage{amssymb}
\usepackage{amsthm}
\usepackage{amsmath}
\usepackage{amsfonts}
\usepackage{amsthm}
\usepackage{pifont}
\usepackage{amsmath}
\usepackage{soul}
\usepackage{epsfig}
\usepackage{latexsym}
\usepackage{amssymb}
\usepackage{color}
\usepackage{amscd}
\usepackage{multirow}
\usepackage{graphicx}
\usepackage{lscape}
\usepackage{varioref}
 \usepackage{tabularx}
\usepackage{float}
\usepackage{xcolor}
\usepackage{bm}
\usepackage{tikz-cd}
 \makeatletter
    \let\stdchapter\section
    \renewcommand*\section{%
    \@ifstar{\starchapter}{\@dblarg\nostarchapter}}
    \newcommand*\starchapter[1]{%
        \stdchapter*{#1}
        \thispagestyle{fancy}
        \markboth{\MakeUppercase{#1}}{}
    }
    \def\nostarchapter[#1]#2{%
        \stdchapter[{#1}]{#2}
        \thispagestyle{fancy}
    }
\makeatother

  \newcommand{\widesim}[2][1.5]{
  \mathrel{\overset{#2}{\scalebox{#1}[1]{$\sim$}}}
}

\newtcolorbox{boxA}{
    fontupper = \bf,
    boxrule = 1.5pt,
    colframe = black  
}

\makeatletter
\@addtoreset{equation}{section} \@addtoreset{equation}{subsection}
\makeatother

\newtheorem{theorem}{Theorem}[section]
\newtheorem*{theorem*}{Theorem}
\newtheorem{lemma}[theorem]{Lemma}
\newtheorem{proposition}[theorem]{Proposition}
\theoremstyle{definition}
\newtheorem{definition}[theorem]{Definition}
\theoremstyle{corollary}

\newtheorem{remark}[theorem]{Remark}
\newtheorem{example}[theorem]{Example}
\theoremstyle{conclusion}

\usepackage{amsmath}
\usepackage{amssymb}
\usepackage{amsfonts}
\usepackage{esvect}
\usepackage{physics}
\usepackage{mathrsfs}
\usepackage{geometry}
\usepackage{abstract}
\usepackage{xcolor}

 \usepackage{color}

\begin{document}

\thispagestyle{empty}
\begin{center}
	\Large{{\bf  Polynomial Poisson Algebras and Superintegrable Systems from Cartan centralisers of Types $B_3$, $C_3$ and $D_3$  }}
\end{center}
\vskip 0.5cm
\begin{center}
	\textsc{Rutwig Campoamor-Stursberg$^{1,\star}$, Danilo Latini$^{2,*}$, Ian Marquette$^{3,\bullet}$, Junze Zhang$^{4,\dagger}$ 
   \\ and Yao-Zhong Zhang$^{4,\ddagger}$}
\end{center}
\begin{center}
	$^1$ Instituto de Matem\'{a}tica Interdisciplinar and Dpto. Geometr\'{i}a y Topolog\'{i}a, UCM, E-28040 Madrid, Spain
\end{center}

\begin{center}
	$^2$ Dipartimento di Matematica “Federigo Enriques”, Università degli Studi di Milano, Via C. Saldini 50, 20133 Milano, Italy \& INFN Sezione di Milano, Via G. Celoria 16, 20133 Milano, Italy
\end{center}

\begin{center}
	$^3$ Department of Mathematical and Physical Sciences, La Trobe University, Bendigo, VIC 3552, Australia
\end{center}

\begin{center}
	$^4$ School of Mathematics and Physics, The University of Queensland, Brisbane, QLD 4072, Australia
\end{center}
\begin{center}
	\footnotesize{$^\star$\textsf{rutwig@ucm.es} \hskip 0.25cm$^*$\textsf{danilo.latini@unimi.it,} \hskip 0.25cm $^\bullet$\textsf{i.marquette@latrobe.edu.au} \hskip 0.25cm
 $^\dagger$\textsf{junze.zhang@uq.net.au} \hskip 0.25cm
 $^\ddagger$\textsf{yzz@maths.uq.edu.au}}
\end{center}
\vskip  0.5cm

\begin{abstract}
\noindent In this work, we construct explicit formulas for the generators of the Cartan centralisers of complex semisimple Lie algebras $B_n,C_n$ and $D_n$, the case $A_n$ being already known \cite{campoamor2023algebraic}.   The precise structures for the cases of rank-three simple Lie algebras ($B_3,C_3$ and $D_3$) are provided, and the inclusion relations between the corresponding polynomial Poisson algebras (finitely generated Poisson algebras over $\mathbb{C}[\mathfrak{h}^*]$) are illustrated. We develop the idea of constructing algebraic superintegrable systems and their integrals from the generators of these polynomial Poisson algebras. In particular, we explicitly present the algebraic superintegrable systems corresponding to the Cartan reduction chains $\mathfrak{h} \subset \mathfrak{so}(6,\mathbb{C})$, $\mathfrak{h} \subset \mathfrak{so}(7,\mathbb{C})$, and $\mathfrak{h} \subset \mathfrak{sp}(6,\mathbb{C})$.
\end{abstract}

\section{Introduction}

\noindent Superintegrable and integrable systems occupy a pivotal position within mathematical physics and symplectic geometry, due to their rich algebraic and geometric structures \cite{MR187763,nehorovsev1968action,MR3119484,MR3493688,MR3942135}, as well as their wide applications in both mathematics \cite{MR4071113,MR4282982,MR4644061} and physics \cite{MR4584218,MR4771337}, respectively.   They further provide a natural connection with the theory of special functions and orthogonal polynomials, in particular, the Askey-Wilson scheme and the Painlev{\'e} transcendents \cite{MR4656047,dunkl2023b_2,MR4188777,MR4410687}. It is well known that superintegrable systems are intimately related to finitely generated Poisson algebras \cite{MR3493688,MR1814439,MR2337668,MR2226333,MR2804560}, a structure that has been observed and exploited in many problems of mathematical physics \cite{MR4644061,post2024racah,BOR,PER}.  The mathematical study of classical and quantum superintegrable systems can be traced back to the 1960s \cite{MR187763}, while the modern description of classical superintegrable systems was first introduced in \cite{nehorovsev1968action}. These systems generalize Liouville-integrable Hamiltonian systems on a $2n$-dimensional symplectic manifold in the case where isotropic (invariant) submanifolds have a dimension less than $n$. In the classical case, a superintegrable system is generally considered a deformation of the Poisson centers of a symplectic manifold. That is, the disjoint union of symplectic leaves forms a non-commutative (non-Abelian) algebra via Poisson projections.  By making the phase space explicit, it can also be constructed through the embedding chain of some associative algebras \cite{MR3493688}.  

The problem of classifying $n$-dimensional quadratically (i.e., only second-order integrals) superintegrable systems was investigated very recently using new geometrical approaches in \cite{MR4720612}. The present paper will not be restricted to quadratic integrals and will develop algebraic approaches based on commutants and classical Lie algebras.
 The construction of superintegrable systems based on Lie algebras is known to be deeply related to their invariants (generalized Casimir invariants), as the co-adjoint orbits of dual Lie algebras can be naturally viewed as a symplectic leaf \cite{TRO,MR0402819}. Such systems are important for understanding the structural properties of the Calogero-Moser and Ruijsenaars systems, among others. On the other hand, by lifting a Lie algebra to its universal enveloping algebra, many studies have revealed that non-commutative polynomial structures are hidden in the centraliser of certain subalgebras of universal enveloping algebras \cite{campoamor2023algebraic,MR4660510,CampoamorStursberg2021,MR4355741,MR4710584,MR3884869}. 
 
 Concerning explicit models,  purely algebraic constructions of superintegrable systems over symmetric algebras or universal enveloping algebras of finite-dimensional Lie algebras have been studied for many different hierarchies of systems. For example, this approach was successfully implemented for the Lie algebra $ \mathfrak{gl}(3)$ \cite{CampoamorStursberg2021} and $\mathfrak{su}(3)$ \cite{Correa2020}, by applying an algorithm on centralisers of subalgebras. More recent examples that connect with different contexts of mathematical physics can be found in \cite{MR4660510, MR4355741,MR4581473}. In particular, in a recent paper \cite{campoamor2023algebraic}, a similar algorithm was also applied to the centraliser of a Cartan subalgebra of type $A_n$ and certain non-semisimple algebras \cite{MR4355741}. These polynomial Poisson algebras (finitely generated Poisson algebras) associated with the centraliser of Cartan subalgebras can all be realized in terms of high-rank Racah-type algebras. 
 
 In this paper, we extend the results of \cite{campoamor2023algebraic} to other semisimple Lie algebras and provide a classification of generators of Cartan centralisers, described explicitly via their respective root systems. The purpose is to characterize the related polynomials explicitly and to provide a scheme to obtain superintegrable systems, without using specific realizations or geometric settings.

\medskip
The structure of this paper is as follows. In Section \ref{sec:prelim}, we review the formal definition of commutants, i.e., the centraliser of a subalgebra in the universal enveloping algebra or the symmetric algebra of a finite-dimensional Lie algebra. In order to find explicit formulas for invariant homogeneous polynomials, we provide an algorithm and define polynomial Poisson algebras in commutants (or centralisers). This method, as mentioned, has been found to be of interest in the construction of superintegrable systems in universal enveloping algebras (see, e.g., \cite{campoamor2023algebraic} and references therein). In Section \ref{sec:conscc}, we construct a finite set consisting of the generators for Cartan commutants, which generates a finitely generated Poisson algebra, called polynomial Poisson algebras. In Section \ref{sec:construction}, using an ordered basis that satisfies the Serre-Chevalley relations, we provide explicit formulas for the generators in the polynomial Poisson algebras $\mathcal{Q}_{B_n}(d)$, $\mathcal{Q}_{C_n}(d'')$, and $\mathcal{Q}_{D_n}(d')$ for fixed integers $d,d'$ and $d''$, where $B_n,C_n$ and $D_n$ are classical semisimple Lie algebras for $n \geq 1$. The inclusion relations and degrees of these polynomial Poisson algebras are also provided. The formal proof of the classification is rather nontrivial, and details on it will be given in a forthcoming paper \cite{abc}. As an illustration of the classification, we compute the indecomposable homogeneous generators in the Cartan centralisers for rank three semisimple Lie algebras of non-exceptional types. In particular, generators and algebraic structures of $\mathcal{Q}_{B_3}(5),\mathcal{Q}_{C_3}(5)$ and $\mathcal{Q}_{D_3}(3)$, which are themselves of interest, albeit $D_3$ is isomorphic to $A_3$.  The rank of these polynomial Poisson algebras and the inclusion relation will also be provided in Subsections \ref{subsec:so5}, \ref{subsec:so6}, and \ref{subsec:sp5}, respectively. In particular, we have also studied the superintegrable systems corresponding to these polynomial Poisson algebras. Finally, in Section \ref{sec:conclusion}, some conclusions and future perspectives will be drawn.

\section{Preliminaries}
\label{sec:prelim}


\noindent Let $\mathfrak{g}$ be a $n$-dimensional semisimple or reductive Lie algebra over a field $\mathbb{F}$, with the commutator $[\cdot,\cdot]$, and let $\beta_\mathfrak{g} = \{X_1,\ldots,X_n\}$ be an ordered basis. Non-trivial brackets satisfy the relations $[X_i,X_j] = \sum_{k=1}^n C_{ij}^kX_k$, where $ C_{ij}^k $ denotes the structure constants of $\mathfrak{g}$ with respect to the given basis. Let $\mathfrak{g}^*$ denote the dual space of $\mathfrak{g}$, which admits a canonical Poisson-Lie bracket $\{\cdot,\cdot \}$ is defined by 
\begin{equation*}
    \{f,g\} (\xi)  = \langle \xi, [d_\xi f,d_\xi g]  \rangle ,\quad f,g \in C^\infty(\mathfrak{g}^{*}),\quad \xi \in \mathfrak{g}^{*}, \label{eq:poiss}
\end{equation*} 
where $d_\xi f, d_\xi g \in  T_\xi^*\mathfrak{g}^*\cong \left(\mathfrak{g}^*\right)^* \cong \mathfrak{g}$, and $\langle \cdot,\cdot \rangle : \mathfrak{g}^*\times \mathfrak{g} \rightarrow \mathbb{R}$ is the dual pair between $\mathfrak{g}^*$ and $\mathfrak{g}$. We define the corresponding coordinate functions $  \big(x_1,\ldots,x_n\big)$ on $\mathfrak{g}^*$ by $x_i(\xi) = \langle \xi,X_i\rangle$ with $i = 1,\ldots,n$ (see \cite[Chapter 7]{MR2906391} for details). 

Let $U(\mathfrak{g})$ be the universal enveloping algebra of $\mathfrak{g}$. As is well-known \cite{MR1451138}, the symmetric algebra $S(\mathfrak{g}) \cong \mathbb{F} [\mathfrak{g}^*]$ of $\mathfrak{g}$ is endowed with the Poisson-Lie bracket $\{\cdot,\cdot\} : S(\mathfrak{g}) \times S(\mathfrak{g}) \rightarrow S(\mathfrak{g})$, for all $ p,q \in S(\mathfrak{g})$, defined by
 \begin{align}
     \{p,q\}    = \sum_{j,k,l=1}^n C_{jk}^l x_l \dfrac{\partial p}{\partial x_j} \dfrac{\partial q}{\partial x_k}. \label{eq:poi}
 \end{align} Hence, $\left(S(\mathfrak{g}),\{\cdot,\cdot\}\right)$ defines a Poisson algebra \cite[Chapter 6, Section 6.1]{olver1993applications}. 

 We are interested in the (co)adjoint action of $\mathfrak{g}$ on $U(\mathfrak{g})$ and $S(\mathfrak{g})$, respectively. We recall that the adjoint representation $ \mathrm{ad}: \mathfrak{g} \rightarrow \mathrm{Aut}(U(\mathfrak{g}))$ and the coadjoint representation $ \mathrm{ad}^*: \mathfrak{g}  \rightarrow \mathrm{Aut}\left(S(\mathfrak{g})\right)$ induce the following actions \cite{MR0760556}
 \begin{align}
       & \quad  P\left( X_1,\ldots,X_n\right) \in U(\mathfrak{g})  \mapsto  [X_j,P] \in U(\mathfrak{g}) \label{eq:lie}, \\
     & \quad p(x_1,\ldots,x_n) \in S(\mathfrak{g})   \mapsto \{x_j,p  \}  = \widetilde{X}_j(p(x_1,\ldots,x_n) ) = \sum_{l,k} C_{jk}^lx_l \dfrac{\partial p}{\partial x_k} \in S(\mathfrak{g}), \label{eq:dual}
  \end{align}
where $\widetilde{X}_j = \sum_{l,k} C_{jk}^l x_l \dfrac{\partial}{\partial x_k}$ is the first-order differential operator associated with the generator $X_j$ of $\mathfrak{g}$.   Using the Leibniz rule, it can be shown by recursion that for any $\prod_i f_i, \prod_j g_j$ with $f_i,g_j \in S(\mathfrak{g})$, the following identity is satisfied:
\begin{align}
     \left\{\prod_{i=1}^m f_i,\prod_{j=1}^n g_j\right\}   = \sum_{i_s,j_r} \{f_{i_s},g_{j_r}\}   \prod_{i \neq i_s} f_i \prod_{j \neq j_r} g_j. \label{eq:muli}
 \end{align}
This formula, when applied to monomials, provides a systematic procedure to compute the Poisson bracket of homogeneous polynomials in a systematic way.

\medskip
We now shift our focus to the concept of the commutant in relation with subalgebras in both the universal enveloping algebra and the symmetric algebra of $\mathfrak{g}$.  The actions given in \eqref{eq:lie} and \eqref{eq:dual} suggest the following definition.

\begin{definition}
\label{2.1}
 Let $\mathfrak{a}$ be a subalgebra of $\mathfrak{g}$ with an ordered basis $\beta_{\mathfrak{a}} = \{X_1,\ldots,X_s\}$, and let $\mathfrak{a}^*$ be its dual. The \textit{commutants} (or \textit{centralisers}) $U(\mathfrak{g})^{\mathfrak{a}}$ and $S(\mathfrak{g})^\mathfrak{a}$ are defined, respectively, as the centraliser of $\mathfrak{a} $ and $\mathfrak{a}^*$ in $U(\mathfrak{g})$ and $S(\mathfrak{g})$:
\begin{align*}
     U(\mathfrak{g})^{\mathfrak{a}}  = &\left\{ P \in U(\mathfrak{g}): \text{ } [X,P] = 0  \quad \forall X \in \mathfrak{a}\right\},  \\
      S(\mathfrak{g})^\mathfrak{a}   =& \left\{p \in S(\mathfrak{g}): \text{ } \{x,p\} = 0  \quad \forall x \in \mathfrak{a}^*\right\},
  \end{align*} where $P$ and $p$ are polynomials in terms of $X_j$ and $x_j$ for all $1 \leq j \leq n$.
\end{definition}

\begin{remark}
(i) Since $\left\{S(\mathfrak{g})^\mathfrak{a},S(\mathfrak{g})^\mathfrak{a}\right\} \subset S(\mathfrak{g})^\mathfrak{a}$ and $ \{S(\mathfrak{g})^\mathfrak{a},\mathfrak{a}\} = 0$, it is clear that $S(\mathfrak{g})^\mathfrak{a}$ inherits a Poisson structure from $S(\mathfrak{g})$. In this context, we will consider $S(\mathfrak{g})^\mathfrak{a}$ as a Poisson algebra characterized by the Poisson bracket $\{\cdot,\cdot\}$. An analogous argument applies to $U(\mathfrak{g})^\mathfrak{a}$.

(ii) The $\textit{Poisson center}$ of $\left(S(\mathfrak{g}),\{\cdot,\cdot\}\right)$ is the set of all $\mathfrak{g}$-invariant polynomials, i.e., \begin{align*}
    S(\mathfrak{g})^\mathfrak{g}  =\left\{ p \in S(\mathfrak{g}) : \text{ } \{p,x\} = 0 \quad \forall x \in \mathfrak{g}^*\right\}.
\end{align*} It consists of all the Casimir invariants of $\mathfrak{g}$.
\end{remark}

For any $h \in \mathbb{N}_0 := \mathbb{N} \cup \{0\}$, we define
\begin{equation*}
U_h(\mathfrak{g}) = \mathrm{span} \{X_1^{i_1} \cdots X_n^{i_n}: i_1+ \ldots + i_n \leq  h\}
\end{equation*}
as the linear subspace of $U(\mathfrak{g})$ spanned by polynomials of degree at most $h$ on the basis of $\mathfrak{g}$. The degree $\delta$ of an arbitrary element $P \in U(\mathfrak{g})$ is defined as $\delta(P) := {\rm inf}\{\ell: P \in U_{\ell}(\mathfrak{g})\}$. Furthermore, there is a natural filtration in $U(\mathfrak{g})$ given by the following relations \cite{MR1451138} \begin{align}
    U_{0}(\mathfrak{g}) = \mathbb{F}, \quad U_h(\mathfrak{g})U_k(\mathfrak{g}) \subset U_{h+k}(\mathfrak{g}), \quad U_h(\mathfrak{g}) \subset U_{h+k}(\mathfrak{g}),\quad \forall k\geq 1. \label{eq:filt}
\end{align}  From the natural filtration relation \eqref{eq:filt}, one can define a vector space $U^k (\mathfrak{g}) := U_k (\mathfrak{g}) /U_{k-1} (\mathfrak{g})  $. Note that $U^k(\mathfrak{g})$ is not a subalgebra for each $k$, since the product of two homogeneous elements of degree $k$ and $\ell$ lies in degree $k+ \ell$. That is, for any $p \in S^k(\mathfrak{g}) $ and $q \in S^{\ell}(\mathfrak{g})$, we have \begin{align*}
    \deg \left( \Lambda(pq)\right) = k + \ell , \qquad   \Lambda(pq) -  \Lambda(p) \Lambda(q) \in U_{k+ \ell-1}(\mathfrak{g}),
\end{align*} where $\Lambda (pq) = \Lambda (p) \Lambda(q) + \text{lower order terms}$. We now pass from the filtered algebra $U(\mathfrak{g})$ to its associated graded algebra \begin{align}
    \mathrm{gr} \, U (\mathfrak{g}) = \bigoplus_{k \geq 0 } U^k(\mathfrak{g}),
\end{align}  where $U^0(\mathfrak{g}) = \mathbb{F}$.

\bigskip
We now focus our attention on the non-commutative Poisson centralisers in $\left(S(\mathfrak{g}),\{\cdot,\cdot\}\right)$. For similar constructions, see \cite{MR191995,marquette2023algebraic} and references therein. In this paper, the term non-commutative refers to the Poisson bracket relations (non-trivial) rather than to algebra. Let $S^k(\mathfrak{g})$ be the subspace of $S(\mathfrak{g})$ formed by homogeneous polynomials of degree $k$. Clearly, this induces the decomposition
\begin{align}
   S(\mathfrak{g}) = \bigoplus_{k \geq 0} S^k(\mathfrak{g})  .
\end{align} It follows that, for any $p \in S(\mathfrak{g})$, the polynomial decomposes as $p = \sum_{k \geq 0} p_k$, where $p_k \in S^k(\mathfrak{g})$ for all $k \geq 0$. We define the vector space of $\mathfrak{a}$-invariant $k$-homogeneous polynomials as  \begin{align*}
 S^k(\mathfrak{g})^{\mathfrak{a}}  = \left\{p_k \in S^k(\mathfrak{g}): \{x,p_k\} = 0  \quad \forall x \in \mathfrak{a}^*\right\},
\end{align*}  where $p_k(x_1,\ldots,x_n)$ is a homogeneous polynomial of degree $k \geq 0$ with the generic form \begin{align}
    p_k(x_1,\ldots,x_n) = \sum_{i_1 + \ldots + i_n = k} \Gamma_{i_1,\ldots, i_n}\, x_1^{i_1} \cdots x_n^{i_n}, \quad  \Gamma_{i_1,\ldots,  i_n} \in \mathbb{F}. \label{eq:ci}
\end{align} Thus, $S(\mathfrak{g})^\mathfrak{a} = \bigoplus_{ k \geq 0}S^k(\mathfrak{g})^\mathfrak{a}$. The condition $p \in S(\mathfrak{g})^\mathfrak{a}$ is equivalent to the system of partial differential equations $\{x_j,p_k\} = 0$ for all $X_j \in \mathfrak{a}_j$. By the  definition of commutants, to find a suitable finitely generating set for centraliser subalgebras, all $\mathfrak{a}$-invariant (linearly independent) indecomposable homogeneous polynomial solutions of the system of partial differential equations
\begin{align}
    \widetilde{X}_j(p_k)(x_1,\ldots,x_n) = \{x_j,p_k\}  = \sum_{1 \leq l,i \leq n} C_{ji}^lx_l \dfrac{\partial p_k}{\partial x_i} = 0, \text{ }\quad 1 \leq j \leq \dim  \mathfrak{a} = s \label{eq:func}
\end{align}
must be found.

\medskip
Note that for an arbitrary (finite-dimensional) Lie algebra, a linear basis in $ S(\mathfrak{g})^\mathfrak{a}$ is not necessarily canonical. In other words, it cannot be guaranteed that the associated Poisson algebra is finitely-generated. However, for the case of semisimple Lie algebras $\mathfrak{g}$, it can be shown that $S(\mathfrak{g})$ is Noetherian \cite[Chapter 2]{MR1451138}. Since $A$ is assumed to be a reductive subgroup of $G$, the invariant Poisson subalgebra $S(\mathfrak{g})^\mathfrak{a}$ is also finitely-generated by the Hilbert-Nagata theorem \cite[Theorem 6.1]{Dolgachev03}. Hence, we always assume that the system \eqref{eq:func} admits an integrity basis formed by polynomials. This implies that once a maximal set of indecomposable polynomials $\left\{p_{k_1},\ldots ,p_{k_s}\right\}$ has been found, there always exists some integer $g_j \in \mathbb{N}$ such that $p_{k_j+g_j}$ is decomposable for all $k \geq 1$. By decomposability of a polynomial $p \in S(\mathfrak{g}) $, we mean that there exists some polynomial $p'  \in  S(\mathfrak{g}) $ of lesser degree such that $p \equiv 0 \mod p'$, i.e., $p'$ divides $p$. We also observe that the linear independence of the basis elements (in the centraliser of a subalgebra) generally does not imply algebraic independence. As a matter of fact, if $\mathfrak{a} = \mathfrak{g}$, the maximal number of functionally independent solutions of \eqref{eq:func} is known to be given by \cite{MR4660510,MR191995,MR2515551,MR2276736,MR0204094,MR0411412}.
\begin{align}
    \mathcal{N} (\mathfrak{a}) = \dim  \mathfrak{g} - \mathrm{rank}\left(C_{ji}^l x_l\right), \quad 1 \leq j  \leq s, \text{ } 1 \leq l,i \leq \dim \, \mathfrak{g}, \label{eq:maximum}
\end{align} where $\left(C_{ji}^l x_l\right)$ denotes  the matrix associated with the commutator table of the Lie algebra $\mathfrak{g}$. Also, \eqref{eq:maximum} holds if $\mathfrak{a}$ is Abelian.

\bigskip
 Let $\mathfrak{a}$ be a Lie subalgebra of $\mathfrak{g}$ with an ordered basis defined above. We propose an algorithm for the construction of indecomposable and linearly independent monomials in $S(\mathfrak{g})^\mathfrak{a}$. This procedure will allow us to build a finitely-generating set for ${S^k(\mathfrak{g})}^\mathfrak{a}$ systematically.  Starting from degree one, it is clear that ${S^1(\mathfrak{g})}^\mathfrak{a} = \mathfrak{a}$ if $\mathfrak{a}$ is a maximal Abelian subalgebra, otherwise ${S^1(\mathfrak{g})}^\mathfrak{a} $ coincides with the centraliser of $\mathfrak{a}$ in $\mathfrak{g}$.  Without loss of generality, suppose that all indecomposable degree-one monomials are given by $$ \textbf{q}_1 = \left\{p_{1,1},\ldots,p_{1,m_1}\right\}. $$ We now proceed with the construction of indecomposable quadratic solutions of \eqref{eq:func}. It is clear that for any $p_1,p_1' \in \textbf{q}_1$, the product $p_2 = p_1 p_1'$ is decomposable, therefore any quadratic solutions that are the product of elements in $\textbf{q}_1$ must be discarded.  It follows that indecomposable quadratic homogeneous polynomials belonging to the centraliser must depend on generators of $\mathfrak{g}$ that do not belong to the subalgebra. Thus,
\begin{align*}
     \textbf{q}_2 = \left\{p_{2,1},\ldots,p_{2,m_2}\right\}.
 \end{align*}
  Iterating the procedure and discarding all elements that can be expressed as products of lower-order polynomials, all indecomposable homogeneous polynomials up to a certain degree $\zeta $ that are solutions to \eqref{eq:func} can be deduced:
 \begin{align}
\textbf{Q}_\zeta = \bigsqcup_{k=1}^\zeta\textbf{q}_k,\quad  \textbf{q}_k =\{p_{k,1} ,\ldots,p_{k,m_k} \}    .\label{eq:bas}
\end{align}
Hence, by construction, $\zeta$ denotes the maximum degree of an indecomposable homogeneous polynomial in the set of all commutants $\textbf{Q}_\zeta$, i.e., any homogeneous polynomial of degree greater than $\zeta$ is decomposable and therefore not contained in $\textbf{Q}_\zeta$.  It is clear that $ \mathbb{F}\langle\textbf{q}_k\rangle \subset S^k(\mathfrak{g})^\mathfrak{a}$ forms a vector space for each $k$, and thus $\mathbb{F}\langle\textbf{Q}_\zeta\rangle$ is also a vector space, where $\mathbb{F}$ is an arbitrary field. Notice that $\mathbb{F}\langle\textbf{Q}_\zeta\rangle$ is infinite-dimensional (as a vector space). It is straightforward to verify that we have the filtration
\begin{equation*}
  \mathbb{F} \subset \mathbb{F}\langle\textbf{Q}_1\rangle \subset \ldots \subset \mathbb{F}\langle\textbf{Q}_\zeta\rangle,\quad \dim_{FL}\mathbb{F}\langle\textbf{Q}_\zeta\rangle= m_1 + \ldots + m_\zeta.
\end{equation*}  Here, $\dim_{FL}$ denotes the number of indecomposable monomials that generate $\mathbb{F}\langle\textbf{Q}_\zeta\rangle$.  As $\mathbb{F}\langle\textbf{Q}_\zeta\rangle$ encompasses all indecomposable and linearly independent solutions of this system, for any $p_h \in \textbf{q}_h$ and $p_\ell \in \textbf{q}_\ell$, define the bilinear map $\{\cdot,\cdot\}  :  \mathbb{F} \langle  \textbf{Q}_\zeta \rangle \times \mathbb{F} \langle  \textbf{Q}_\zeta \rangle\rightarrow \mathbb{F} \langle  \textbf{Q}_\zeta \rangle$ by
\begin{equation}
  \{p_h,p_\ell\}  =  \sum_{m_{r_1}+ \ldots + m_{r_k} = \ell + h -1} \Gamma_{h,\ell}^{r_1,\ldots,r_k} p_{r_1}^{m_{r_1}} \cdots p_{r_k}^{m_{r_k}}, \label{eq:newpoi}
\end{equation}
where $\Gamma_{h,\ell}^{r_1,\ldots,r_k} \in\mathbb{F}$ and $1\leq r_i\leq \zeta$ for $1\leq i\leq k$. Here $m_{r_j}$ is the degree of the polynomial $p_{r_j}$.  Moreover, since $\mathbb{F}\langle\mathbf{Q}_\zeta\rangle  $ is closed under the restriction of the Lie-Poisson bracket, the Jacobi identity holds, and \eqref{eq:newpoi} endows $\mathbb{F}\langle\mathbf{Q}_\zeta\rangle$ with a Poisson algebra structure, in conjunction with additional polynomial relations $ P(\textbf{q}_1,\ldots,\textbf{q}_\zeta) = 0$.
For convenience, we denote the algebra generated by $\mathbb{F}\langle\textbf{Q}_\zeta\rangle$ with a Poisson bracket $\{\cdot,\cdot\}$ as $ \mathcal{Q}_\mathfrak{g}(d) $, where the degree $d$ is defined as the maximal number of elements appearing in the decomposition \eqref{eq:newpoi}, i.e.,   \begin{align*}
d := \max\left\{m_{r_1}+\ldots+m_{r_k}: p_{r_j}^{m_{r_j}}\notin\mathcal{Z},\,1\leq m_{r_j}\leq\zeta,\,j=1,\dots,k\right\}.
 \end{align*} denote the degree of $\mathcal{Q}_\mathfrak{g}(d)$. Here, $\mathcal{Z} := \{p \in \textbf{Alg}\left\langle \textbf{Q}_\zeta\right\rangle :  \{p,q\} = 0, \text{ } \text{ for all } q \in \textbf{Alg}\left\langle \textbf{Q}_\zeta\right\rangle \}$ is the center of $\textbf{Alg}\left\langle \textbf{Q}_\zeta\right\rangle$.  In particular, if $d=0$, then $\mathcal{Q}_\mathfrak{g}(0)$ is an Abelian algebra.

\begin{remark}
\label{rem:algebra}
   In general, a polynomial algebra (a.k.a. a polynomial ring) is a special case of a finitely generated algebra, as it does not contain any polynomial relations. Although $S(\mathfrak{g})^\mathfrak{a}$ may contain a polynomial relation, it is therefore not necessarily a polynomial ring. However, as we constructed above via polynomial ansatz, the Poisson bracket relations in $S(\mathfrak{g})^\mathfrak{a}$ are closed in a polynomial way, see \eqref{eq:newpoi}. Regarding whether it contains polynomial relations $P(\textbf{q}_1,\ldots,\textbf{q}_\zeta) = 0$, we always call $S(\mathfrak{g})^\mathfrak{a}$ a polynomial Poisson algebra. Hence, by a polynomial Poisson algebra, we mean a finitely generated Poisson $\mathbb{C}[\mathfrak{h}^*]$-algebra whose underlying commutative algebra is a quotient of a polynomial ring by a Poisson ideal.
\end{remark}

Finally, we focus on the rank of $S(\mathfrak{g})^A$. Recall that for a finitely-generated integral domain $D$ over the base field $\mathbb{F}$, the rank of $D$ is the transcendence degree defined by \begin{align}
    \mathrm{rank}_\mathbb{R} D: = \mathrm{trdeg}_\mathbb{R} D. \label{eq:rank}
\end{align} Throughout this paper, the rank of a finitely-generated algebra is defined in \eqref{eq:rank}.

Note that the observation regarding the terminology of \textit{polynomial Poisson algebra} and the rank also applies to the quantization case via the symmetrization map defined in \eqref{eq:symme} below.

 \medskip
 We now consider the setting in the universal enveloping algebra $U(\mathfrak{g})$, which can be regarded as the quantized setting. Recall that there is a well-defined canonical isomorphism (see \cite{MR432819} for details)
 \begin{align}
\nonumber
      \Lambda: S(\mathfrak{g})&\rightarrow U(\mathfrak{g}), \\
     p\left(x_1,\ldots,x_n\right) & \mapsto \Lambda\left(p\left(x_1,\ldots,x_n\right)\right) =P\left(X_1,\ldots,X_n\right)  ,
 \label{eq:symme}
\end{align} in terms of the monomial basis of $S(\mathfrak{g})$ given by
\begin{align}
    \Lambda(x_{i_1} \cdots x_{i_n}) = \frac{1}{n!} \sum_{\sigma \in S_n} X_{i_{\sigma(1)}} \cdots X_{i_{\sigma(n)}}
\end{align}
with $S_n$ being the permutation group  on the set $\{1,\ldots,n\}$, such that $\Lambda\big( \tilde{X} (p(x_1,\ldots,x_n) \big) = [X,\Lambda(p)]$ for any $X \in \mathfrak{g}$. In particular, for any $p \in S^k(\mathfrak{g}) $ and $q \in S^\ell(\mathfrak{g})$, we have \begin{align*}
    \deg \left( \Lambda(pq)\right) = k + \ell , \qquad   \Lambda(pq) -  \Lambda(p) \Lambda(q) \in U_{k+ \ell-1}(\mathfrak{g}).
\end{align*}
It follows that $\Lambda(p_h(x_1,\ldots,x_n)) =P_h(X_1,\ldots,X_n)$ will be non-commutative, functionally independent polynomials in $U(\mathfrak{g}),$ where $h$ is the degree of the generators.  Moreover, it is an isomorphism of filtered algebras. That is,   $ \Lambda_k \left( S^k(\mathfrak{g}) \right) =  U^k(\mathfrak{g})$, where $\Lambda_k = \Lambda\vert_{S^k(\mathfrak{g})} $ and $ U^k(\mathfrak{g}) = \mathrm{span}  \left\{X_1^{i_1} \cdots X_n^{i_n}: i_1+ \ldots + i_n = k\right\}$. From the Poincar\'e-Birkhoff-Witt theorem (PBW in short), it can be easily deduced that the dimension of each filtration block is
\begin{align}
    \dim  U^k(\mathfrak{g}) = \dim \frac{U_k(\mathfrak{g})}{ U_{k-1}(\mathfrak{g})} = \binom{\dim \mathfrak{g} +k-1}{k}.
\end{align}
Define \begin{align*}
    U^k(\mathfrak{g})^{\mathfrak{a}}  &= \left\{ Y \in U^k(\mathfrak{g}): [X,Y] = 0  \quad \forall X \in \mathfrak{a}\right\}.
\end{align*}
By construction, it follows that $  U (\mathfrak{g})^\mathfrak{a} = \bigoplus_{k \geq 0}   U^k(\mathfrak{g})^{\mathfrak{a}}  $ and $ S (\mathfrak{g})^\mathfrak{a} = \bigoplus_{k \geq 0}  S^k(\mathfrak{g})^{\mathfrak{a}}  . $ In particular, as mentioned above, $\Lambda_k$ also induces an algebra isomorphism between $ U^k (\mathfrak{g})^\mathfrak{a} $ and $S^k (\mathfrak{g})^\mathfrak{a}$.

Let $\hat{\textbf{q}}_k = \left\{P_{k,1} ,\ldots, P_{k,m_k} \right\} = \Lambda \big(\textbf{q}_k\big)$ be the set of all indecomposable homogeneous representatives of degree $k$, that is, $P_{k,j} = \Lambda(p_{k,j})$ for all $1 \leq j \leq m_k$, and let $\zeta$ be the maximal degree of homogeneous representatives. Define the set of indecomposable polynomials by $$  \tilde{\textbf{Q}}_\zeta = \bigsqcup_{k=1}^\zeta\tilde{\textbf{q}}_k. $$  As before, we can consider the linear space spanned by $\tilde{\textbf{Q}}_\zeta  $ with dimension $ m_1 + \ldots + m_\zeta$.   In analogy to the commutative case,  for any $P_h  \in \tilde{\textbf{q}}_h $ and $ P_\ell  \in \tilde{\textbf{q}}_\ell$, we define the commutator by
\begin{align}
 [P_h ,P_\ell ]   = \sum_{m_{r_1}+ \ldots + m_{r_k}= \ell + h -1}  \Gamma_{h,\ell}^{r_1,\ldots,r_k} P_{r_1}^{m_{r_1}} \cdots P_{r_j}^{m_{r_j}}    , \label{eq:rec}
\end{align}
where $ \Gamma_{h,\ell}^{r_1,\ldots,r_k}$ are constants. We denote the algebra generated by $\tilde{\textbf{Q}}_\zeta$ with the commutator $[\cdot,\cdot]$ as $\tilde{\mathcal{Q}}_\mathfrak{g}(d)$, where $d$ is the degree of $\tilde{\mathcal{Q}}_\mathfrak{g}(d)$. Similar to Remark \ref{rem:algebra}, we refer to $\tilde{\mathcal{Q}}_\mathfrak{g}(d)$ as a polynomial associated algebra.

From these constructions, we can define algebraic Hamiltonians and their corresponding superintegrable systems in $S\left(\mathfrak{g}\right)$, which can be considered classical superintegrable systems. Then, applying the symmetrization mapping $\Lambda$, we can deduce a quantized version of these systems.

\begin{definition}
\label{H} \cite{MR2515551}.
  Let $\mathfrak{a}^* \subset \mathfrak{g}^*$ be a Lie subalgebra with an ordered basis $\beta_{\mathfrak{a}^*} = \{x_1,\ldots,x_s\} $, and let $ \mathcal{Q}_\mathfrak{g}(d)  $ be a polynomial Poisson algebra in $S(\mathfrak{g}) $.  An algebraic Hamiltonian with respect to $\mathcal{Q}_\mathfrak{g}(d)$ is given by  \begin{align}
      \mathcal{H}  =\sum_{i_1,\ldots,i_k}^{s} \Gamma_{i_1,\ldots,i_k}x_{i_1}\cdots x_{i_k}   + \sum_t \gamma_t \mathcal{C}_t \in \mathcal{Z}\left(\mathcal{Q}_\mathfrak{g}(d)\right),\label{eq:Hamil}
 \end{align}  where $ 1 \leq i_1,\ldots,i_k \leq s$, $\Gamma_{i_1,\ldots,i_k},\gamma_t \in \mathbb{F}$, and $\mathcal{C}_t$ are Casimir invariants of $\mathfrak{g}^*$, and $\mathcal{Z}\left(\mathcal{Q}_\mathfrak{g}(d)\right)$ is the center of $\mathcal{Q}_\mathfrak{g}(d)$.
\end{definition}

\begin{remark}
\label{3.7}
 Using the symmetrization map $\Lambda$, for any $p  \in  S(\mathfrak{g})^\mathfrak{a}$, we have $P  = \Lambda(p ) \in U(\mathfrak{g})^\mathfrak{a}$ such that $[\tilde{\mathcal{H}},P ] = \Lambda(\{\mathcal{H},p   \}) = 0 $, where $\tilde{\mathcal{H}}$ is the Hamiltonian, with respect to the ordered basis of $\mathfrak{g}$, adopts the form
 \begin{align}
      \tilde{\mathcal{H}} = \sum_{i_1,\ldots,i_k}^s \Gamma_{i_1,\ldots,i_k}X_{i_1}\cdots X_{i_k}+   \sum_t \sigma_t C_t + {\rm L.O.T.}, \label{eq:quHa}
\end{align}
where $  1 \leq i_1,\ldots,i_k \leq s$, $\Gamma_{i_1,\ldots,i_k},\sigma_t\in\mathbb{F}$, $C_t$ denote the Casimir operators of $\mathfrak{g}$, and L.O.T. refers to the lower-order terms arising from expressing $\Lambda(\mathcal{H})$ on the ordered basis of $U(\mathfrak{g})$. The number of functionally independent integrals of motion related to $\tilde{\mathcal{H}}$ can be found using \eqref{eq:func}. In a physical framework, since Hamiltonians are typically associated with quadratic differential operators tied to a Schrödinger equation, $\tilde{\mathcal{H}}$ must be a quadratic differential operator.  Hence, appropriate realizations of the generators $X_i$ can be selected to ensure that the first summation in $\tilde{\mathcal{H}}$ does not produce differential operators of order greater than 2 \cite{CampoamorStursberg2021}.
\end{remark}

\section{Construction of Cartan centralisers of semisimple Lie algebras}
\label{sec:conscc}

\noindent In the following, we assume that $\mathfrak{g}$ is a rank $n$ complex semisimple Lie algebra with a Cartan subalgebra $\mathfrak{h}$.  Hence, $\mathfrak{g}$ is isomorphic to $A_n \cong \mathfrak{sl}(n+1,\mathbb{C})$, $ B_n \cong \mathfrak{so}(2n+1,\mathbb{C})$, $ C_n \cong \mathfrak{sp}(2n,\mathbb{C})$, $ D_n \cong \mathfrak{so}(2n,\mathbb{C})$, or one of the exceptional Lie algebras $G_2$, $ F_4$, $E_6, E_7$, or $E_8$. We categorize elements within the Cartan commutants for all classical semisimple Lie algebras, structuring them on the ordered basis derived from the root space decomposition.  \footnote{Commutants of the nilradical of Borel subalgebras of semisimple Lie algebras, which are deeply related to decompositions of universal enveloping algebras, have been considered in \cite{MR4710584}.}  

Let $\Phi$ be the root system associated with the complex semisimple algebra $\mathfrak{g}$, and let $\Delta = \{\beta_1, \ldots, \beta_n\}$ be a set of simple roots. In this context, the triangular decomposition of $\mathfrak{g}$ is given by $\mathfrak{g} = \mathfrak{h} \oplus \mathfrak{g}^+ \oplus \mathfrak{g}^-$, where $\mathfrak{g}^\pm = \bigoplus_{\beta \in \Phi^\pm} \mathfrak{g}_\beta$ being a one-dimensional space generated by a root vector $\mathfrak{g}_\beta$. Let $\mathfrak{h} = \mathrm{span}\{H_1,\ldots,H_n\}$. We choose root vectors $E_\beta \in \mathfrak{g}_\beta$ for each positive root $\beta \in \Phi^+$ and denote $\mathfrak{g}^+ :=\mathrm{span} \{E_1,\ldots,E_{|\Phi^+|}\}$. Similarly, $\widehat{E}_j$ for negative roots. By the Serre-Chevalley relations, we have the following:
\begin{align}
\nonumber
    [H_i,H_j] =& 0, \text{ } [E_i,E_j] = \delta_{ij}H_i,\text{ } [H_i,E_j] = a_{ij} E_j,\text{ } [H_i,\widehat{E}_j] = -a_{ij} \widehat{E}_j, \\
   &  \left(\mathrm{ad} E_i\right)^{-a_{ij} + 1}E_j =   \left(\mathrm{ad} \widehat{E}_i\right)^{-a_{ij} +1}\widehat{E}_j = 0 \label{eq:18}
\end{align}
for all $1 \leq i\neq j \leq n$.

\medskip
In what follows,  we denote the Cartan commutant by $U(\mathfrak{g})^\mathfrak{h}$. Notice that, by the PBW theorem, $U(\mathfrak{g}) \cong U(\mathfrak{h}) \otimes U^+ \otimes U^-$ is a vector space isomorphism, where $$U^+ = U(\mathfrak{g}^+) = \bigotimes_{\beta \in \Phi^+} U\left(\mathfrak{g}_\beta\right) \text{ and } U^- = U(\mathfrak{g}^-) = \bigotimes_{\beta \in \Phi^+} U\left(\mathfrak{g}_{-\beta}\right).$$ The subalgebras $U^+$ and $U^-$ are generated by $E_{\gamma_j} $ and  $\widehat{E}_{w_j(\gamma_j)}$, denoted by $E_j$ and $\widehat{E}_j$, respectively, where $w_j \in \mathcal{W}$ such that $w_j(\gamma_j) = -\gamma_j$.  By the PBW-theorem, we can write
\begin{align}
    U(\mathfrak{g}) = \mathrm{span}\left\{ H_1^{k_1} \cdots H_n^{k_n} E_1^{l_1}\cdots E_t^{l_t} \widehat{E}_1^{s_1} \cdots \widehat{E}_t^{s_t}: l_i, k_j, s_i \in \mathbb{N}_0  \right\}. \label{eq:PBW}
\end{align} Here $\mathbb{N}_0 = \mathbb{N} \sqcup \{0\}.$  Let $\boldsymbol{x} = \left(H_1,\ldots,H_n, E_1,\ldots,E_{|\Phi^+|},\widehat{E}_1,\ldots,\widehat{E}_{|\Phi^-|}\right)$. Then for any $P(\boldsymbol{x}) \in U(\mathfrak{g})$, one can write \begin{align*}
    P(\boldsymbol{x}) = & \sum_{ k_j, s_t, l_t  \in \mathbb{N}} \Gamma_{l_1,\ldots,l_t,s_1,\ldots,s_t}^{k_1,\ldots,k_n}  H_1^{k_1} \cdots H_n^{k_n} E_1^{l_1}\cdots E_t^{l_t} \widehat{E}_1^{s_1} \cdots \widehat{E}_t^{s_t},
\end{align*} where  $\Gamma_{l_1,\ldots,l_t,s_1,\ldots,s_t}^{k_1,\ldots,k_n} $ are constants and $t = |\Phi^\pm|$. As stated in Section \ref{sec:prelim}, determining the indecomposable generators of the centraliser subalgebra is easier to compute within $S(\mathfrak{g})$ due to the commutative nature of the dual elements. 

Denote the coordinates of a dual basis by $$\boldsymbol{x} = \big(h_1,\ldots,h_n,\varepsilon_1,\ldots,\varepsilon_{|\Phi^+|},\widehat{\varepsilon}_1, \ldots,\widehat{\varepsilon}_{|\Phi^-|}\big).$$  Observe that $S(\mathfrak{g}) \cong S(\mathfrak{h}) \otimes S(\mathfrak{u})$ is a vector space isomorphism, where $\mathfrak{u} = \mathfrak{g}^+ \oplus \mathfrak{g}^-$, and that $S(\mathfrak{g})^\mathfrak{h} \cong S(\mathfrak{h}) \otimes S(\mathfrak{u})^\mathfrak{h}$, as $S(\mathfrak{h})$, admits a trivial $\mathfrak{h}$-action, and $S(\mathfrak{u})$ is a $\mathfrak{h}$-module.  By definition, the kernel of the coadjoint action of $\mathfrak{h}$ on $S(\mathfrak{g})$ is given by
\begin{align}
    S(\mathfrak{g})^\mathfrak{h} =  \left\{ p \in S(\mathfrak{g}) : \text{ } \{p, h\}  = 0  \text{ for all } h \in \mathfrak{h}^* \right\}. \label{eq:PDE20}
\end{align}
 Since $S(\mathfrak{h})$ is commutative, we have that $p \in S(\mathfrak{h})$ implies $p \in S(\mathfrak{g})^\mathfrak{h}$. We now consider $p \in S(\mathfrak{u})^\mathfrak{h}$. In the analytical frame, the PDEs system corresponding to \eqref{eq:PDE20} becomes
\begin{align}
   \widetilde{H}_j(p) (\boldsymbol{x}) = \sum_{\beta \in \Phi^+} \beta  (h_j) \varepsilon_\beta \dfrac{\partial}{\partial \varepsilon_\beta} - \sum_{\beta \in \Phi^+}   \beta  (h_j) \varepsilon_{-\beta} \dfrac{\partial}{\partial \varepsilon_{-\beta}} = 0, \label{eq:PDEs}
\end{align}
where $\beta(h_j)$ is the weight of the roots $\beta$ with respect to the Cartan generators. We determine the (homogeneous) polynomial solutions of the system \eqref{eq:PDEs} of degree greater than $1$ as follows:
\begin{align}
   p(\boldsymbol{x}) =\sum_{k_j,s_t,l_t \in \mathbb{N}} \Gamma_{l_1,\ldots,l_t,s_1,\ldots,s_t}^{k_1,\ldots,k_n}  h_1^{k_1} \cdots h_n^{k_n} \varepsilon_1^{l_1}\cdots \varepsilon_t^{l_t} \widehat{\varepsilon}_1^{s_1} \cdots \widehat{\varepsilon}_t^{s_t}    \in S(\mathfrak{g})   . \label{eq:function}
\end{align}
Here, $\Gamma_{l_1,\ldots,l_t,s_1,\ldots,s_t}^{k_1,\ldots,k_n} $ are constants. Using the explicit expression of the Poisson bracket in \eqref{eq:muli} and the root decomposition,  a monomial $  \varepsilon_{i_1} \cdots \varepsilon_{i_r} \in S(\mathfrak{u})$ is a solution of \eqref{eq:PDEs} if and only if
\begin{align}
        \left(l_1  - s_1\right) a_{1j} + \cdots + \left( l_t - s_t \right)a_{tj} = 0 . \label{eq:wei}
     \end{align}
That is, the weight of the monomial with respect to each generator of the Cartan subalgebras must be zero. We consider a setting that exploits root systems instead of weights. In this context, we observe that $p  \in S(\mathfrak{g})^\mathfrak{h} $ is linearly independent and indecomposable if and only if there exists a root $\gamma_k$ such that $ \mathrm{lgh}(\gamma_k) = \max_{1 \leq j \leq r} \left\{\mathrm{lgh}(\gamma_j)\right\}$ and $\sum_j^r \gamma_j = 0$, where $\mathrm{lgh}:\Phi \rightarrow \mathbb{N}_0$ is given by $\gamma \mapsto \sum_{k=1}^r |m_k|$ with $\gamma = \sum_{k=1}^r m_k\beta_k$ being the length of a root in $\Phi$. For example, in $A_2$, $\varepsilon_\alpha \widehat{\varepsilon}_\alpha$ is indecomposable of degree $2$, $\varepsilon_\alpha \varepsilon_\beta \widehat{\varepsilon}_{\alpha + \beta}$ is indecomposable of degree $3$, etc. In fact, we find $\deg p  = \mathrm{lgh}(\gamma_k) + 1$ such that the maximal root height essentially controls the total degree. In the subsequent discussion, for any $p = \varepsilon_{i_1} \cdots \varepsilon_{i_r} \in S(\mathfrak{u})$, let the associated roots for each root vector within $p$ be denoted by $R(p) = R(\varepsilon_{i_1}) + \ldots + R(\varepsilon_{i_r}) = \gamma_{i_1} + \ldots + \gamma_{i_r}$. In particular, if $p \in S(\mathfrak{g})^\mathfrak{h}$, from \eqref{eq:wei}, $R(p) = 0$. 

\begin{definition}
 For a degree $h$ polynomial $p_h  \in S(\mathfrak{g})^\mathfrak{h}$, we call a homogeneous invariant polynomial $p_h $ $\textit{decomposable}$ if there exists some  $p_s'  \in S(\mathfrak{g})^\mathfrak{h}$ with $s < h$ such that $p  = \prod_{s \in J} p_s'  $ with an index set $|J| =h < \infty $ and $R\left(p_s' \right) = 0$.
\end{definition}
Let $\textbf{q}_k$ be the set that contains all indecomposable monomials of degree $k$ that satisfy $\{h_i,p_k \} = 0$.  We now construct the monomials for each degree. Starting with $\deg p  = 2$, for any $\beta \in \Phi^+$, it is clear that $ \varepsilon_\beta \widehat{\varepsilon}_\beta$ are degree $2$ monomials, which are contained in the set $\textbf{q}_2,$ where $\widehat{\varepsilon}_\beta := \varepsilon_{-\beta}$.  Moreover, we note that $ \left| \textbf{q}_2\right|$ is equal to $\left|\Phi^+\right|$, where $\left|\cdot\right|$ denotes the cardinality of a set. If $p$ belongs to $\textbf{q}_3$, it is clear that decomposable monomials of the form $h_i \varepsilon_\beta \widehat{\varepsilon}_\beta$ must be excluded for any $\beta \in \Phi^+$ and $1 \leq i \leq \mathrm{rank}\,\mathfrak{g}$. Thus, for any $\gamma_1, \gamma_2 \in \Phi$, an indecomposable degree $3$ monomial must be expressed as $\varepsilon_{\gamma_1} \varepsilon_{\gamma_2} \widehat{\varepsilon}_{\gamma_1 + \gamma_2}$. By recursion, for any $p  \in \textbf{q}_k$, there exists a root of maximal length $\varepsilon_{\gamma_j}$ with $\gamma_j = \sum_{i=1}^{j-1} c_i \gamma_i $ such that $p= \varepsilon_{\gamma_1} \cdots \varepsilon_{\gamma_{j-1}}\widehat{\varepsilon}_{\gamma_j}$, where $\gamma_i \in \Phi$ and $c_i$ are constants for all $i$. Moreover, if $p  \in \textbf{q}_k$, then $\hat{p}(\boldsymbol{x}) := p(\hat{\boldsymbol{x}}) \in \textbf{q}_k$ for all $k$, where $\boldsymbol{x}$ is a commuting coordinate of $\mathfrak{g}^*$. In this context, $\hat{\cdot}$ represents an operator that converts a positive root vector to its corresponding negative root vector.  Therefore, it is sufficient to express $\textbf{q}_k$ as follows:
\begin{align}
    \textbf{q}_k =    \left\{\varepsilon_{\gamma_1}^{l_1} \cdots \varepsilon_{\gamma_j}^{l_j}: \text{ } \sum_i^j l_i\gamma_i = 0, \text{ } \sum_i^j l_i = k \text{ and } l_i \in \{0,1\} \right\}  \label{eq:subs}
\end{align}
for all $k \geq 2$. In other words, for each generator $\varepsilon_{\gamma_1} \cdots \varepsilon_{-\gamma_1 - \ldots - \gamma_{k-1}} \in \textbf{q}_k$, the following relation holds$$R\left(\varepsilon_{\gamma_1} \ldots \widehat{\varepsilon}_{\gamma_1 + \ldots + \gamma_{k-1}} \right) = \gamma_1 + \gamma_2 + \ldots + \left(-\gamma_1 - \ldots - \gamma_{k-1}\right) = 0 .$$ 
 Let $\gamma$ be the highest root of $\Phi^+$. Then we have $\gamma = \lambda_1 \beta_1 + \ldots + \lambda_\ell \beta_\ell \in \Phi^+ $ such that \begin{align}
    p_{\mathrm{lgh}(\gamma) +1} = \left(\widehat{\varepsilon}_{\beta_1}\right)^{\lambda_1} \cdots  \left(\widehat{\varepsilon}_{ \beta_\ell}\right)^{\lambda_\ell} \varepsilon_{\gamma} \in S(\mathfrak{g})^\mathfrak{h}.
\end{align} Here $\lambda_i \in \mathbb{N}_0 , \beta_i \in \Delta$.  We conclude that if $\gamma_1, \ldots,\gamma_r$ are roots such that $r \geq \mathrm{lgh}(\gamma)+2$ and $\gamma_1 + \ldots + \gamma_r =0$, then there exists a reordering $\gamma_1',\ldots,\gamma_{r'}',\gamma_{r'+1}',\ldots,\gamma_r'$ such that $\gamma_1' + \ldots + \gamma_r' = 0$ and $\gamma_{r'+1}'+ \ldots + \gamma_r' = 0$.   Hence, no indecomposable generators have a degree of more than $\mathrm{lgh}(\gamma) + 1$. Therefore,  there exists a finite integer $\zeta$ such that the composition of sets
\begin{align}
   \textbf{Q}_\zeta = \left\{h_i\right\}_{i=1}^n \sqcup \bigsqcup_{k \geq 2}^\zeta\textbf{q}_k  =  \left\{ h_1,\ldots,h_n,  \varepsilon_{\gamma} \varepsilon_{-\gamma} ,\varepsilon_{\gamma_1} \varepsilon_{\gamma_2} \varepsilon_{-\gamma_1-\gamma_2},\ldots, \varepsilon_{\gamma_1} \cdots \varepsilon_{-\gamma_1 - \ldots - \gamma_{k-1}} \right\}  \label{eq:id26}
\end{align}
provides a finitely generating set for the Cartan centraliser. Let $\mathbb{C}\langle \textbf{Q}_\zeta\rangle$ denote the vector space consisting of all linearly independent homogeneous polynomial solutions of \eqref{eq:PDEs}. Using \eqref{eq:muli} and \eqref{eq:newpoi}, for any $ p = \varepsilon_{\gamma_1} \cdots \varepsilon_{\gamma_t}\widehat{\varepsilon}_{\gamma_{t+1}} \in \textbf{q}_t$ and $q = \varepsilon_{\theta_1} \cdots \varepsilon_{\theta_s}\widehat{\varepsilon}_{\theta_{s+1}} \in \textbf{q}_s$, the bilinear product
\begin{align}
    \left\{p,q\right\} = \sum_{k,l =1}^{t,s}  \left\{\varepsilon_{\gamma_k},\varepsilon_{\theta_l}\right\}\varepsilon_{\gamma_i} \prod_{i \neq k,j \neq l}\varepsilon_{\theta_j}\widehat{\varepsilon}_{\theta_{s+1}} \widehat{\varepsilon}_{\gamma_{t+1}} + \ldots \label{eq:prok}
\end{align}  is well defined and can be expressed as the sums of the products of the elements in $\mathbb{C}\langle \textbf{Q}_\zeta\rangle$, satisfying the constraint $$ R \left(\prod_{i \neq k,j \neq l} \{\varepsilon_{\gamma_k},\varepsilon_{\theta_l}\}\varepsilon_{\gamma_i} \varepsilon_{\theta_j}\widehat{\varepsilon}_{\theta_{s+1}} \widehat{\varepsilon}_{\gamma_{t+1}}\right) =0 $$ for each $k,l$, showing that the latter generates a polynomial Poisson algebra.  We denote $\mathcal{Q}_\mathfrak{g}(d)$ as the polynomial Poisson algebra formed by $\mathbb{C}\langle \textbf{Q}_\zeta \rangle$, incorporating the bilinear form $\{\cdot, \cdot\}$ described in \eqref{eq:prok}, where $d$ represents the maximum number of factors in the product (\eqref{eq:prok}). In particular, given a chain of Lie algebras $\mathfrak{g}_1 \subset \mathfrak{g}_2 \subset \ldots \subset \mathfrak{g}_t$, their corresponding polynomial Poisson algebras induce the following subspace chain $\mathcal{Q}_{\mathfrak{g}_1}(d_1) \subset \mathcal{Q}_{\mathfrak{g}_2}(d_2) \subset \ldots \subset \mathcal{Q}_{\mathfrak{g}_t}(d_t) $ with $d_1 \leq d_2 \leq \ldots \leq d_t \in \mathbb{N}$. \footnote{ See Section \ref{sec:examples} for the explicit construction of these algebra chains.}

\section{Explicit generators for \texorpdfstring{$B_n, C_n$ and $D_n$}{Bn, Cn and Dn} Cartan centralisers}
\label{sec:construction}

\noindent In this Section, we provide explicit formulae for the  centraliser generators for Lie algebras of types $B_n,C_n$ and $D_n$. In order to distinguish different polynomial Poisson algebras, we denote the polynomial Poisson algebras of type $B_n$, $D_n$, and $C_n$ by $\mathcal{Q}_{B_n}(d), \mathcal{Q}_{D_n}(d')$ and $\mathcal{Q}_{C_n}(d'')$, respectively, where $d, d',d'' \in \mathbb{N}$ denotes the degree of the corresponding polynomial Poisson algebra. We also determine the degree of $\mathcal{Q}_\mathfrak{g}(d)$ and the maximum degree of indecomposable monomials in each case. We begin our analysis with the Cartan centraliser of $B_n$ and classify all indecomposable generators of the polynomial Poisson algebra $\mathcal{Q}_{B_n}(d)$. We then observe that $\mathcal{Q}_{B_n}(d)$ contains all the Cartan commutant generators of $A_n$ and $D_n$. Similarly, $\mathcal{Q}_{C_n}(d'')$ contains the commutant generators of $A_n$. In other words, the following inclusion relations hold:
\begin{align*}
     \mathcal{Q}_{A_{n-1}}(n-1) \subset \mathcal{Q}_{D_n}(d') \subset \mathcal{Q}_{B_n}(d),\quad  \mathcal{Q}_{A_{n-1}}({n-1}) \subset \mathcal{Q}_{C_n}(d'') .
\end{align*}
In particular, $\dim_{FL} \mathcal{Q}_{B_n}(d)  = \dim_{FL}  \mathcal{Q}_{C_n}(d'')  $. From \cite{campoamor2023algebraic}, it is sufficient to deduce the higher rank Racah algebras contained in $\mathcal{Q}_{B_n}(d)$, $ \mathcal{Q}_{C_n}(d')$, and $\mathcal{Q}_{C_n}(d'')$. In order to form a superintegrable system, we still need to compute the functionally independent generators for these subalgebras. By Definition \ref{H}, the possible Hamiltonians have the form of \begin{align}
    \mathcal{H} = \sum_{i_1+\cdots + i_n \leq k} \Gamma_{i_1,\ldots, i_n} h_1^{i_1}\cdots h_n^{i_n} \in S(\mathfrak{h}),
\end{align} where $\Gamma_{i_1,\ldots,i_n}$ are constants. In the following, both the terminology and the notations are adapted from \cite{MR1920389}.

\subsection{Explicit generators of \texorpdfstring{$S(B_n)^\mathfrak{h}$}{S(Bn)h}}
\label{subsec:consono}

The Cartan subalgebra $\mathfrak{h}$ of $B_n$ is formed by elements $H = \sum_{i=1}^n a_i\left( E_{i\,i} - E_{n+i\,n+i}\right)$, where $E_{i\,i}$ is the standard basis in $GL_{2n+1}(\mathbb{R})$. In particular, the basis elements of $\mathfrak{h}$ are
\begin{align*}
H_i = E_{i\,i}- E_{n+i\,n+i} - E_{i+1\,i+1} + E_{n+i+1\,n+i+1},\quad 2 \leq i \leq n.
\end{align*}

The dual space $\mathfrak{so}^*(2n+1,\mathbb{C})$ admits the coordinates
\begin{align*}
\boldsymbol{x}_B = \left(h_1,\ldots,h_n,\ \varepsilon_{12}^-, \ldots,\varepsilon_{n-1n}^-, \  \varepsilon_{12}^+,\ldots,\widehat{\varepsilon}_{n-1n}^+,\  \varepsilon_1,\ldots,\widehat{\varepsilon}_n\right).
\end{align*}
Recall that within the root system $B_n$ in $\mathbb{R}^n$, positive roots consist of $n$ short roots $\alpha_i$ and long roots $ \alpha_i  \pm \alpha_j$ for $1 \leq i<  j \leq n$. In addition, the negative roots are represented by $\hat{\alpha}_{ij}^\pm = -\left(\alpha_i \pm \alpha_j\right)$ and $\hat{\alpha}_k$. In particular, $\hat{\alpha}_{ij}^- = \alpha_{ji}^-$. Notice that the simple roots of $B_n$ are given by $\Delta_{B_n} = \{\alpha_{ss+1}^- ,\alpha_n  : \text{ } 1 \leq s \leq n -1\} $.   The roots corresponding to those root vectors are given by \begin{align}
\nonumber
   R\left( \varepsilon_{ij}^-\right)   = & \, \alpha_{ij}^-:= \sum_{s=i}^{j-1}\alpha_{ss+1}^-  , \\
   R\left( \varepsilon_{ij}^+\right) = & \,  \alpha_{ij}^+:= \sum_{s=i}^{j-1} \alpha_{ss+1}^-+ 2\sum_{s=j}^{n-1}\alpha_{ss+1}^-+ 2 \alpha_n, \label{eq:roots}\\
   \nonumber
   R\left( \varepsilon_j\right) := & \, \sum_{s=j}^{n-1}\alpha_{ss+1}^- +   \alpha_n.
\end{align}Moreover, $\alpha_{in}^+ = \alpha_{in}^- + 2 \alpha_n$ for all $1 \leq i <n$. Since all roots are $\mathrm{ad}^* (H_i)$-diagonalizable for all $H_i \in \mathfrak{h} $, from \eqref{eq:roots}, we will deduce the following PDEs \begin{align}
     \{h_i,  \varepsilon_{jk}^-\} = \mu_{j,k}^i  \varepsilon_{jk}^-\text{ and } \{h_i,\varepsilon_j\} = \left(\mu_{j,n}^i + \mu_n^i\right) \varepsilon_j =\mu_{j,j}^i \varepsilon_j , \label{eq:mon}
 \end{align} where \begin{align}
     \mu_{j,j+1}^{j\pm 1} = - 1, \quad \mu_{j,j+1}^j = 2,  \quad \mu_n^{n-1} = -2 = -\mu_n^n,  \quad \mu_{j,j+1}^l = 0   \label{eq:val0}
 \end{align} for any $l \neq j, j\pm 1$ and $ 1 \leq j \leq n-1$.

 We now proceed to classify the generators in $\mathcal{Q}_{B_n}(d)$ for all $ n\geq 1 $.  From \cite{campoamor2023algebraic}, we know that the indecomposable monomials consisting of the root vectors in $\varepsilon_{ii+1}^-$ can be realized as  $k$-cycles in $S_n$, which are equivalent classes in the Weyl group of $A_n$. Since $\Phi_{B_n}$ contains extra long and short roots, it is reasonable to consider the Weyl group of $B_n$. That is, $\mathcal{W}_{B_n}:=S_n \rtimes \left(\mathbb{Z}_2\right)^n$. This provides an alternative approach to analyze the Cartan centraliser in $A_n$ and to classify indecomposable monomials in $\mathcal{Q}_{B_n}(d)$. Taking into account the symmetric algebra of $B_n$ for any $n \geq 1$, according to the PBW theorem and the discussion in Section \ref{sec:conscc}, $\mathcal{Q}_{B_n}(d)$ is generated by the Cartan subalgebra and the set    \begin{align}
 \nonumber
   \mathcal{Q}_{B_n}(d) = \mathfrak{h} \oplus \mathrm{span} & \left\{  \prod_{c,w,u,a,b} \Gamma_{c,w,u,a,b} \left(\varepsilon_{i_cj_c}^-\right)^{l_c}     \left(\varepsilon_{k_wl_w}^+\right)^{t_w} \left(\widehat{\varepsilon}_{k_ul_u}^+\right)^{\hat{t}_u}    \left(\varepsilon_{s_a}\right)^{\xi_a}   \left(\widehat{\varepsilon}_{s_b}\right)^{\hat{\xi}_b} : \right. \\
   &\left.\sum_{c,w,u,a,b}  \Gamma_{c,w,u,a,b} R\left(\varepsilon_{i_cj_c}^- \varepsilon_{k_wl_w}^+ \widehat{\varepsilon}_{k_ul_u}^+ \varepsilon_{s_a} \widehat{\varepsilon}_{s_b}\right)  =0\right\},  \label{eq:qd}
\end{align} where $i_s\neq j_s,k_w \neq l_w,k_u\neq l_u$ and $s_a \neq s_b$ are in $\{1,\ldots,n\}$. Here $\gamma_i \in \Phi_{B_n} $ for all $i$. Taking into account the Weyl group $\mathcal{W}_{B_n}$, we have the following definition.

\begin{definition}
\label{def:span}
   Let $\boldsymbol{x}_1 = \left(\varepsilon_{12}^-, \ldots,\varepsilon_{n-1n}^-\right)$ and $\boldsymbol{x}_2  = \left( \varepsilon_{12}^+,\ldots,\right.$ $\left.\widehat{\varepsilon}_{n-1n}^+, \varepsilon_1,\ldots,\widehat{\varepsilon}_n\right) $. Define
  \begin{align}
\nonumber
       S_n \text{-span} := & \text{ } \mathrm{span} \left\{  \left(\varepsilon_{12}^-\right)^{l_1}\cdots \left(\varepsilon_{n-1 n}^-\right)^{l_{n(n-1)}}:   \textbf{l}= \left(l_1,\ldots,l_{n(n-1)}\right) \in \mathbb{N}_0^{n(n-1)}\right\}, \\
     \mathbb{Z}_2 \text{-span}:= & \text{ }\mathrm{span} \left\{\left(\varepsilon_{12}^+ \right)^{t_1} \cdots   \left(\widehat{\varepsilon}_{n-1n}^+ \right)^{t_{2n(n-1)}}\prod_{a,b}\left(\varepsilon_a\right)^{\ell_a} \left(\widehat{\varepsilon}_b\right)^{\ell_b}: \text{ } t_w,t_u,\ell_a,\ell_b \in \mathbb{N}_0\right\}.\label{eq:id30}
\end{align}   The polynomials in $\mathcal{Q}_{B_n}(d)$ have the form of $p(\boldsymbol{x}_B) = f(\textbf{h})g(\boldsymbol{x}_1)m(\boldsymbol{x}_2) $, where $g(\boldsymbol{x}_1) \in S_n$-span and $m(\boldsymbol{x}_2) \in \mathbb{Z}_2$-span. We call the root vectors in $g(\boldsymbol{x}_1)$ as $\textit{permutation type of roots}$ and $m(\boldsymbol{x}_2)$ component as $\textit{$\mathbb{Z}_2$-type of roots}$.
\end{definition}

\begin{remark}
\label{r4.2}
 (i)  Notice that we can split the $\mathbb{Z}_2$-span into two parts. By the PBW-theorem, we can write $\mathbb{Z}_2$-span $:= \mathbb{Z}_2^L\text{-span} \otimes \mathbb{Z}_2^S\text{-span}$, where
 \begin{align*}
      \mathbb{Z}_2^S\text-\mathrm{span} =& \text{ }  \mathrm{span} \left\{\prod_{a,b}\left(\varepsilon_a\right)^{\ell_a}\left(\widehat{\varepsilon}_b\right)^{\ell_b}: \text{ }\ell_a,\ell_b \in \mathbb{N}_0 \right\}; \\
      \mathbb{Z}_2^L\text-\mathrm{span} =& \text{ } \mathrm{span} \left\{ \left(\varepsilon_{12}^+ \right)^{t_1} \cdots \left(\varepsilon_{n-1 n}^+ \right)^{t_{n(n-1)}} \left(\widehat{\varepsilon}_{12}^+ \right)^{t_{n(n-1)+1}}\cdots \left(\widehat{\varepsilon}_{n-1n}^+ \right)^{t_{2n(n-1)}}: \text{ } t_w,t_u \in \mathbb{N}_0\right\}.
  \end{align*}
 Hence, we can further decompose the root vectors of $\boldsymbol{x}_2$ into $\boldsymbol{x}_L = \left(\varepsilon_{12}^+,\ldots,\varepsilon_{n-1n}^+,\widehat{\varepsilon}_{12}^+,\ldots, \widehat{\varepsilon}_{n-1n}^+\right)$ and $\boldsymbol{x}_S = \left(\varepsilon_1,\ldots,\varepsilon_n,\widehat{\varepsilon}_1,\right.$ $\left.\ldots, \widehat{\varepsilon}_n\right)$ such that $m(\boldsymbol{x}_2) = m_1(\boldsymbol{x}_L) m_2(\boldsymbol{x}_S)$, where $m_1(\boldsymbol{x}_L) \in \mathbb{Z}_2^L$-span  and $m_2(\boldsymbol{x}_S) \in \mathbb{Z}_2^S$-span.

  (ii) Suppose that $p\left(\varepsilon_{ij}^-,\boldsymbol{x}_2\right) = \varepsilon_{ij}^- m(\boldsymbol{x}_2) \in \mathcal{Q}_{B_n}(d)$ with $m(\boldsymbol{x}_2)  \in \mathbb{Z}_2$-span and $1 \leq i \neq j \leq n$. It turns out that there exists a partition $\left(\lambda_1,\ldots,\lambda_l\right) $ such that $\varepsilon_{i\lambda_1}^-  \varepsilon_{ \lambda_1\lambda_2}^-   \cdots \varepsilon_{\lambda_lj}^- m(\boldsymbol{x}_2)  \in \mathcal{Q}_{B_n}(d)$ is indecomposable.  In what follows, we say that $g_1,g_2 \in S_n$-span is $S_n$-equivalent if $g_1 \in [g_2]$ and $p = [g_2]m$ is indecomposable. For example, take $g = \varepsilon_{ij}^-$, then all elements in \begin{align}
        \left[\varepsilon_{ij}^-\right]: = \left\{\varepsilon_{i\lambda_1}^-  \varepsilon_{ \lambda_1\lambda_2}^-   \cdots \varepsilon_{\lambda_lj}^-: 1 \leq  \lambda_j \leq n, \text{ } \lambda_j \in \mathbb{N} \right\}.
    \end{align} make $[g]m$ indecomposable that are $S_n$-equivalent to $g$. In particular, if we take the example $B_3$, we have \begin{align*}
        [13^-] \in \left\{\varepsilon_{13}^-, \varepsilon_{12}^-\varepsilon_{23}^-\right\},
    \end{align*} and   $p_{[13^-];\,12^+,\hat{23}^+;\,\widehat{1},\widehat{2}}$ means \emph{any} indecomposable product obtained by replacing $[13^-]$ with one of its representatives: \begin{align*}
        \varepsilon_{13}^- \varepsilon_{12}^+\hat{\varepsilon}_{23}^+\ \hat{\varepsilon}_1\hat{\varepsilon}_2 \text{ or }
\left(\varepsilon_{12}^-\varepsilon_{23}^-\right)\varepsilon_{12}^+\ \hat{\varepsilon}_{23}^+  \hat{\varepsilon}_1  \hat{\varepsilon}_2.
    \end{align*}  See more detailed construction in Section \ref{sec:examples}. 

(iii) By Definition \ref{def:span}, under a coordinate $\boldsymbol{x}_B = \left(h_1,\ldots,h_n,\boldsymbol{x}_1,\boldsymbol{x}_2\right)$ of $B_n$, an element in $ \mathcal{Q}_{B_n}(d)$ has the generic form of $p(\boldsymbol{x}_B) = f(\textbf{h}) g(\boldsymbol{x}_1)m_1(\boldsymbol{x}_L)m_2(\boldsymbol{x}_S)$. However, from the algebraic structures of $\mathcal{Q}_{B_n}(d)$, $S(\mathfrak{h})$ is the first layer in $\mathcal{Q}_{B_n}(d)$. Hence, if $p(\boldsymbol{x}_B)$ contains $f(\textbf{h})$ in its expression, it must be decomposable. Therefore, throughout this paper, we will study the form of $p(\boldsymbol{x}_B') = g(\boldsymbol{x}_1) m_1(\boldsymbol{x}_L)m_2(\boldsymbol{x}_S)$, where $ \boldsymbol{x}_B' =\left( \varepsilon_{12}^-, \ldots,\varepsilon_{n-1n}^-,\varepsilon_{12}^+,\ldots,\widehat{\varepsilon}_{n-1n}^-, \varepsilon_1,\ldots,\widehat{\varepsilon}_n\right) $.To avoid an abuse of notation, we  take $ \boldsymbol{x}_B':=\boldsymbol{x}_B  $ such that $p(\boldsymbol{x}_B') := p(\boldsymbol{x}_B)$. Analogous polynomial expressions are applied in the classes of $D_n$ and $C_n$. That is, we will exclude the Cartan component in the generators for the remaining cases.
\end{remark}

Define the partition of indices by \begin{align}
    S^+ = \left\{(i_s,j_s): i_s < j_s, \text{ }  1 \leq i_s,j_s \leq n\right\}  \text{ and }  S^- = \left\{(i_s,j_s): i_s > j_s, \text{ }  1 \leq  i_s,j_s \leq n\right\}. \label{eq:indices}
\end{align}  These ordered sets in \eqref{eq:indices} help us distinguish the positive or negative roots that appear in the $S_n$-span.  Using the properties of the root system, it is easy to observe the following facts:

\begin{lemma}
\label{lem:bs}
    Let $\mathcal{Q}_{B_n}(d)$ be the polynomial Poisson algebra defined above.  Following properties hold:

  (i)  For all $1 \leq i,j \leq n$, the monomials $ \varepsilon_{ij}^-\varepsilon_{ji}^-$, $\varepsilon_{ij}^+\widehat{\varepsilon}_{ij}^+$ and $\varepsilon_i \widehat{\varepsilon}_i$ are indecomposable generators that form $\textbf{q}_2$.

  (ii)  For some $r,q,t \in \mathbb{N}_0$ and all pairs in $ S^\pm$, the monomials $\prod_s^r \varepsilon_{i_sj_s}^-$, $\prod_a^q\varepsilon_{s_a}$, $\prod_w^t\varepsilon_{k_wl_w}^+$, $ \prod_a^q \varepsilon_{s_a}\left(\widehat{\varepsilon}_n\right)^q$ and $\prod_w^t\varepsilon_{k_wl_w}^+\left(\widehat{\varepsilon}_n\right)^{2t}$ are in $S(\mathfrak{g}^+)$. Therefore, they are not in $\mathcal{Q}_{B_n}(d)$.

(iii)  For any $p = g m  \in \mathcal{Q}_{B_n}(d)$, we deduce that $\alpha_n \notin R\left(p\right)$. Hence $\deg m \in 2\mathbb{N}$.
\end{lemma}

\begin{proof}
For any monomial $p$ in root coordinates, let $R(p)$ denote the sum of the corresponding roots. Then, by the definition of Cartan commutants, $p\in S(\mathfrak{g})^{\mathfrak{h}}$ if and only if $R(p)=0$. In what follows, we use \eqref{eq:roots} with the usual expressions of $\alpha_{ij}^-,\alpha_{ij}^+,\alpha_s$ in terms of the simple roots of $B_n$.

\smallskip
We first show part (i). By checking the root summations of the quadratic indecomposable generators,  we have \begin{align*}
R(\varepsilon_{ij}^-\varepsilon^-_{ji})=\alpha_{ij}^-+(-\alpha_{ij}^-)=0,\quad R(\varepsilon^+_{ij}\widehat{\varepsilon}^+_{ij})=\alpha_{ij}^++(-\alpha_{ij}^+)=0,\quad R(\varepsilon_i\widehat{\varepsilon}_i)=\alpha_i+(-\alpha_i)=0.
\end{align*} Hence, each of the generators lies in the Cartan commutant. Any nontrivial factorization of such a quadratic invariant inside $S(\mathfrak{g})^{\mathfrak{h}}$ must contain degree-one invariants in root coordinates. Hence, these elements are indecomposable and form $\mathbf{q}_2$.

\smallskip

We now show the second part by checking that each displayed monomial has $R(\cdot)\neq 0$.

(1) For $p_1 = \prod_{s=1}^r \varepsilon^-_{i_s j_s}$, the root summation is given by \begin{align*}
R (p_1)=\sum_{s=1}^r \alpha^-_{i_s j_s},
\end{align*} which is a nonnegative combination of $\{\alpha_{t,t+1}^-\}_{t=1}^{n-1}$ with no $\alpha_n$ component. Thus, we deduce $R(p_1)\neq 0$.

(2) For $p_2 = \prod_{a=1}^q \varepsilon_{s_a}$,   \begin{align*}
R(p_2)=\sum_{a=1}^q \alpha_{s_a}
\end{align*} has a positive $\alpha_n$ component and nonnegative $\alpha_{t,t+1}^-$ components, hence $R(p_2)\neq 0$.

(3) For $p_3 = \prod_{w=1}^t \varepsilon^+_{k_w \ell_w}$,
\begin{align*}
R(p_3)=\sum_{w=1}^t \alpha^+_{k_w \ell_w}
\end{align*}
has $\alpha_n$-component $2t>0$, hence $R(p_3)\neq 0$.

(4) For $p_4= \big(\prod_{a=1}^q \varepsilon_{s_a}\big)(\widehat{\varepsilon}_n)^q$, the $\alpha_n$-component in $R(p_4)$ is canceled, but the projection onto $\mathrm{span}\{\alpha_{t,t+1}^-\}$ makes $R(p_4)$ strictly positive. Hence $p_4 \notin \mathcal{Q}_{B_n}(d)$.

(5) For $p_5 = \big(\prod_{w=1}^t \varepsilon_{k_w \ell_w}^+\big)(\widehat{\varepsilon}_n)^{2t}$, the $\alpha_n$-component is again canceled, but summation of terms $\{\alpha_{t,t+1}^-\}_{t=1}^{n-1}$ are nonzero and positive. Hence $R(p_5)\neq 0$.

Therefore, none of these monomials is in $\mathcal{Q}_{B_n}(d)$.

\smallskip
Finally, we show $\alpha_n \notin R(p)$ and the parity in the $\mathbb{Z}_2$-span.  Write $p=g\,m$ with $g \in S_n$-span and $m = m_1  m_2 \in \mathbb{Z}_2$-span. By Definition \ref{def:span}, $R(g)$ contains only permutation roots $\{\alpha_{t,t+1}^-\}_{t=1}^{n-1}$. Thus, the $\alpha_n$-component of $R(p)$ equals that of $R(m)$.

Let \begin{align*}
L^+(m)= & \,\{ \varepsilon^+_{ij} :  \varepsilon^+_{ij} \text{ in } m_1\},\quad L^-(m)=\{ \widehat{\varepsilon}^+_{ij}:  \widehat{\varepsilon}^+_{ij} \text{ in } m_2\}; \\
S^+(m)= & \,\{ \varepsilon_s: \text{ } \varepsilon_s\text{ in } m_2\},\quad \text{ } \text{ } S^-(m)=\{ \widehat{\varepsilon}_s: \widehat{\varepsilon}_s \text{ in } m_2\}.
\end{align*} Then the $\alpha_n$-component of $R(m)$ is
\begin{align}
2\left(\lvert L^+(m)\rvert-\lvert L^-(m)\rvert\right)+\left(\lvert S^+(m)\rvert-\lvert S^-(m)\rvert\right). \label{eq:coef}
\end{align} If $p\in\mathcal{Q}_{B_n}(d)$, then $R(p)=0$. Hence, the expression \eqref{eq:coef} vanishes. Applying the reduction modulo $2$, we observe that \begin{align*}
\lvert S^+(m)\rvert - \lvert S^-(m)\rvert \equiv 0 \pmod{2} \quad \text{implies that} \quad
\lvert S^+(m)\rvert + \lvert S^-(m)\rvert \equiv 0 \pmod{2}.
\end{align*}This implies that the parity of both the sum and the difference in cardinality of $S^+(m)$ and $S^-(m)$ is even.  Define
\begin{align*}
\deg (m(\boldsymbol{x}_2)):=2\left(\lvert L^+(m)\rvert+\lvert L^-(m)\rvert\right)+\left(\lvert S^+(m)\rvert+\lvert S^-(m)\rvert\right).
\end{align*}
Hence, $\deg (m(\boldsymbol{x}_2))\in 2\mathbb{N}$. This shows $\alpha_n\notin R(p)$.
\end{proof}

From Definition \ref{def:span}, the generic form of a monomial in $\mathcal{Q}_{B_n}(d)$ takes the form of  \begin{align}
    p\left(\varepsilon_{i_1j_1}^-,\ldots, \widehat{\varepsilon}_{k_tl_t}^+, \varepsilon_{s_1},\ldots,\widehat{\varepsilon}_{s_q}\right) = \prod_{c=1}^r \varepsilon_{i_cj_c}^-   \prod_{w=1}^{\mu} \varepsilon_{k_wl_w}^+ \prod_{u = \mu +1}^t \widehat{\varepsilon}_{k_ul_u}^+  \prod_{a=1}^{\nu} \varepsilon_{s_a} \prod_{b=\nu+1}^q \widehat{\varepsilon}_{s_b}  .  \label{eq:41}
\end{align} In particular, define  \begin{align*}
    g\left(\varepsilon_{i_1j_1}^-,\ldots,\varepsilon_{i_rj_r}^-\right) = & \text{ } \prod_{c=1}^r \varepsilon_{i_cj_c}^-\in S_n \text{-span};
    \\   m_1\left(\varepsilon_{k_1l_1}^+,\ldots,\varepsilon_{k_\mu l_\mu}^+,\widehat{\varepsilon}_{k_{\mu+1}l_{\mu+1}},\ldots,\widehat{\varepsilon}_{k_tl_t}^+\right) = & \text{ } \prod_{w=1}^{\mu} \varepsilon_{k_wl_w}^+ \prod_{u = \mu +1}^t \widehat{\varepsilon}_{k_ul_u}^+  \in \mathbb{Z}_2^L \text{-span} ;\\
m_2\left(\varepsilon_{s_1},\ldots,\varepsilon_{s_\nu},\widehat{\varepsilon}_{\nu+1},\ldots,\widehat{\varepsilon}_q\right) =
 & \text{ } \prod_{a=1}^{\nu} \varepsilon_{s_a} \prod_{b=\nu+1}^q \widehat{\varepsilon}_{s_b}  \in \mathbb{Z}_2^S \text{-span}.
\end{align*}
Here, $r,t,q,\mu,\nu \in \mathbb{N}_0$ are undetermined constants. Roughly speaking, these constants are constrained by the ranks of Lie algebras. The precise relation between $\mathrm{rank} \,\mathfrak{g}$ and these constants will be displayed later. For any $p \in \mathcal{Q}_{B_n}(d)$, the corresponding roots have the form 
\begin{align}
\nonumber
    R\left(p \right) = & \, \sum_{c = 1}^r \alpha_{i_cj_c}^- + \sum_{w=1}^{\mu} \left(\alpha_{k_wn}^- + \alpha_{l_wn}^-\right) + \sum_{w=\mu+1}^t \left(\alpha_{nk_w}^- + \alpha_{nl_w}^-\right)  \\
    & \, + \sum_{a=1}^{\nu} \alpha_{s_an}^- +  \sum_{b= \nu+1}^q \alpha_{ns_b}^- + \left(2t - 4 \mu + 2\nu - q  \right) \alpha_n = 0. \label{eq:r41ida}
\end{align}  with the constraint $2t - 4 \mu + 2\nu - q  = 0$.  Therefore, to satisfy the identity \eqref{eq:r41ida}, we have to evaluate all possible linear combinations of roots so that $\alpha_n$ is canceled. This amounts to determine \begin{align}
  \alpha_n \notin R(m) = \sum_{w=1}^t\gamma (m_w)\alpha_{k_w l_w}^+ +   \sum_{a=1}^q\gamma(\ell_a)\alpha_{s_a}      \label{eq:all}
\end{align} such that $\sum_{c=1}^r\alpha_{i_c j_c}^-  =-  \left( \sum_{w=1}^t\gamma(m_w)\alpha_{k_w l_w}^+ +   \sum_{a=1}^q\gamma(\ell_a)\alpha_{s_a} \right) $ holds, where the terms $\gamma(m_w) , \gamma(\ell_a) \in \mathbb{F} $.

To complete the classification, five distinct scenarios for $p   \in \mathcal{Q}_{B_n}(d) $ have to be considered:

\smallskip
\begin{center}
    \textbf{(a)} $p$ equals to either $g$, $m_1$ or $m_2;$ \quad \textbf{(b)} $p = gm_1;$ \quad \textbf{(c)} $p =g m_2$;\quad  \textbf{(d)} $p = m_1 m_2$; \quad \textbf{(e)} $p = g m_1 m_2$.
\end{center}

\smallskip
To establish the classification, it is essential to confirm that $R(p) = 0$ holds for any $p$ in the cases above. Moreover, the constraint \eqref{eq:all} provides the generic forms of $m_1$ and $m_2$, respectively (see, e.g., \eqref{eq:classif} and \eqref{eq:classif2} in Section \ref{sec:examples}.) Depending on the form of $p$, we need to examine all possible permutation roots that encompass the roots in $$R(m_1) = \sum_{w=1}^{\mu} \left( \alpha_{k_wn}^- + \alpha_{l_wn}^- \right) + \sum_{w=\mu+1}^t \left( \alpha_{nk_w}^- + \alpha_{nl_w}^- \right) + (2\mu - 2t) \alpha_n , \text{ }  R(m_2) = \sum_{a=1}^{\nu} \alpha_{s_an}^- + \sum_{b=\nu+1}^q \alpha_{ns_b}^- + (\nu - q)\alpha_n,$$ and $R(m) = R(m_1) +R(m_2)$, respectively.

\medskip
Alternatively to using the coordinate argument to express indecomposable monomials, we now introduce a representative formula using the indices $(i_c j_c),(k_w l_w)$, $(k_u l_u)$,  $s_a$, and $s_b$ in each of the corresponding root vectors. That is, the generic formula of indecomposable monomials can be rewritten in the indices argument as follows:  \begin{align}
    p_{[i_1j_1^-],\ldots,[i_rj_r^-];\ k_1l_1^+,\ldots, \widehat{k_tl_t}^+;\ s_1,\ldots,s_\nu,\widehat{s}_1,\ldots,\widehat{s}_q}. \label{eq:indrep}
\end{align} Here, $[i_1j_1^-],\ldots,[i_rj_r^-]$ represents all the $S_n$-equivalent permutation root vectors, $k_1l_1^+,\ldots, \widehat{k_tl_t}^+$ and $s_1,\ldots,s_\nu$, $\hat{s}_1,\ldots,\hat{s}_q$ represent long and short root vectors, respectively (see Remark \ref{r4.2} (ii)). Note that the double comma $;$ in \eqref{eq:indrep} divides each type of root vector. If the indecomposable monomial $p$ contains the root vectors (the same index label), we express them  by $ p_{(\cdot)^k;\ldots;\ldots}$, where $k$ is the number of repetitions of the root vector. Moreover, the expression \eqref{eq:indrep} is well-defined, as the order of indices in each section does not change the form of $p$.  The following Table \ref{tab:indexstrings} illustrates the situation for $1 \leq i,j,k \leq n$.

\begin{table}[htbp]
\begin{tabular}{|c|c|}
    \hline
    \textbf{Index string} & \textbf{Actual product} \\ \hline
    $p_{ij^-,ji^-}$ & $\varepsilon_{ij}^-\,\varepsilon_{ji}^-$ \\
    \hline
    $p_{ij^+,\widehat{ij}^+}$ & $\varepsilon_{ij}^+\,\hat{\varepsilon}_{ij}^+$ \\
    \hline
    $p_{i,\hat{i}}$ & $\varepsilon_i\,\hat{\varepsilon}_i$ \\
    \hline
      $p_{[l i^-];ij^+,\widehat{kl}^+;\hat{j},\hat{k}}$  &$\left[\varepsilon_{li}^-\right] \varepsilon_{ij}^+\widehat{\varepsilon}_{kl}^+\widehat{\varepsilon}_j\widehat{\varepsilon}_k$ \\
      \hline
  \end{tabular}
  \vskip 0.2cm
    \caption{Dictionary between index strings $p_{\cdots}$ and the corresponding  monomials in root coordinates.}
  \label{tab:indexstrings}
\end{table}  
In what follows, we will use the indices argument to express indecomposable monomials.

 \begin{proposition}
 \label{prop:pcase}
    Let\begin{align}
        p\left(\varepsilon_{i_1j_1}^-,\ldots,\varepsilon_{k_tl_t}^+,\ldots,\widehat{\varepsilon}_n\right) = \prod_{c=1}^r \varepsilon_{i_cj_c}^- \prod_{w=1}^t \varepsilon_{k_wl_w}^+ \prod_{a=1}^q \varepsilon_{s_a} \left(\widehat{\varepsilon}_n\right)^{q +2t}  \in \mathcal{Q}_{B_n}(d)    \label{eq:as1.5}
      \end{align} with $s_a,k_w,l_w \neq n$ for all $w,a$. Then $p $ is decomposable if one of the following conditions holds:
     \begin{center}
            (a) $t = 0$ and $q > 2;$ \qquad (b) $t > 1 $ and $q =0;$  \qquad  (c)  all $r,t,q \neq 0$ with $2t + q > 2 $.
     \end{center} In particular, \begin{align}\
      \begin{matrix}
            p_{[i_1i_r^-], i_ri_1^-}   , \text{ } p_{nj_1^-,nj_2^-;\widehat{j_1j_2}^+;\hat{n}^2}  \text{ and } p_{[s_1n^-];s_1,\hat{n}}
      \end{matrix}
      \end{align}  are indecomposable. Here, $    1 \leq i_1,j_1,j_2,i_r,s_1,s_2  \leq n $.
\end{proposition}

\begin{proof}
By definition,  \begin{align}
R\left(p\left(\varepsilon_{i_1j_1}^-,\ldots,\varepsilon_{k_tl_t}^+,\ldots,\widehat{\varepsilon}_n\right)\right) = \sum_{c=1}^r \alpha_{i_cj_c}^-  +  \sum_{w=1}^t \left(\alpha_{k_wn}^-+    \alpha_{l_wn}^-\right)  +  \sum_{a=1}^q\alpha_{s_an}^-    = 0. \label{eq:as1}
\end{align} Here $r,t ,q \in \mathbb{N}_0$ are finite. In what follows, we restrict the constants $r,t,q$ in the identity \eqref{eq:as1}.

(i) If $t = q =0$ and $r \neq 0$, that is, $ p=  \prod_{c=1}^r \varepsilon_{i_cj_c}^- \in \mathcal{Q}_{B_n}(d) \cap S_n$-span. By \cite[Lemma 1]{campoamor2023algebraic}, there exist a partition of indices $i_1 < j_1 = i_2 < j_2 = i_3 < \ldots < j_{r-1} = i_r  < j_r $ such that $p$ are non-Cartan parts of generators for $\mathcal{Q}_{A_{n-1}}(n-1)$, and $p =  p_{[i_1i_r^-], i_ri_1^-} $. In contrast, if $r = 0$ and $t, q \neq 0$, then the $p  = \prod_{w=1}^t \varepsilon_{k_wl_w}^+ \prod_a^q \varepsilon_{s_a} \left(\widehat{\varepsilon}_n\right)^{q +2t}$. Lemma \ref{lem:bs} implies that $p   \notin \mathcal{Q}_{B_n}(d)$.

(ii) If $t=0$ and $q,r \neq 0$, then identity \eqref{eq:as1.5} becomes  \begin{align}
        p  = \prod_{c=1}^r \varepsilon_{i_cj_c}^-   \prod_a^q \varepsilon_{s_a} \left(\widehat{\varepsilon}_n\right)^q  \quad  \Longrightarrow R\left(p \right) = R\left(\prod_{c=1}^r \varepsilon_{i_cj_c}^-  \right) + \sum_{a}^q \alpha_{s_a,n}^- = 0.  \label{eq:as22}
      \end{align}  Now assume that $q = 1$. We then have $ R(p) = \sum_{c=1}^r \alpha_{i_cj_c}^- + \alpha_{s_1n}^- = 0$. Without loss of generality, we find a partition of indices as follows: \begin{align}
          j_1 = s_1 <  i_1 = j_2 < i_2 = \ldots = j_r < i_r = n.  \label{eq:partition}
      \end{align}   Returning to identity \eqref{eq:as1}, we deduce $-\alpha_{s_1n}^- + \alpha_{s_1} - \alpha_n = 0$. Hence, $p_{[s_1n]^-;s_1,\hat{n}} \in \mathcal{Q}_{B_n}(d)$.  We now show that \eqref{eq:as22} leads to a decomposable monomial if $q \geq 2$. Assume that $q =2$. From the above argument, there exists a partition similar to \eqref{eq:partition} in the pair $ \{(j_ci_c): 1 \leq j_c < i_c \leq n\}  $ covering $(s_1n) $ and $(s_2n) $ such that $p  =p_{[s_1n]^-;s_1,\hat{n}}\ p_{[s_2n]^-;s_2,\hat{n}} $.  Hence, $p$ is decomposable if $q \geq 3$.

 (iii)  If $q =0$ and $t,r \neq 0$. Then $ p\big(\boldsymbol{x}_1,\boldsymbol{x}_L^+,\widehat{\varepsilon}_n\big) = \prod_{c=1}^r \varepsilon_{i_cj_c}^-   \prod_w^t \varepsilon_{k_wl_w}^+ \left(\widehat{\varepsilon}_n\right)^{2t} \in \mathcal{Q}_{B_n}(d) $ and   \begin{align}
    R\left(  p\big(\boldsymbol{x}_1,\boldsymbol{x}_L^+,\widehat{\varepsilon}_n\big)\right) = \sum_{c=1}^r \alpha_{i_cj_c}^- + \sum_w^t\left(\alpha_{k_wn}^- + \alpha_{l_wn}^-\right) = 0.\label{eq:a}
\end{align}Take $t =1$. We need to find suitable values in pairs $(i_c j_c)$ such that $ \sum_{c=1}^r \alpha_{i_cj_c}^- +\alpha_{k_1n}^- + \alpha_{l_1n}^- = 0$ and $p\big(\boldsymbol{x}_1,\boldsymbol{x}_L^+,\widehat{\varepsilon}_n\big)$ are indecomposable. Using Remark \ref{r4.2}  (ii), the only possible choice of $(i_c j_c)$ that satisfies these conditions is \begin{align*}
   k_1 = j_{r'} < i_c \leq j_{c-1} < \ldots \leq i_2 < i_1 = n \text{ and } l_1  =j_r < i_r \leq j_{r-1} < \ldots \leq j_{k+1} < i_{r'+1} = n,
\end{align*} where $1 \leq r' \leq r$. So, $p_{[nj_1^-],[nj_2^-];j_1j_2^+;\hat{n}^2} \in \mathcal{Q}_{B_n}(d)$. Also let $t \geq 2$. According to \eqref{eq:a}, we have $R\left(p\right) = \sum_{c=1}^r \alpha_{i_cj_c}^- + \left(\alpha_{k_1n}^- + \alpha_{l_1n}^-\right) + \ldots + \left(\alpha_{k_tn}^- + \alpha_{l_tn}^-\right) = 0$. As a consequence of this, for each $ \alpha_{k_wn}^- + \alpha_{l_wn}^-$, we are able to find a partition in $\sum_{c=1}^r \alpha_{i_cj_c}^-$ such that $p = \prod_{w=1}^t p_{[nk_w]^-,[nl_w]^-;k_wl_w^+;\hat{n}^2}$.

(iv) Suppose that both $r \neq  0$ and $t,q \neq 0$. From the identity \eqref{eq:as1}, it is clear that $(k_w n),(l_w n), (s_a n) \in S^+$ for all $w,a$. Thus, if $(i_c j_c) \in S^+$ for all $ 1 \leq c \leq r$, $p \notin \mathcal{Q}_{B_n}(d)$. Hence, there must exist some pairs $(i_c j_c) \in S^- $ such that \begin{align}
     \sum_{c=1}^{r'} \alpha_{i_cj_c}^-  +  \sum_{w=1}^t\alpha_{k_wn}^-+   \sum_{w=1}^t\alpha_{l_wn}^-  +  \sum_{a=1}^q\alpha_{s_an}^-    =  \sum_{c=r'+1}^r \alpha_{j_ci_c }^-  , \label{eq:39}
\end{align} where $r' = \left|S^-\right|$. Notice that if $r' \neq 0$, there must exist a negative root $\alpha_{j_{c'}i_{c'}}^-$ such that $ \alpha_{j_{c'}i_{c'}}^- = \sum_{c=1}^{r'} \alpha_{i_cj_c}^-$. By Lemma \ref{lem:bs}, we deduce a $p'  = \left[\varepsilon_{i_1j_{r'}}^-\right] \varepsilon_{j_{r'}i_1}^-  $ such that $p  \equiv 0 \mod p' $. Hence $r' =0$, there exists a partition of the negative permutation roots such that the positive long and short roots are fully covered. In other words, we induce a partition of indices such that \begin{align*}
\begin{matrix}
     k_w = j_1 < \ldots < i_{c_1} = n , \quad l_w = j_{c_1} < \ldots < i_{c_2} =n, \quad s_a = j_{c_2} < \ldots < i_r =n.
\end{matrix}
\end{align*} Here $c_i \neq c_j \in \{1,\ldots,r\}$.   From the conclusions in (ii) and (iii) above, we conclude that the identity \eqref{eq:39} contains the following indecomposable monomials \begin{align*}
  p_{[ns_a^-]; s_a,\hat{n}} =  \left[\varepsilon_{ns_a}^-\right]\varepsilon_{s_a}  \widehat{\varepsilon}_n, \quad \quad \text{ } p_{[nk_w^-],[nl_w^-]; k_wl_w^+ ;\hat{n}^2} = \left[\varepsilon_{nk_w}^-\right] \left[\varepsilon_{nl_w}^-\right] \varepsilon_{k_wl_w}^+\left(\widehat{\varepsilon}_n\right)^2
\end{align*}  for all $w,a$. Therefore, condition (c) holds.
\end{proof}

The five types \textbf{(a)}-\textbf{(e)} of monomials $p\in \mathcal{Q}_{B_n}(d)$  constitute a tool to separate permutation-type monomials $g(\boldsymbol{x}_1)$, built from $\varepsilon_{ij}^-$, from the $\mathbb{Z}_2$-type monomials $m_1(\boldsymbol{x}_L),m_2(\boldsymbol{x}_s)$, built from long and short roots, respectively (Definition \ref{def:span} and Remark \ref{r4.2}). The $\mathfrak{h}$-invariance property reduces to the root-balance equation $R(p)=0$, which forces the $\alpha_n$-component to cancel and gives the constraint \begin{align*}
2t-4\mu+2\nu-q=0
\end{align*} for $p=g\,m_1\,m_2$ with the notations \eqref{eq:indices} to \eqref{eq:41}. Thus, the general classification reduces to a combinatorial problem on $W_{B_n}=S_n\rtimes(\mathbb{Z}_2)^n$-orbits: We must partition the permutation indices such that $R(m_1)$ (long roots) and $R(m_2)$ (short roots) are exactly covered while excluding products that factor through lower-degree generators. The same algorithm applies to $D_n$, where the balance condition is enforced by $t=2\mu$, and to $C_n$ (see Section \ref{subsec:conson} below). Although Section \ref{sec:examples} provides a complete list of three rank levels and illustrates the inclusion relations in $\mathcal{Q}_{A_{n-1}}(n-1), \mathcal{Q}_{D_n}(d'), \mathcal{Q}_{B_n}(d) $ and $ \mathcal{Q}_{C_n}(d'')$, a complete all-rank classification demands a systematic treatment of these integer-lattice constraints, together with the $S_n/W_{B_n}$-orbit structure. Here $d, d',d''$ are the degrees of the polynomial Poisson algebras. Hence, the analysis of the classification of indecomposable monomials is technically involved and will be presented in a forthcoming work \cite{abc}.

\subsection{Explicit generators of \texorpdfstring{$S(C_n)^{\mathfrak{h}}$}{S(Cn)h} and \texorpdfstring{$S(D_n)^{\mathfrak{h}}$}{S(Dn)h}} \label{subsec:conson}

As in Subsection \ref{subsec:consono}, we fix Cartan centralisers and express any element as $p=g\,m_L\,m_S$, where $g$ is a product of permutation root vectors, $m_L$ is a product of long root vectors, and $m_S$ is a product of short-root vectors (indices always lie in $\{1,\dots,n\}$ with obvious distinctness conditions). In this parameterization, the exponents are taken within finite ranges \begin{align*}
   t,\mu,\nu,q\in\mathbb{N}_{0},\qquad 0\leq \mu\leq t,\qquad 0\leq \nu\leq q,\qquad \deg p=r+t+q\leq \zeta .
\end{align*} So, in particular, $t+q\leq \zeta $, where $\zeta $ is the maximal degree of indecomposable generators in $S(\mathfrak{g})^{\mathfrak{h}}$. Hence, only finitely many indecomposable monomials occur. For type $B_n$, the zero-weight balance reads: \begin{align*}
    2t-4\mu+2\nu-q=0,
\end{align*} while for type $C_n$, it becomes
\begin{align*}
2t-4\mu+4\nu-2q=0,
\end{align*}
due to the double weight of the short roots. The $D_n$ classification is contained in the $B_n$ case, restricting to the long root subsystem forces $q=\nu=0$ and $t=2\mu$, so the same construction applies literally. With these bounds and constraints, the five families obtained for $B_n$ transfer to $C_n$ through the standard identification of the permutation and $\mathbb{Z}_2$ roots, while $D_n$ is the long root cut within $B_n$. Thus, the lists above give a complete way to deduce the finite set of indecomposable generators of $S(D_n)^{\mathfrak{h}}$ and $S(C_n)^{\mathfrak{h}}$. We begin with the case $S(D_n)^\mathfrak{h}$.

\subsubsection{Generators of $S(D_n)^\mathfrak{h}$}
A Cartan subalgebra of $D_n$ is defined by $$\mathfrak{h} = \left\{\mathrm{diag}(a_1,\ldots,a_n,-a_1, \ldots, -a_n): a_i \in \mathbb{R},\text{ } 1 \leq i \leq n \right\}.$$  For any $H \in \mathfrak{h}$, we can write $H = \sum_{j=1}^n a_j\left( E_{j\,j} -  E_{n+j\,n+j} \right).$ A direct computation shows that \begin{align}
\begin{matrix}
     [H_i,E_{i\,i+1}]  = 2E_{i\,i+1} , \text{ } [H_{i-1},E_{i  \,i+1}] = -E_{i \,i+1} , \text{ } [H_{i+1},E_{i\,i+1}] = -E_{i\,i+ 1} ,\\
 [H_{n-2},E_{n-2\,n-1}]   = -E_{n-2\,n-1} ,\text{ }    [H_{n-2},E_{n-1\,n }]  = -E_{n-1\,n } , \text{ } [H_n,E_{n-1\,n}] = 0 = [H_j,E_{i\,i+1}] ,
\end{matrix} \label{eq:eigen22}
\end{align} where $j \neq i \pm 1$ and $i = 1,\ldots,n-2.$ The dual space $\mathfrak{so}^*(2n,\mathbb{C})$ admits a coordinate $$\boldsymbol{x}_D = \left(h_1,\ldots,h_n,\varepsilon_{12}^-,\ldots,\varepsilon_{n-1n}^-,\varepsilon_{12}^+,\ldots,\widehat{\varepsilon}_{n-1n}^-\right).$$   The corresponding roots to the root vectors are
\begin{align}
    R\left(\varepsilon_{ij}^-\right) = \sum_{s =i }^{j-1} \alpha_{ss+1}^-\text{ and } R\left(\varepsilon_{ij}^+\right) =   \alpha_{in-1}^- + \alpha_{jn-1}^- + \alpha_{n-1n}^- +  \alpha_{n-1} + \alpha_n. \label{eq:so2n}
\end{align}
We observe that $\alpha_{n-1 n}^+ = \alpha_{n-1} +  \alpha_n$ and $\alpha_{i n-1}^+ = \alpha_{n-1 n}^- + \alpha_{i n}^-+\alpha_{n-1}    +\alpha_n $.  Notice that $\Delta_{D_n} = \left\{\alpha_{ss+1}^-,\alpha_n + \alpha_{n-1}\right\}_{s=1}^{n-1}$ forms the basis for the root system of $D_n$.  Considering the $\mathfrak{h}$-invariant symmetric space of $D_n$, we need to find the set of generators such that the Cartan commutant of $D_n$ is of the form:
\begin{align}
    \mathcal{Q}_{D_n}(d') = \mathfrak{h}^* \oplus \mathrm{span} \left\{  \left(\varepsilon_{i_sj_s}^-\right)^\textbf{l}     \left(\varepsilon_{k_wl_w}^+\right)^{\textbf{m}} \left(\widehat{\varepsilon}_{k_ul_u}^+\right)^{\hat{\textbf{m}}}      : \sum_{s,w,u,a,b}  \Gamma_{s,w,u} R\left(\varepsilon_{i_sj_s}^- \varepsilon_{k_wl_w}^+ \widehat{\varepsilon}_{k_ul_u}^+  \right)  =0\right\},\label{eq:so2nge}
\end{align}      where $ \textbf{l},\textbf{m} $ and $\hat{\textbf{m}} $ are integer tuples.  Hence, similar to the permutations and long roots in $\Phi_{B_n}$, we separate the monomials into the tensor product of distinct root vectors.

\begin{definition}
\label{spanso}
   Let $\boldsymbol{x}_1= \left(\varepsilon_{12}^-,\ldots,\varepsilon_{nn-1}^-\right)$ and $\boldsymbol{x}_L = \left(\varepsilon_{12}^+,\ldots,\widehat{\varepsilon}_{n-1n}^+\right)$ be the coordinates. For any $p(\boldsymbol{x}_D) = g(\boldsymbol{x}_1)m_1(\boldsymbol{x}_L)$ $  \in \mathcal{Q}_{D_n}(d)$, suppose that $g(\boldsymbol{x}_1) \in S_n$-span and $$   \mathbb{Z}_2^L\text-\mathrm{span} = \text{ } \mathrm{span} \left\{ \left(\varepsilon_{12}^+ \right)^{t_1} \cdots \left(\varepsilon_{n-1 n}^+ \right)^{t_{n(n-1)}} \left(\widehat{\varepsilon}_{12}^+ \right)^{t_{n(n-1)+1}}\cdots \left(\widehat{\varepsilon}_{n-1n}^+ \right)^{t_{2n(n-1)}}: \text{ } t_w,t_u \in \mathbb{N}_0\right\}.  $$ We call the subspaces spanned by $m_1(\boldsymbol{x}_L)$ as $\textit{$\mathbb{Z}_2^L$-span}$.
\end{definition}

By Definition \ref{spanso} and \eqref{eq:so2nge}, the monomial in $\mathcal{Q}_{D_n}(d')$ has the form of \begin{align}
p\left(\varepsilon_{i_1j_1}^-,\ldots,\varepsilon_{i_rj_r}^-,\varepsilon_{k_1l_1}^+,\ldots ,\widehat{\varepsilon}_{k_tl_t}^+\right)  = g\left(\varepsilon_{i_1j_1}^-,\ldots,\varepsilon_{i_rj_r}^-\right)   m_1\left(\varepsilon_{k_1l_1}^+,\ldots, \widehat{\varepsilon}_{k_tl_t}^+\right) ,
\end{align} where $g\left(\varepsilon_{i_1j_1}^-,\ldots,\varepsilon_{i_rj_r}^-\right)  =\prod_{s=1}^r \varepsilon_{i_sj_s}^- \in S_n$-span and $m_1\left(\varepsilon_{k_1l_1}^+\ldots \widehat{\varepsilon}_{k_tl_t}^+\right) =  \prod_{w=1}^{\mu}\varepsilon_{k_wl_w}^+\prod_{u=\mu+1}^t \widehat{\varepsilon}_{k_ul_u}^+ \in \mathbb{Z}_2^L$-span. Since $R(p) = 0$, we must have $R(g) = -R(m)$. Hence, \begin{align*}
    R(g ) = -\sum_{w=1}^{\mu} \left(\alpha_{k_wn-1}^- + \alpha_{l_wn-1}^-\right) + \sum_{u=\mu+1}^t \left(\alpha_{k_un-1}^- + \alpha_{l_un-1}^-\right) + (t - 2\mu) \left(\alpha_{n-1n}^- + \alpha_{n-1} + \alpha_n\right).
\end{align*}  Analogously to Lemma \ref{lem:bs} in the $B_n$ case, we must determine every possible combination of permutation roots in $R(m_1)$ that results in the vanishing of the constant term $\alpha_{n-1n}^- + \alpha_{n-1} + \alpha_n$. Hence $t = 2\mu$.  Similar to Section \ref{subsec:consono}, all types of indecomposable monomials of type \textbf{(a)} and a partial classification of type \textbf{(b)} are given below:

\begin{theorem}
   For any $n\geq 2$, $p\in\mathcal{Q}_{D_n}(d')$ admits a decomposition \begin{align*}
    p=g\,m_1,\qquad g\in S_n\text{-span}, \text{ }  m_1\in\mathbb{Z}_2^L\text{-span},
\end{align*} as stated in Definition \ref{spanso}. Then, the following is satisfied: 
 \begin{itemize}
\item[(a)] If $p\in\mathcal{Q}_{D_n}(d')\cap S_n\text{-span}$ or $p\in\mathcal{Q}_{D_n}(d')\cap \mathbb{Z}_2^L\text{-span}$, then the inclusion $\mathcal{Q}_{A_{n-1}}(n-1)\subset \mathcal{Q}_{D_n}(d')$ holds. Moreover, for any integer $\mu\geq 1$ and $t=2\mu$ and any indices $k_1,l_2,\ldots,l_t\in\{1,\dots,n\}$, \begin{align*}
m_1 \left(\varepsilon_{k_1l_2}^+ ,\ldots ,\widehat{\varepsilon}_{l_tk_1}^+ \right) = \varepsilon_{k_1l_2}^+ \,\widehat{\varepsilon}_{l_2l_3}^+\cdots \varepsilon_{l_{t-1}l_t}^+\,\widehat{\varepsilon}_{l_tk_1}^+ \in \mathbf{q}_{2h}
\end{align*} lies in $\mathcal{Q}_{D_n}(d')\cap \mathbb{Z}_2^L\text{-span}$ and is indecomposable.
\item[(b)] Let $\mathcal{SK}_1:=\left(S_n\text{-span}\otimes \mathbb{Z}_2^L\text{-span}\right)\cap \mathcal{Q}_{D_n}(d')$ be a vector subspace. Suppose that $m_1 \in \mathbb{Z}_2^L$-span is indecomposable, and further assume that $R(p) = 0$ with $ p= gm_1 $ and $0 \neq g \in S_n$-span. Then, for every $n\geq 3$, there exist integers $r\geq 1$, $\mu\geq 1$ (with $t=2\mu$) and indices \begin{align*}
    1\leq  i_s,j_s,k_w,l_w \leq n \quad \text{with } 1\leq s\leq r, 1\leq w\leq 2\mu
\end{align*} such that \begin{align}
p = \left(\prod_{s=1}^r \varepsilon_{i_sj_s}^- \right)\left(\prod_{w=1}^{\mu}\prod_{u=\mu+1}^{2\mu}   \varepsilon_{k_wl_w}^+  \widehat{\varepsilon}_{k_ul_u}^+ \right)\in \mathcal{SK}_1 \label{eq:gm1factor}
\end{align} is indecomposable. 
\end{itemize}

\end{theorem}

 \begin{proof}
     We use the root summation formula $R(\,\cdot\,)$. For $m_1$ with long root vectors unhatted (positive) $\mu$ and $t-\mu$ hatted (negative), we have the explicit identity \begin{align*}
R(m_1) =\sum_{w=1}^{\mu}\left(\alpha_{k_w n-1}^-+\alpha_{l_w,n-1}^-\right) -\sum_{u=\mu+1}^t\left(\alpha_{k_u n-1}^-+\alpha_{l_u n-1}^-\right) +(t-2\mu)\,\left(\alpha_{n-1,n}^-+\alpha_{n-1}+\alpha_n\right).
\end{align*} Thus, the Cartan commutant condition $R(p)=R(g)+R(m_1)=0$ forces $t=2\mu$, as $R(g)$ does not have a constant $C:=\alpha_{n-1,n}^-+\alpha_{n-1}+\alpha_n$ component.

\medskip

We first show that part (a) holds. This splits into two parts: On the one hand, if $p\in S_n\text{-span}$, then $m_1=1$ and $R(p)$ is a linear combinations of permutation roots only. The commutant is exactly the $A_{n-1}$-type, and hence $\mathcal{Q}_{A_{n-1}}(n-1)\subset \mathcal{Q}_{D_n}(d')$.

On the other hand, if $p\in \mathbb{Z}_2^L\text{-span}$, then $g=1$ and $t=2\mu$. We first show that $m_1\in \mathbb{Z}_2^L\text{-span}$ has no nontrivial monomial inside $\mathbb{Z}_2^L\text{-span}$.  Fix integers $t,\mu$ with $t=2\mu\geq 2$, and indices \begin{align*}
1\le k_1,l_2,l_3,\dots,l_t\le n,\qquad l_{t+1}:=k_1,
\end{align*} such that all pairs of indices $(k_1,l_2),\,(l_2,l_3),\dots, (l_t,k_1)$ are allowed. Define the alternating monomial by
\begin{align}
m_1\left(\varepsilon_{k_1l_2}^+,\ldots,\widehat{\varepsilon}_{l_tk_1}^+\right):=\varepsilon_{k_1l_2}^+\,\widehat{\varepsilon}_{l_2l_3}^+\cdots \varepsilon_{l_{t-1}l_t}^+\,\widehat{\varepsilon}_{l_tk_1}^+. \label{eq:m1factor}
\end{align}
 Consider the decomposition of $R(m_1)$ into the $\alpha^-_{\bullet,n-1}$-roots (permutation roots). Using the identities \eqref{eq:so2n} and applying them to \eqref{eq:m1factor}, we directly check that the $\alpha_{\bullet,n-1}^-$-projection of $R(m_1)$ telescopes as follows:
\begin{align}
\left(\alpha^-_{k_1,n-1}+\alpha^-_{l_2,n-1}\right)-\left(\alpha^-_{l_2,n-1}+\alpha^-_{l_3,n-1}\right)+\cdots+\left(\alpha^-_{l_{t-1},n-1}+\alpha^-_{l_t,n-1}\right)-\left(\alpha^-_{l_t,n-1}+\alpha^-_{k_1,n-1}\right)=0. \label{eq:telescoping}
\end{align}
Together with $t=2\mu$, this yields $R(m_1)=0$, and hence $m_1\in \mathcal{Q}_{D_n}(d')\cap \mathbb{Z}_2^L\text{-span}$.  By construction, each index $l_j$ with $2\leq j\leq t$ occurs exactly twice among the root vectors $\varepsilon_{ij}^+$ and $\widehat{\varepsilon}_{kl}^+$.   Hence $R(m_1) = 0$, and therefore $m_1\in \mathcal{Q}_{D_n}(d')\cap \mathbb{Z}_2^L\text{-span}$.

To prove the indecomposability of $m_1$ in \eqref{eq:m1factor}, suppose that $m_1=u\cdot v$ with both $u,v\in \mathbb{Z}_2^L\text{-span}$ are non-units. Let $I\subset\{1,\ldots,t\}$ be the index set of the root vectors of $m_1$ that appear in $u$. Since $R$ is additive on products and the $\alpha_{\bullet,n-1}^-$-contribution of each root vector of $m_1$ is $\alpha_{a,n-1}^-+\alpha_{b,n-1}^-$ or $-(\alpha_{a,n-1}^-+\alpha_{b,n-1}^-)$ depending on whether the root vector is $\varepsilon_{ab}^+$ or $\widehat{\varepsilon}_{ab}^+$, the root $R(u)$ in terms of $\alpha_{\bullet,n-1}^-$ is equal to: \begin{align}
\sum_{s\in I}\sigma_s\left(\alpha^-_{a_s,n-1}+\alpha^-_{b_s,n-1}\right), \label{eq:rupro}
\end{align} where $\sigma_s\in\{+1,-1\}$ represents the type of $s$-th root vector in $m_1$. Examining the components of \eqref{eq:rupro} one-by-one shows that only the first and last root vectors of $m_1$ include $\alpha_{k_1,n-1}^-$, with coefficients $+1$ and $-1$. Thus, $\alpha_{k_1,n-1}^-$ disappears in \eqref{eq:rupro} if and only if either of these two root vectors is included in $u$ or both are excluded. In the former case, looking at the $\alpha_{l_2,n-1}^-$-coordinate forces the inclusion of the second factor of $m_1$. This forces the inclusion of the third factor by looking at $\alpha^-_{l_3,n-1}$, etc.   Recursively, we obtain $I=\{1,\ldots,t\}$, i.e., $u=m_1$, and $v=1$. In the latter case, the same induction starting from $\alpha_{l_2,n-1}^-$ shows $I=\emptyset$, i.e., $u=1$, and $v=m_1$. In either case, one root vector is a unit, contrary to the assumption. Therefore, $m_1$ is indecomposable in $\mathcal{Q}_{D_n}(d')\cap \mathbb{Z}_2^L\text{-span}$.

\medskip
We now start with part (b). Note that the existence of $m_1$ in the form of \eqref{eq:gm1factor} can be verified by a case examination. This can be seen as follows: For $n\geq 3$, choose the distinct $a,b,c\in\{1,\dots,n\}$ and set $\mu=1$, choose indices of the order $a<b<c$, and set:
\begin{align*}
g:=\varepsilon^-_{b,b+1}\,\varepsilon^-_{b+1,b+2}\cdots \varepsilon^-_{c-1,c}\in S_n\text{-span},\qquad
m_1':=\varepsilon^+_{ab}\,\widehat{\varepsilon}{}^+_{ac}\in \mathbb{Z}_2^L\text{-span}.
\end{align*}
Then
\begin{align*}
R(m_1')&=\left(\alpha^-_{a,n-1}+\alpha^-_{b,n-1}\right)-\left(\alpha^-_{a,n-1}+\alpha^-_{c,n-1}\right)=\alpha^-_{b,n-1}-\alpha^-_{c,n-1}\\
&=\sum_{s=b}^{c-1}\alpha^-_{s,s+1}=R(g),
\end{align*} so $R(gm_1')=0$ and $z:=gm_1'\in \mathcal{SK}_1$. Let $\mu>1$. Using $D_n$ root assignments \eqref{eq:so2n} and the additivity of $R$, we obtain the following:
\begin{align*}
R(m_1)=\sum_{w=1}^{\mu}\left(\alpha^-_{k_w,n-1}+\alpha^-_{l_w,n-1}\right)-\sum_{u=\mu+1}^{2\mu}\left(\alpha^-_{k_u,n-1}+\alpha^-_{l_u,n-1}\right).
\end{align*}
To obtain the existence in $\mathcal{SK}_1$ for arbitrary long root indices, fix a bijection between the two sets of indices:
\begin{align*}
\{\mu+1,\ldots,2\mu\} \longrightarrow \{1,\ldots,\mu\},\qquad u\longmapsto w(u).
\end{align*}Define $g\in S_n\text{-span}$ by reordering the indices $k$ and $l$ between the two components:\begin{align*}
g := \left(\prod_{u=\mu+1}^{2\mu}\varepsilon_{\min\{k_{w(u)},k_u\}^-,\,\max\{k_{w(u)},k_u\}} \right) \cdot \left(\prod_{u=\mu+1}^{2\mu}\varepsilon_{\min\{l_{w(u)},l_u\},\,\max\{l_{w(u)},l_u\}}^-\right).
\end{align*} Without loss of generality, take $k_{w(u)}=k_u$ or $l_{w(u)}=l_u$, the corresponding root vector is omitted. For each $u$, the root assignment is given as follows
\begin{align*}
R\left(\varepsilon_{\min\{k_{w(u)},k_u\}^-,\,\max\{k_{w(u)},k_u\}}\right) =\alpha_{k_u\ n-1}^- -\alpha_{k_{w(u)}^-\ n-1},\qquad R\left(\varepsilon_{\min\{l_{w(u)},l_u\}^-,\,\max\{l_{w(u)},l_u\}}\right)
=\alpha_{l_u  n-1}^--\alpha_{l_{w(u)} n-1}^-,
\end{align*}
hence $R(g)=-R(m)$. Therefore, $R(gm)=0$ and, since $g\in S_n$-span and $m\in\mathbb{Z}_2^L\text{-span}$,
\begin{align*}
p:=g\,m \in \left(S_n\text{-span}\otimes \mathbb{Z}_2^L\text{-span}\right)\cap \mathcal{Q}_{D_n}(d')=\mathcal{SK}_1.
\end{align*}    Write the given element in PBW order as  $p  =  g\,m$  with   $g\in S_n\text{-span}$, and $m\in \mathbb{Z}_2^L\text{-span}$. Assume that the long root block $m$ is indecomposable in $\mathbb{Z}_2^L$-span. Define the $A$-height on the $S_n$-part by \begin{align*}
\mathrm{ht}_A(g') := \sum_{s=1}^{n-1} c_s\quad\text{when}\quad R(g')=\sum_{s=1}^{n-1} c_s\,\alpha^-_{s,s+1}\ (c_s\in\mathbb{N}_0).
\end{align*} Since $R$ is additive and $R(m)$ does not have a $\alpha^-_{s,s+1}$-component, $\mathrm{ht}_A$ is additive and $\mathrm{ht}_A(m)=0$. Moreover, $R(g)\neq 0$ unless $g=1$, so $\mathrm{ht}_A(p)=\mathrm{ht}_A(g)\geq 1$.

Suppose that $p$ has a non-trivial decomposition $p=p_1p_2$ with $p_i\in\mathcal{SK}_1$ non-units. In the PBW basis splitting, write  $p_1=g_1m_1, p_2=g_2m_2$, $g_i\in S_n$-span, $m_i\in \mathbb{Z}_2^L$-span. By additivity,  \begin{align}
\mathrm{ht}_A(p)=\mathrm{ht}_A(p_1)+\mathrm{ht}_A(p_2)=\mathrm{ht}_A(g_1)+\mathrm{ht}_A(g_2)\geq 1. \label{eq:ht}
\end{align} Without loss of generality, choose $p$ with $\mathrm{ht}_A(p)$ minimal. Equivalently, one may assume $\mathrm{ht}_A(p)=1$. Then in \eqref{eq:ht}, we observe that at least one factor has zero $A$-height. Otherwise, both $\mathrm{ht}_A(g_i)\geq 1$ and hence $\mathrm{ht}_A(p)=\mathrm{ht}_A(g_1)+\mathrm{ht}_A(g_2)\geq 2$ contradict the minimality of $\mathrm{ht}_A(p)$. Thus, we have $\mathrm{ht}_A(p_1)=0$, which implies that $g_1=1$ and $p_1=m_1\in \mathbb{Z}_2^L$-span.

Comparing the long root parts of $p=p_1p_2=(m_1)(g_2m_2)$ in the PBW basis yields, \begin{align}
m_1\,m_2 = m. \label{eq:m1m2}
\end{align} Since $m$ is indecomposable, the equality \eqref{eq:m1m2} forces either $m_1=1$ or $m_2=1$. If $m_1=1$ then $p_1$ is a unit, contrary to the assumption. If $m_2=1$ then $p_2=g_2$ lies in the $S_n$-span with \begin{align*}
    \mathrm{ht}_A(p_2)=\mathrm{ht}_A(g_2)=\mathrm{ht}_A(p)-\mathrm{ht}_A(p_1)=\mathrm{ht}_A(p)\geq 1,
\end{align*} so $p_2$ is a nontrivial $S_n$-part but does not have a long root factor, again contradicting the assumed nontrivial factorization $p=p_1p_2$ with $p_1$ already completely in $\mathbb{Z}_2^L$-span. Both cases lead to a contradiction. Hence $p$ does not admit a nontrivial factorization in $\mathcal{SK}_1$, i.e., $p$ is indecomposable.

 This completes the proof of \textup{(a)} and \textup{(b)}.
 
 \end{proof}

\begin{remark}
(i) All the above products are finite: $\mu\in\mathbb{N}$, $t=2\mu\in2\mathbb{N}$, $r\in\mathbb{N}$, and all the indices are in $\{1,\dots,n\}$. The indices in $S_n$-span and $\mathbb{Z}_2^L$-span are nonnegative integers (Definition \ref{spanso}, Remark \ref{r4.2} (i)), so there are no infinite products.

(ii) When $t=2\mu>2$, finding all the indecomposable generators such that $R(p) = 0$ is equivalent to classifying all indecomposable $m\in\mathbb{Z}_2^L$-span with $R(m)=0$ requires combining cancellations of the permutation roots $\alpha_{\bullet,n-1}^-$ under the constraint $t=2\mu$. Consequently, admissibility is characterized not by a single condition within either $R(g)$ or $R(m)$, but instead by how the long root vectors $2\mu$ are distributed over the different index arrangements. Even for $t=2$, there are several different indexing choices, and only specific paths produce indecomposable generators, see Section \ref{sec:examples}. For $t>2$, these choices of indices grow inductively, and the number of admissible arrangements grows rapidly. Together with PBW splitting $p=g\,m_1$, $g\in S_n$-span, $m_1\in\mathbb{Z}_2^L$-span, and the examination of indecomposability, it is impossible to provide a complete classification of indecomposable generators in a short closed list. Consequently, for $\mu>1$, we proceed constructively:  We first list all monomials $m$ in $\mathbb{Z}_2^{L}\text{-span}$ such that $R(m)=0$.  Among these, we discard every $m$ that admits a non-trivial factorization $m=m_1m_2$ with $R(m_1)=R(m_2)=0$.  For each remaining monomial $m$, we then choose permutation monomials $g\in S_n\text{-span}$ with prescribed $A$-height such that $gm\in \mathcal{SK}_1$ is indecomposable.

\end{remark}

Notice that if $n = 2$, we will have that $\mathcal{Q}_{D_2}(0)$ is Abelian, as shown in the example below.

  \begin{example}
Consider the commutant of the Cartan subalgebra in $\mathfrak{so}(4,\mathbb{C})$, the roots decomposition will give us the following basis $\beta_{\mathfrak{so}^*(4,\mathbb{C})} = \left(h_1,h_2, \varepsilon_{12}^-,   \varepsilon_{12}^+,  \varepsilon_{21}^-,   \widehat{\varepsilon}_{12}^+\right)$. Under the basis $\beta_{\mathfrak{so}^*(4)}$, all the indecomposable homogeneous polynomial solutions in the form of \eqref{eq:PDEs} to the PDEs $\{\mathfrak{h},S(B_2)\} = 0$ are
\begin{align*}
    A_1 =   h_1, \text{ } A_2 = h_2 \in \textbf{q}_1; \text{ } \quad \text{ }   p_{12^-,21^-} = \varepsilon_{12}^- \varepsilon_{21}^-, \text{ } p_{12^+;\widehat{12}^+} = \varepsilon_{12}^+ \widehat{\varepsilon}_{12}^+ \in \textbf{q}_2 ,
\end{align*}
which generate a commutative Poisson algebra $\mathcal{Q}_{D_2}(0) = \mathbb{C}\langle \textbf{Q}_2 \rangle $, where $\textbf{Q}_2 = \textbf{q}_1 \sqcup \textbf{q}_2$. This implies that $\mathcal{Q}_{D_2}(0) $ is isomorphic to a polynomial ring $\mathbb{C}\left[A_1,A_2, p_{12^-,21^-}  ,  p_{12^+;\widehat{12}^+} \right]$. It is finitely-generated and forms an integrable system with the families of Hamiltonians $\mathcal{H} = \Gamma_1 h_1 + \Gamma_2 h_2 + \Gamma_{12} h_1 h_2$, where $\Gamma_1,\Gamma_2$ and $\Gamma_{12} \in \mathbb{R}$.
\end{example}

\subsubsection{Generators of $S(C_n)^\mathfrak{h}$}
\label{subsubsec:sp}
For the symplectic Lie algebra $\mathfrak{sp}(2n,\mathbb{C})$, the Cartan subalgebra is defined by $$  \mathfrak{h} = \left\{H \in \mathfrak{sp}(2n,\mathbb{C}) : H = \sum_{j =1}^n a_j \left( E_j - E_{n + j\,n+j} \right), \text{ } h_j(H) = a_j  \text{ } h_j \in \mathfrak{h}^*\right\}.$$ Positive roots are given by
$\alpha_{j k}^\pm =   \alpha_j \pm  \alpha_k,   \text{ for all } 1 \leq j < k \leq n$,   together with $    \alpha_{i,i} =\alpha_i, $ for all $  1 \leq i \leq n$,   where $\alpha_i,\alpha_j$ are linear functionals on $\mathfrak{h}^*$, that are generated by simple roots $
   \Delta_{C_n} = \{  \alpha_{12}^-, \ldots,  \alpha_{n-1n}^-,    2\alpha_n \}$. Moreover, the correspondence between the root vectors $E_{k\,j}$ and their related roots is \begin{align*}
    \begin{matrix}
        2\alpha_j \rightarrow E_{j \, n+j} & 1 \leq j \leq n \\
        -2\alpha_j \rightarrow  E_{n+j\,j}  & 1 \leq j \leq n \\
   \alpha_j - \alpha_k \rightarrow E_{j \,k} - E_{n+j \,n+k} & 1 \leq j < k \leq n\\
   \alpha_j + \alpha_k \rightarrow E_{j\, k+n} - E_{k \,j+n} & 1 \leq j < k \leq n \\
    -(\alpha_j + \alpha_k) \rightarrow E_{j+n\, k} - E_{n+k \,j} & 1 \leq j < k \leq n
     \end{matrix}
\end{align*}The dual space $\mathfrak{sp}^*(2n ,\mathbb{C})$ admits the coordinates
\begin{align*}
\boldsymbol{x}_C = \left(h_1,\ldots,h_n,\varepsilon_{12}^-, \ldots,\varepsilon_{n-1n}^-, \right.  \left.\varepsilon_{12}^+,\ldots,\widehat{\varepsilon}_{n-1n}^-, \varepsilon_1,\ldots,\widehat{\varepsilon}_n\right).
\end{align*} In particular, the roots in $\Phi_{C_n}$ satisfy the following decomposition \begin{align}
\nonumber
 R\left(\varepsilon_{jk}^-\right) := & \,\alpha_{jk}^- =\sum_{l = j}^{k-1} \alpha_{ll+1} ;\\    R\left(\varepsilon_j\right) := & \, \alpha_j = 2\sum_{l = j}^{n-1} \alpha_{ll+1} + 2\alpha_n;\label{eq:root2}  \\
 \nonumber
 R\left(\varepsilon_{jk}^+\right) := & \,\alpha_{jk}^+ =  \alpha_{jk}^- + \alpha_k = \alpha_{jn}^- + \alpha_{kn}^- + 2 \alpha_n
\end{align} with $1 \leq j < k \leq n$.   For all $ H_i \in \mathfrak{h} $, the $\mathrm{ad}^*(H_i)$-action on the root vector will generate the following diagonalizable terms \begin{align*}
    \{h_i,\varepsilon_{jk}^-\} = \mu_{j,k}^i \varepsilon_{jk}^-   \text{ and } \{h_i, \varepsilon_j\} = \left(2 \mu_{j,n}^i + \mu_n^i \right)\varepsilon_j =  \mu_{j,j}^i \varepsilon_j ,
\end{align*} where \begin{align}
         \mu_{j,j+1}^{j\pm 1} = - 1, \quad \mu_{j,j+1}^j = 2,  \quad \mu_{n-1,n-1}^n = -2,  \quad \mu_{j,j+1}^l = 0 , \text{ } j \neq   l \pm 1, l = 1,\ldots,n. \label{eq:val4}
\end{align}

  We now turn our attention to the classification of explicit generators for polynomial Poisson algebra $\mathcal{Q}_{C_n}(d'')$. Define a mapping \begin{align}
      f: = (f_1,f_2,f_3): \Phi_{B_n\vert_{\mathfrak{g}^+}} \rightarrow \Phi_{C_n\vert_{\mathfrak{g}^+}} \text{   given by } (\alpha_{ij}^-,\alpha_{kl}^+,\alpha_p) \mapsto ( \alpha_{ij}^-,\alpha_{kl}^+  ,2\alpha_p ) \label{eq:identi}
  \end{align}  from the root vectors to their corresponding roots, which establishes the relation between $\Phi_{B_n}$ and $\Phi_{D_n}$. It is clear that $f_1 = f_2 = \mathrm{id} $ and $f_3 =\mathrm{id}$, where $\mathrm{id}$ is the identity map. We can establish the analogous correspondence for negative roots. Consequently, the permutation roots and long roots of $\Phi_{B_n}$ and $\Phi_{C_n}$ coincide.  As in Section \ref{subsec:consono}, the monomials within the Cartan commutant can again be categorized into five classes. Namely,

\qquad \qquad \textbf{(a)} $p$ equals $g$, $m_1$ or $m_2$, \text{ } \textbf{(b)} $p=gm_1$, \text{ } \textbf{(c)} $p=gm_2$, \text{ } \textbf{(d)} $p=m_1m_2$, and \textbf{(e)} $p=gm_1m_2$.

Indeed, the decomposition of generators into the permutation-type of root vectors $g$ (built from $\varepsilon_{ij}^-$ and controlled by the $S_n$-action) and the $\mathbb{Z}_2$-type root vectors $m=m_1m_2$ (built from long root variables $\varepsilon^+_{ij},\widehat{\varepsilon}^+_{ij}$ and short-root variables $\varepsilon_i,\widehat{\varepsilon}_i$) is purely combinatorial and depends only on the common Weyl group $W_{B_n}\cong W_{C_n}\cong S_n\ltimes(\mathbb{Z}_2)^n$. Under the standard identification of the two root systems that fix the permutation and long root sectors and rescale the short roots via $\alpha_i\mapsto 2\alpha_i$ (equivalently, see \eqref{eq:identi}, $(\alpha_{ij}^-,\alpha^+_{kl},\alpha_p)\mapsto(\alpha_{ij}^-,\alpha^+_{kl},2\alpha_p)$), the sets of admissible monomials in each sector correspond bijectively. Thus, the cases split \textbf{(a)}-\textbf{(e)} proposed above remain unchanged.  What changes is only the linear balance in the constraint of $\alpha_n$ cancellation in Lemma \ref{lem:bs}, but this affects the numerical exponents, e.g., the relation that enforces $R(p)=0$ and does not affect the structure of the classification itself. The same three building blocks ($g$, $m_1$, $m_2$) and the same five combinations \textbf{(a)}-\textbf{(e)} exhaust all possible types of indecomposable generators in $C_n$, as in $B_n$.   The relationship between these root systems implies that the monomial $p(\textbf{x}_C)$, expressed in terms of permutation root vectors or $\mathbb{Z}_2^L$-span, will resemble the monomials in cases \textbf{(a)} and \textbf{(b)}, as previously noted. From the root system in \eqref{eq:root2}, the elements in the Cartan commutant $S\left(C_n\right)^\mathfrak{h}$ should exhibit a form similar to that in \eqref{eq:41}. Moreover, the assignment of roots corresponding to $p(\textbf{x}_C) \in \mathcal{Q}_{C_n}(d'')$ is given by  \begin{align}
 \nonumber R(p(\boldsymbol{x}_C)) = & \, \sum_{c = 1}^r \alpha_{i_cj_c}^- + \sum_{w=1}^{\mu} \left(\alpha_{k_wn}^- + \alpha_{l_wn}^-\right) + \sum_{w=\mu+1}^t \left(\alpha_{nk_w}^- + \alpha_{nl_w}^-\right) \\
  & \,+ 2 \left(\sum_{a=1}^{\nu} \alpha_{s_an}^-\right) +  2\left(\sum_{b= \nu+1}^q \alpha_{ns_b}^-\right) +  \left(2t - 4 \mu + 4\nu - 2q  \right) \alpha_n = 0  \label{eq:r41id}
\end{align}  with $ 2t - 4 \mu + 4\nu - 2q =0$ and $\mu,\nu,r,t,q \in \mathbb{N}_0$.   Therefore, we can disregard the classifications within the spaces $\mathcal{Q}_{C_n}(d'') \cap S_n$-span, $ \mathcal{Q}_{C_n}(d'') \cap \mathbb{Z}_2^L$-span, $ \mathcal{Q}_{C_n}(d'') \cap \mathbb{Z}_2^S$-span, and $\mathcal{SK}_1$.

\section{Non-commutative polynomial Poisson algebras of rank \texorpdfstring{$3$}{3} }
\label{sec:examples}

\noindent In this section, we present the classification of the generators for Cartan centralisers for specific ranks ($3$ in particular). The classification of Lie algebras with rank $< 2$ is straightforward, as for rank one Lie algebras, the centralizer reduces to the Casimir element. As in illustration of rank two, we consider the case $\mathfrak{g} \cong B_2$.

\begin{example}
\label{r5.1}
     In $B_2$, the set of roots is given by $\Phi_{B_2} = \left\{\alpha_{12}^-,\alpha_{21}^-,\alpha_{12}^+,\hat{\alpha}_{12}^+, \alpha_1,\hat{\alpha}_1,\alpha_2,\hat{\alpha}_2\right\}$ with a set of simple roots $\Delta_{B_2} = \left\{\alpha_{12}^-,\alpha_2\right\}$. In this case, the indecomposable polynomial solutions of the system of the corresponding PDEs $\{p ,h_i\} = 0$, for all $i =1,2$, consist of all the Cartan elements and \begin{align}
    \nonumber
    & p_{12^-,21^-} =  \varepsilon_{12}^- \varepsilon_{21}^-, \quad p_{12^+,\widehat{12}^+} = \varepsilon_{12}^+\widehat{\varepsilon}_{12}^+, \quad p_{1,\widehat{1}} =  \varepsilon_1\widehat{\varepsilon}_1, \quad p_{2,\widehat{2}} =  \varepsilon_2\widehat{\varepsilon}_2  \in \textbf{q}_2 \\
     &  p_{12^-;2,\widehat{1}} = \varepsilon_{12}^- \varepsilon_2 \widehat{\varepsilon}_1, \text{ }      p_{21^-;1,\widehat{2}} = \varepsilon_{21}^- \varepsilon_1\widehat{\varepsilon}_{2}, \text{ }    p_{12^+;\widehat{1},\widehat{2}} = \varepsilon_{12}^+ \widehat{\varepsilon}_1\widehat{\varepsilon}_{2}, \text{ }  p_{\widehat{12}^+;1,2} =   \widehat{\varepsilon}_{12}^+ \varepsilon_{1}\varepsilon_2  \in \textbf{q}_3  \label{eq:generators} \\
     \nonumber
       & p_{12^-;12^+;\widehat{2}^2}  = \varepsilon_{12}^-\varepsilon_{12}^+\left(\widehat{\varepsilon}_{1}\right)^2, \text{ }  p_{12^-;\widehat{12}^+;2^2} = \varepsilon_{12}^-\widehat{\varepsilon}_{12}^+\left(\varepsilon_{2}\right)^2, \text{ } p_{21^-;12^+;\widehat{2}^2} = \varepsilon_{21}^-\varepsilon_{12}^+\left(\widehat{\varepsilon}_{2}\right)^2,  \text{ } p_{21^-;\widehat{12}^+;1^2} = \varepsilon_{21}^-\widehat{\varepsilon}_{12}^+\left(\varepsilon_{1}\right)^2  \in \textbf{q}_4.
\end{align} They form a cubic algebra $\mathcal{Q}_{B_2}(d)$ with $\dim_{FL} \mathcal{Q}_{B_2}(d) = 14$. Using the Chevalley basis relation, the commutator relations are given by \begin{align}
      [\mathfrak{h},\mathfrak{g}_{\pm \alpha}] \subset \mathfrak{g}_{\pm \alpha}, \quad [\mathfrak{g}_\alpha,\mathfrak{g}_{-\alpha}] \subset \mathfrak{h}, \quad [\mathfrak{g}_{\alpha},\mathfrak{g}_\beta] = \left\{\begin{matrix}
          \mathfrak{g}_{\alpha + \beta} & \text{ if } \alpha + \beta \in \Phi \\
          0  & \text{ if } \alpha + \beta \notin \Phi
      \end{matrix}\right. \label{eq:comrot}
  \end{align} The explicit commutator relations are:
   \begin{align*}
\left\{p_{12^-;21^-},p_{12^+;\widehat{12}^+}\right\}&=0,\\
\left\{p_{12^-;21^-},p_{1;\widehat{1}}\right\}&=p_{21^-;1,\widehat{2}}-p_{12^-;2,\widehat{1}}, &
\left\{p_{12^-;21^-},p_{2,\widehat{2}}\right\}&=p_{21^-;1,\widehat{2}}-p_{12^-;2,\widehat{1}},\\
\left\{p_{12^+;\widehat{12}^+},p_{1;\widehat{1}}\right\}&=2\,\left(p_{12^+;\widehat{1},\widehat{2}}+p_{\widehat{12}^+;1,2}\right), &
\left\{p_{12^+;\widehat{12}^+},p_{2,\widehat{2}}\right\}&=2\,\left(p_{\widehat{12}^+;1,2}-p_{12^+;\widehat{1},\widehat{2}}\right),\\
\left\{p_{1;\widehat{1}},p_{2,\widehat{2}}\right\}&=p_{12^-;2,\widehat{1}}-p_{21^-;1,\widehat{2}}+2\,p_{12^+;\widehat{1},\widehat{2}}-2\,p_{\widehat{12}^+;1,2}.
\end{align*} Using the same Lie-Poisson rule together with the Leibniz property, the mixed degree 2 and degree 3 Poisson brackets are
\begin{align*}
\left\{p_{12^-;21^-},p_{12^-;2,\widehat{1}}\right\}&=h_1\,p_{12^-;2,\widehat{1}}+p_{12^-;21^-}\left(p_{1;\widehat{1}}+p_{2,\widehat{2}}\right), \\
\left\{p_{12^-;21^-},p_{21^-;1,\widehat{2}}\right\}&=h_1\,p_{21^-;1,\widehat{2}}-p_{12^-;21^-}\left(p_{1;\widehat{1}}+p_{2,\widehat{2}}\right),\\
\left\{p_{12^-;21^-},p_{12^+;\widehat{1},\widehat{2}}\right\}&=p_{21^-;12^+;\widehat{2}^2}-p_{12^-;12^+;\widehat{1}^2}, \\
\left\{p_{12^-;21^-},p_{\widehat{12}^+;1,2}\right\} & =p_{21^-;\widehat{12}^+;1^2}-p_{12^-;\widehat{12}^+;2^2},\\[2pt]
\left\{p_{12^+;\widehat{12}^+},p_{12^-;2,\widehat{1}}\right\}&=2\,p_{12^-;\widehat{12}^+;2^2},  \quad \left\{p_{12^+;\widehat{12}^+},p_{21^-;1,\widehat{2}}\right\} =2\,\left(p_{21^-;12^+;\widehat{2}^2}+p_{21^-;\widehat{12}^+;1^2}\right),\\
\left\{p_{12^+;\widehat{12}^+},p_{12^+;\widehat{1},\widehat{2}}\right\}&=2\,h_2\,p_{12^+;\widehat{1},\widehat{2}}+2\,p_{12^+;\widehat{12}^+}\left(p_{1;\widehat{1}}+p_{2,\widehat{2}}\right),   \\
\left\{p_{12^+;\widehat{12}^+},p_{\widehat{12}^+;1,2}\right\} & =2\,h_2\,p_{\widehat{12}^+;1,2}+2\,p_{12^+;\widehat{12}^+}\,p_{2,\widehat{2}},\\[2pt]
\left\{p_{1;\widehat{1}},p_{12^-;2,\widehat{1}}\right\}&=h_1\,p_{12^-;2,\widehat{1}}+2\,p_{12^-;12^+;\widehat{1}^2}-p_{12^-;21^-}p_{1;\widehat{1}}-p_{1;\widehat{1}}p_{2,\widehat{2}}, \\
\left\{p_{1;\widehat{1}},p_{\widehat{12}^+;1,2}\right\}&=-\,h_1\,p_{\widehat{12}^+;1,2}-p_{21^-;\widehat{12}^+;1^2}-2\,p_{1;\widehat{1}}p_{2,\widehat{2}}+2\,p_{12^+;\widehat{12}^+}p_{1;\widehat{1}},\\
\left\{p_{2,\widehat{2}},p_{12^-;2,\widehat{1}}\right\}&=h_2\,p_{12^-;2,\widehat{1}}+2\,p_{12^-;\widehat{12}^+;2^2}+p_{12^-;21^-}p_{2,\widehat{2}}-p_{1;\widehat{1}}p_{2,\widehat{2}}.
\end{align*} Note that the Poisson brackets of degree $3$ generators are quite cumbersome. We merely present one representative $\{\textbf{q}_3,\textbf{q}_3\}$ bracket as follows:
\begin{align*}
\left\{p_{12^-;2,\widehat{1}},p_{21^-;1,\widehat{2}}\right\}
&= h_1\,p_{1;\widehat{1}}p_{2,\widehat{2}} + h_1\,p_{12^-;21^-}p_{2,\widehat{2}} + h_2\,p_{12^-;21^-}p_{1;\widehat{1}} \;-\; 2\,p_{12^-;21^-}\,\left(p_{12^+;\widehat{1},\widehat{2}}+p_{\widehat{12}^+;1,2}\right).
\end{align*} Also, from the commutator relations above, we further conclude that $d = 2$.

Note that the monomials in \eqref{eq:generators} are not functionally independent. Indeed, using \eqref{eq:maximum}, we deduce that the maximum number of functionally independent generators is $\mathcal{N}(\mathfrak{h}) = \dim \mathfrak{so}(5,\mathbb{C}) - \dim \mathfrak{h} = 8$.  Polynomial dependence relations are given by
 \begin{align*}
p_{12^-;2,\widehat{1}}\,p_{12^+;\widehat{1},\widehat{2}} &= p_{2,\widehat{2}}\,p_{12^-;12^+;\widehat{1}^2}, &\qquad
p_{12^-;2,\widehat{1}}\,p_{\widehat{12}^+;1,2} &= p_{1;\widehat{1}}\,p_{12^-;\widehat{12}^+;2^2},\\
p_{21^-;1,\widehat{2}}\,p_{12^+;\widehat{1},\widehat{2}} &= p_{1;\widehat{1}}\,p_{21^-;12^+;\widehat{2}^2}, &\qquad
p_{21^-;1,\widehat{2}}\,p_{\widehat{12}^+;1,2} &= p_{2,\widehat{2}}\,p_{21^-;\widehat{12}^+;1^2}.
\end{align*} Furthermore, the pure cubic products factor through the quadratic layer as
\begin{align*}
p_{12^-;2,\widehat{1}}\,p_{21^-;1,\widehat{2}} &= p_{12^-;21^-}\,p_{1;\widehat{1}}\,p_{2,\widehat{2}}, &
p_{12^+;\widehat{1},\widehat{2}}\,p_{\widehat{12}^+;1,2} &= p_{12^+;\widehat{12}^+}\,p_{1;\widehat{1}}\,p_{2,\widehat{2}}.
\end{align*} Therefore, the set of functionally independent generators is given by $$\mathcal{F}_{B_2} = \left\{h_1,h_2, p_{12^-,21^-} ,   p_{12^+,\widehat{12}^+}  , p_{1,\widehat{1}} ,   p_{2,\widehat{2}} , p_{12^-;2,\widehat{1}} ,      p_{12^+;\widehat{1},\widehat{2}}\right\}. $$ With the Hamiltonian $\mathcal{H} = \sum_{i_1,i_2} \Gamma_{i_1,i_2} h_1^{i_1} h_2^{i_2} \in S(\mathfrak{h})$, the integrals of motion in $\mathcal{F}_{B_2}$ form a superintegrable system $\mathcal{S}$.

\end{example}

We now focus on the non-exceptional complex semisimple Lie algebras of rank $3$, namely $\mathfrak{g} = \mathfrak{sl}(4,\mathbb{C}) \cong A_3$, $\mathfrak{so}(6,\mathbb{C}) \cong D_3$, $\mathfrak{so}(7,\mathbb{C}) \cong B_3$, and $\mathfrak{sp}(6,\mathbb{C}) \cong C_3$. Notice that the polynomial Poisson algebra $\mathcal{Q}_{A_3}(3)$ has been fully studied in \cite{campoamor2023algebraic}. For this reason, we omit this case. We give explicit expressions for polynomial generators in $B_3,D_3,C_3$-type of Cartan commutants, using the classification results listed in Section \ref{sec:construction}, and therefore construct their corresponding superintegrable systems. We emphasize the fact that, although $A_3$ and $D_3$ are isomorphic, the analysis of $D_3$ is itself of interest, in order to illustrate the structure of the polynomial Poisson algebra for higher ranks.

\subsection{Explicit generators for \texorpdfstring{$\mathcal{Q}_{B_3}(d)$}{Q	extunderscore{B3}(d)}}
\label{subsec:so5}

Starting with the Lie algebra $\mathfrak{so}(7,\mathbb{C})$ and an ordered dual basis 
 \begin{align}
    \beta_{\mathfrak{so}^*(7,\mathbb{C})} = \left\{h_1,h_2,h_3,\varepsilon_{ij}^-,\varepsilon_{ji}^-,\varepsilon_{kl}^+,\widehat{\varepsilon}_{kl}^+,\varepsilon_a,\widehat{\varepsilon}_b: 1 \leq i,j,k,l,a,b \leq 3\right\}, \label{eq:b3coo}
\end{align}  steps \textbf{(a)} to \textbf{(e)} in Section \ref{subsec:consono} allow us to determine the Cartan generators and show that these indecomposable generators form a polynomial Poisson algebra $\mathcal{Q}_{B_3}(d)$ for some $d \in \mathbb{N}_0$. By the definition of the polynomial Poisson algebra, $\mathfrak{h}^* = \mathrm{span} \left\{h_1,h_2,h_3\right\}$ forms the first layer in $\mathcal{Q}_{B_3}(d)$. From observation \textbf{(a)}, $p \in \mathcal{Q}_{B_3}(d)$ has the form of $g$, $m_1 $, or $m_2 $. In detail, we have \begin{align}
 &   p_{12^-,21^-}   = \varepsilon_{12}^-\varepsilon_{21}^-, \quad  p_{[13^-],31^-} = \left[\varepsilon_{13}^-\right]\varepsilon_{31}^-, \quad  p_{[23^-],32^-}    = \left[\varepsilon_{23}^-\right]\varepsilon_{32}^-   , \label{eq:genan} \\
  &  p_{12^+,\widehat{12}^+}  = \varepsilon_{12}^+\widehat{\varepsilon}_{12}^+, \quad  p_{13^+,\widehat{13}^+}   = \varepsilon_{13}^+\widehat{\varepsilon}_{13}^+, \quad  p_{23^+,\widehat{23}^+}   = \varepsilon_{23}^+\widehat{\varepsilon}_{23}^+.
\end{align} and $p_{a,\hat{a}} = \varepsilon_a \widehat{\varepsilon}_a$ for all $1 \leq a \leq 3$, where $\left[\varepsilon_{13}^-\right]:= \left\{\varepsilon_{13}^-,\varepsilon_{12}^-\varepsilon_{23}^-\right\} $ and $\left[\varepsilon_{23}^-\right]:= \left\{\varepsilon_{23}^-,\varepsilon_{21}^-\varepsilon_{13}^-\right\} $. See the notation convention defined in \eqref{eq:indrep}. Note that, with the Cartan elements, the generators in \eqref{eq:genan} form the polynomial Poisson algebra $\mathcal{Q}_{A_2}(2) $.

In observation \textbf{(b)}, the monomial has the form of $ g m_1 \in \mathcal{SK}_1$. To ensure that $gm_1$ is indecomposable, we first note that neither $R(g)$ nor $R(m_1)$ is zero.  Since $\alpha_n \notin R(m_1 )$, we have $\deg m_1  \in 2 \mathbb{N}$. Due to the distinct length of the long roots in the block $\varepsilon_{k_wl_w}^+ \widehat{\varepsilon}_{k_ul_u}^+$, with the constraint $\alpha_n \notin R(m_1)$, the form of the monomial $m_1 $ satisfying $\alpha_n \notin R(m_1)$ corresponds to one of the following three types: 
 \begin{align}
 \nonumber
& \text{ (A1) }    \prod_{w=1}^{\mu} \prod_{u=\mu + 1}^{2\mu} \varepsilon_{k_wl_w}^+ \widehat{\varepsilon}_{k_ul_u}^+ \text{ with $1 \leq k_w \neq l_w \neq k_u \neq l_u \leq 3$} ; \\
& \text{ (A2) }    \prod_{w=1}^{\mu} \prod_{u=\mu + 1}^{2\mu} \varepsilon_{k_wk_u}^+ \widehat{\varepsilon}_{k_ul_u}^+ \text{ with $1 \leq  k_w \neq l_u \leq 3$} ; \label{eq:classif} \\
\nonumber
& \text{ (A3) }  \prod_{w=1}^{\mu} \prod_{u=\mu + 1}^{2\mu} \varepsilon_{k_wl_w}^+ \widehat{\varepsilon}_{k_ul_u}^+ \prod_{1 \leq a \neq b \neq c  \leq 3} \varepsilon_{ab}^+\widehat{\varepsilon}_{bc}^+ \text{ with $1 \leq  k_w \neq l_w \neq k_u \neq l_u\leq 3$}.
\end{align} 
 We observe that the existence of some types is conditioned by the rank of $\mathfrak{g}$. So, if cases (A1) and (A3) are valid, then $n \geq 4$. Hence, the only possible form of $m_1$ is the case (A2), as specified in \eqref{eq:classif}. Explicitly, all potential forms of $m_1 $, as expressed in (A2), are provided by \begin{align}
\begin{matrix}
      m_1^{(a)}(\varepsilon_{12}^+,\widehat{\varepsilon}_{23}^+) =  \varepsilon_{12}^+\widehat{\varepsilon}_{23}^+ , \quad  m_1^{(b)}(\varepsilon_{12}^+,\widehat{\varepsilon}_{13}^+ ) = \varepsilon_{12}^+\widehat{\varepsilon}_{13}^+ , \quad  m_1^{(c)} ( \varepsilon_{13}^+,\widehat{\varepsilon}_{23}^+) = \varepsilon_{13}^+\widehat{\varepsilon}_{23}^+ ; \\
     \hat{m}_1^{(a)}(\widehat{\varepsilon}_{12}^+,\varepsilon_{23}^+) = \widehat{\varepsilon}_{12}^+\varepsilon_{23}^+ , \quad  \hat{m}_1^{(b)}(\widehat{\varepsilon}_{12}^+,\varepsilon_{13}^+) = \widehat{\varepsilon}_{12}^+\varepsilon_{13}^+ , \quad  \hat{m}_1^{(c)} (\widehat{\varepsilon}_{13}^+,\varepsilon_{23}^+) = \widehat{\varepsilon}_{13}^+\varepsilon_{23}^+.
\end{matrix} \label{eq:m1}
\end{align} Then  \begin{align}
\begin{matrix}
    p_{[13^-];12^+,\widehat{23}^+}     = \left[\varepsilon_{31}^-\right] m_1^{(a)} , \quad   p_{[23^-];12^+,\widehat{13}^+}   = \left[\varepsilon_{32}^-\right] m_1^{(b)} , \quad   p_{[12^-];13^+,\widehat{23}^+}   = \left[\varepsilon_{21}^-\right] m_1^{(c)} , \\
    p_{[31^-];23^+,\widehat{12}^+}   = \left[\varepsilon_{13}^-\right] \hat{m}_1^{(a)} , \quad   p_{[32^-];13^+,\widehat{12}^+}   = \left[\varepsilon_{23}^-\right] \hat{m}_1^{(b)} , \quad   p_{[21^-];23^+,\widehat{13}^+} = \left[\varepsilon_{12}^-\right] \hat{m}_1^{(c)} ,
\end{matrix}
\end{align} are indecomposable monomials in either $\textbf{q}_3$ and $\textbf{q}_4$, where \begin{align}
\begin{matrix}
       \left[\varepsilon_{13}^-\right]:= \left\{\varepsilon_{13}^-,\varepsilon_{12}^-\varepsilon_{23}^-\right\} ,  \text{ }\left[\varepsilon_{23}^-\right]:= \left\{\varepsilon_{23}^-,\varepsilon_{21}^-\varepsilon_{13}^-\right\}  \text{ and } \left[\varepsilon_{12}^-\right]:= \left\{\varepsilon_{12}^-,\varepsilon_{13}^-\varepsilon_{32}^-\right\} ;\\
       \left[\varepsilon_{31}^-\right]:= \left\{\varepsilon_{31}^-,\varepsilon_{32}^-\varepsilon_{21}^-\right\} ,  \text{ }\left[\varepsilon_{32}^-\right]:= \left\{\varepsilon_{32}^-,\varepsilon_{31}^-\varepsilon_{12}^-\right\}  \text{ and } \left[\varepsilon_{21}^-\right]:= \left\{\varepsilon_{21}^-,\varepsilon_{23}^-\varepsilon_{31}^-\right\}.
\end{matrix}\label{eq:id5520}
\end{align}

We now look at the monomial from observation \textbf{(c)}. The monomial lives in the subspace $\mathcal{SK}_2 :  = \left(S_n\text{-span} \otimes \mathbb{Z}_2^S\text{-span}\right) \cap \mathcal{Q}_{B_n}(d) $ in the form of $ [ g ]  \prod_{1 \leq a \neq b \leq 3} \varepsilon_a \widehat{\varepsilon}_b$. Using Proposition \ref{prop:pcase}, the monomials that span $\mathcal{SK}_2$ are \begin{align}
\left.\begin{matrix}
    p_{[12^-],\widehat{1},2}  = \left[  \varepsilon_{12}^-\right]\widehat{\varepsilon}_1 \varepsilon_2,\quad  p_{[21^-],1,\widehat{2}}  = \left[  \varepsilon_{21}^-\right]\varepsilon_1 \widehat{\varepsilon}_2 \\
    p_{[13^-],\widehat{1},3}  = \left[  \varepsilon_{13}^-\right]\widehat{\varepsilon}_1 \varepsilon_3,\quad  p_{[31^-],1,\hat{3}}  = \left[  \varepsilon_{31}^-\right]\varepsilon_1 \widehat{\varepsilon}_3 \\
   p_{[23^-],\widehat{2},3}  = \left[  \varepsilon_{23}^-\right]\widehat{\varepsilon}_2 \varepsilon_3,\quad  p_{[32^-],2,\hat{3}}  = \left[  \varepsilon_{32}^-\right]\varepsilon_2 \widehat{\varepsilon}_3
\end{matrix} \right\} \in \textbf{q}_3 \text{ or } \textbf{q}_4.
\end{align}

Next, we construct indecomposable monomials in the subspace $\mathcal{K}_1\mathcal{K}_2 : =    \mathbb{Z}_2\text{-span}  \cap \mathcal{Q}_{B_n}(d)$. As we discussed earlier, to ensure indecomposability, we first observe that neither $R(m_1) = 0$ nor $R(m_2) = 0$ are present, as we have $\alpha_n \notin R(m)$ by Lemma \ref{lem:bs}. As $\alpha_n$ appears in the root assignments of both $m_1$ and $m_2$, it must either be present in both $R(m_1)$ and $R(m_2)$ or absent from both. In this way, $m_1$ is classified into two types: $\alpha_n \notin R\left(m_1 \right)$ and $\alpha_n \in R\left(m_1 \right)$. On the one hand, if $\alpha_n \notin R\left(m_1 \right)$, then $\alpha_n \notin R\left(m_2 \right)$ and $m_1 $ must be of the form in \eqref{eq:classif}. From the argument in observation \textbf{(b)}, all the possible forms of $m_1 $ are given by \eqref{eq:m1}. Then the indecomposable monomials should have the form of \begin{align}
\left.\begin{matrix}
     p_{12^+,\widehat{23}^+;\widehat{1},3}  =  m_1^{(a)}  \widehat{\varepsilon}_1 \varepsilon_3,\quad  p_{23^+,\widehat{12}^+;1,\hat{3}}  =  \hat{m}_1^{(a)}  \varepsilon_1 \widehat{\varepsilon}_3 \\
  p_{12^+,\widehat{13}^+;\widehat{2},3} =   m_1^{(b)}  \widehat{\varepsilon}_2 \varepsilon_3,\quad  p_{13^+,\widehat{12}^+;2,\hat{3}} =  \hat{m}_1^{(b)}  \varepsilon_2 \widehat{\varepsilon}_3\\
  p_{13^+,\widehat{23}^+;\widehat{1},2}  =     m_1^{(c)}  \widehat{\varepsilon}_1 \varepsilon_2, \quad  p_{13^+,\widehat{23}^+;1,\hat{3}}  =   \hat{m}_1^{(c)}  \varepsilon_1 \widehat{\varepsilon}_2
\end{matrix} \right\} \in \textbf{q}_4 .
\end{align} On the other hand, if $ \alpha_n \in R\left(m_1 \right)$, to ensure that $\alpha_n \notin R(m)$, we must have $\alpha_n \in R\left(m_2 \right)$. Based on the results in \eqref{eq:classif}, $m_1 $ can admit only one of the following forms
\begin{align}
\nonumber & \text{ (B1) }    \prod_{w=1}^t\varepsilon_{k_wl_w}^+ \text{ with $ 1 \leq k_w \neq  l_w \leq 3$} ; \\
 \nonumber
& \text{ (B2) }  \prod_{1 \leq i,j \leq 3}\varepsilon_{ij}^+ \prod_{w=1}^{\mu} \prod_{u=\mu + 1}^{2\mu} \varepsilon_{k_wl_w}^+ \widehat{\varepsilon}_{k_ul_u}^+  \text{ with $ 1 \leq k_w ,  l_w   \leq 3$} ; \\
& \text{ (B3) }   \prod_{1 \leq i,j \leq 3}\varepsilon_{ij}^+ \prod_{w=1}^{\mu} \prod_{u=\mu + 1}^{2\mu} \varepsilon_{k_wk_u}^+ \widehat{\varepsilon}_{k_ul_u}^+  \text{ with $1 \leq  k_w \neq l_u   \leq 3$} ; \label{eq:classif2} \\
\nonumber
& \text{ (B4) }       \prod_{1 \leq i,j \leq 3}\varepsilon_{ij}^+\prod_{w=1}^{\mu} \prod_{u=\mu + 1}^{2\mu} \varepsilon_{k_wl_w}^+ \widehat{\varepsilon}_{k_ul_u}^+ \prod_{1 \leq a,b,c  \leq 3} \varepsilon_{ab}^+\widehat{\varepsilon}_{bc}^+ \text{ with $1 \leq   k_w \neq l_w \neq k_u \neq l_u \leq 3$},
\end{align} where $1 \leq i,j,k_w,l_w,k_u,l_u,a,b,c \leq 3$. From the preceding analysis, since the rank of $\mathfrak{so}(7,\mathbb{C})$ is $3$, we deduce that only (B1) and (B3) form $m_1$. We also have the following analysis:

\textbf{(i)} Suppose that $m_1$ is of the form (B1). The indecomposable monomials in $\mathcal{K}_1\mathcal{K}_2$ are $  m_1  \widehat{\varepsilon}_{s_1} \widehat{\varepsilon}_{s_2}$ with $1 \leq s_1 \neq s_2 \leq 3$. A direct counting shows that all the indecomposable monomials are
\begin{align}
\left.\begin{matrix}
        p_{12^+;\widehat{1},\widehat{2}}  = \varepsilon_{12}^+ \widehat{\varepsilon}_1\widehat{\varepsilon}_2, \quad  p_{13^+;\widehat{1},\hat{3}}  = \varepsilon_{13}^+ \widehat{\varepsilon}_1\widehat{\varepsilon}_3, \quad  p_{23^+;\widehat{2},\hat{3}}  = \varepsilon_{23}^+ \widehat{\varepsilon}_2\widehat{\varepsilon}_3; \\
    p_{\widehat{12}^+;1,2}  = \widehat{\varepsilon}_{12}^+ \varepsilon_1\varepsilon_2, \quad   p_{\widehat{13}^+;1,3}  = \widehat{\varepsilon}_{13}^+ \varepsilon_1\varepsilon_3, \quad   p_{\widehat{23}^+;2,3}  = \widehat{\varepsilon}_{23}^+ \varepsilon_2\varepsilon_3   \label{eq:60}
\end{matrix}\right\} \in \textbf{q}_3
\end{align}

\textbf{(ii)} Suppose that the monomial $m_1 $ is of the form (B3) in \eqref{eq:classif2}, then the indecomposable polynomials have the form \begin{align*}
    \begin{matrix}
        p_{13^+,12^+,\widehat{23}^+;\widehat{1}^2} = \varepsilon_{13}^+m_1^{(a)}  \widehat{\varepsilon}_1^2, \quad   p_{23^+,12^+,\widehat{13}^+; \widehat{2}^2}   = \varepsilon_{23}^+ m_1^{(b)}  \widehat{\varepsilon}_2^2, \quad  p_{ 12^+, 13^+,\widehat{23}^+;\widehat{1}^2}   = \varepsilon_{12}^+m_1^{(c)} \widehat{\varepsilon}_1^2, \\
    p_{\widehat{13}^+,23^+,\widehat{12}^+;1^2}   =\widehat{\varepsilon}_{13}^+\hat{m}_1^{(a)} \varepsilon_1^2, \quad   p_{ \widehat{23}^+,13^+,\widehat{12}^+;2^2}   =\widehat{\varepsilon}_{23}^+   \hat{m}_1^{(b)} \varepsilon_2^2, \quad   p_{\widehat{12}^+,23^+,\widehat{13}^+;1^2} =\widehat{\varepsilon}_{12}^+ \hat{m}_1^{(c)} \varepsilon_1^2
    \end{matrix}
\end{align*} are in $\textbf{q}_5$. Here, the monomials $m_1^{(a)},m_1^{(b)}$ and $m_1^{(c)}$ are defined in \eqref{eq:m1}.

Finally, we will list all the indecomposable monomials from observation \textbf{(e)}. From the above discussion, with different choices of $m_1 $, the indecomposable monomials have the following forms: \begin{align}
\nonumber
      p_I  =& \text{ } \left[g_I\right] \varepsilon_{k_wk_u}^+  \widehat{\varepsilon}_{k_ul_u}^+ \varepsilon_a\widehat{\varepsilon}_b \text{ with $1 \leq k_w \neq l_u, a \neq b \leq 3$}; \\
       p_{II} = & \text{ }\left[g_{II}\right] \varepsilon_{kl}^+  \widehat{\varepsilon}_{s_1}\widehat{\varepsilon}_{s_2} \text{ with $ 1 \leq k,l \leq 3$} ; \label{eq:poly} \\
       \nonumber
      p_{II}'  =& \text{ } \left[g_{II}'\right] \varepsilon_{kl}^+ \varepsilon_{k_wk_u}^+ \ \widehat{\varepsilon}_{k_ul_u}^+ \widehat{\varepsilon}_{s_1}\widehat{\varepsilon}_{s_2} \text{ with $1 \leq k_w \neq l_u , k,l \leq 3$,}
\end{align} where $[g] \in S_3$-span contains $S_3$-equivalent monomials such that $p$ is indecomposable.  

\begin{proposition}
    Let $\mathcal{Q}_{B_3}(d)$ be the Cartan commutant $S(B_3)^\mathfrak{h}$. Then $p_{II}' \in \mathcal{Q}_{B_n}(d)$ in \eqref{eq:poly} are decomposable if $s_1 \neq s_2$.
 \end{proposition}

\begin{proof}
    Suppose the contrary. That is, there exists a $p_{II}'  \in \mathcal{Q}_{B_n}(d)$ in \eqref{eq:poly} with $s_1 \neq s_2 $ such that $p_{II}' $ is indecomposable. Since $R\left(p_{II}' \right) = 0$, it follows that $$R(g_{II}') = - \left(\alpha_{k3}^- + \alpha_{l3}^- + \alpha_{k_w3}^- + \alpha_{3l_u}^- + \alpha_{3s_1}^-  + \alpha_{3s_2}^-\right). $$ Then all the possible combinations of positive and negative roots in $R\left(g_{II}'\right)$ are given by \begin{center}
    \begin{tabular}{|c|c|  c c|}
    \hline
         & $(k_w3)$ & $(k3)$ & $(l3)$  \\
         \hline
        $(3l_u)$ & $(k_wl_u)$ & $(kl_w)$ & $(ll_u)$ \\
        \hline
        $(3s_1)$ & $(k_ws_1)$ & $(ks_1)$ & $(ls_1)$ \\
        $(3s_2)$ & $(k_ws_2)$ & $(ks_2)$ & $(ls_2)$   \\
        \hline
    \end{tabular}

\quad

Table 5.1 \label{ta}

\end{center}  Here, index pairs are used to denote the positive and negative roots for each entry in Table \ref{ta}. For example, in the first entry, the pair $(k_wl_u)$ represents the root decomposition $\alpha_{k_w3}^- + \alpha_{3l_u}^- = \alpha_{k_wl_u}^-$. In other words, we may write $(k_w3)+(3l_u) = (k_wl_u)$. The remaining elements in Table \ref{ta} are represented similarly.  Note that it is sufficient to write Table \ref{ta} as a non-singular matrix \begin{align}
    R = \begin{pmatrix}
           (k_wl_u) & (kl_w) & (ll_u) \\
        (k_ws_1) & (ks_1) & (ls_1) \\
         (k_ws_2) & (ks_2) & (ls_2)
    \end{pmatrix} \label{eq:coefficient}
\end{align}
such that all possible combinations of pairs forming $g_{II}'$ can be taken (exactly once) from each column or row of $R$. Since $R$ is not singular, the number of choices of indices to form $R(g_{II}')$ equals $3!= 6$. That is, there are $6$ different pairs that form the root vectors in $g_{II}'$. Specifically, the combinations of pairs are given by \begin{align}
\begin{matrix}
      (k_wl_u) (k s_1) (ls_2), \quad  (k_wl_u) (k s_2) (ls_1), \quad   (kl_u) (k_ws_1) (ls_2), \\
      (kl_u) (ls_1) (k_ws_2) , \quad (ll_u) (ks_1)(k_ws_2), \quad  (ll_u) (k_ws_1)(k s_2)    .
\end{matrix} \label{eq:ind}
\end{align} Here, the combination of indices pairs $(kl_u) (k_ws_1) (ls_2)$ corresponds to the monomial $\varepsilon_{l_uk}^- \varepsilon_{s_1k_w}^- \varepsilon_{s_2l}^- \in S_3$-span, and so on for the rest of the combinations.

Note that not all combinations of \eqref{eq:ind} lead to an indecomposable monomial. Hence, to show $P_{II}'$ with $s_1 \neq s_2$ is decomposable, it suffices to demonstrate that no combinations within \eqref{eq:ind} produce indecomposable monomials. Clearly, as $$R\left(\varepsilon_{l_uk_w}^-\varepsilon_{k_wk_u}^+ \ \widehat{\varepsilon}_{k_ul_u}^+ \right) =0,$$ the combinations $(k_wl_u) (k s_1) (ls_2)$, and $  (k_wl_u) (k s_2) (ls_1) $ make $p$ decomposable. Therefore, we will exclude these first two combinations.   Since the rank of $\mathfrak{so}(7,\mathbb{C})$ is $3$, we deduce $s_1 \neq  s_2,k \neq l \in \{k_w,k_u,l_u\} \subset I_3 := \{1,2,3\}$. Note that only one of the values $s_i$ is equal to $k$ or $l$ for all $i = 1,2$. If, instead, both $s_1 = k$ and $s_2 = l $ are true, then by the identity \eqref{eq:60}, $p_{II}'$ becomes decomposable. Without loss of generality, take $s_1 = l$ and $s_2 \neq l$. Then the rest of the combinations in \eqref{eq:ind} become
\begin{align}
  \begin{matrix}
    (kl_u) (k_wl) (ls_2),  \quad   (kl_u)   (k_ws_2) , \quad (kl)(ll_u) (k_ws_2), \quad  (k_wl)(ll_u) (k s_2)    .
\end{matrix}  \label{eq:ind2}
\end{align}
  Without considering the value of $s_2$, from the fact that $R\left(\varepsilon_{l_uk_w}^-\varepsilon_{k_wk_u}^+   \widehat{\varepsilon}_{k_ul_u}^+ \right) =0$, we observe that \begin{align*}
    R\left(\varepsilon_{l_ul }^-\varepsilon_{lk_w}^-\varepsilon_{k_wk_u}^+   \widehat{\varepsilon}_{k_ul_u}^+ \right) = R\left(\varepsilon_{l_uk_w}^-\varepsilon_{k_wk_u}^+   \widehat{\varepsilon}_{k_ul_u}^+ \right) =0.
 \end{align*} Hence, the choice $(k_wl)(ll_u)(ks_2)$ makes $p_{II}'$ decomposable such that we can discard this combination from \eqref{eq:ind2}. Note that $(k_wl) + (ls_2) = (k_ws_2)$ and $(kl) +(ll_u) = (kl_u)$. Thus, only the combination $(kl_u)  (k_ws_2)$ leads $p_{II}'$ to being indecomposable.  To this extent, we now determine the value of the indices $s_2$. From the assumption $s_1 \neq s_2$, we deduce $s_2 \neq l$. Moreover, if $s_2 \notin \{k,l_u,k_w\}$, then $m_1 $ contains at least $3$ non-equal indices, which leads to a contradiction of the fact that $\mathrm{rank} \, \mathfrak{so}(7,\mathbb{C}) = 3$.  Therefore, $s_2 \in \{k,l_u,k_w\}$. Note that $s_2 \neq k$; otherwise, $(k_wk) + (kl_u) = (k_wl_u)$ will lead to $p_{II}'$ being decomposable. This gives $s_2 = k_w$. Therefore, the indecomposable monomial $p_{II}'$ has the form of  \begin{align}
        p_{l_uk^-;kl^+,k_wk_u^+,\widehat{k_ul_u}^+;\hat{l},\hat{k}_w} = \varepsilon_{l_uk}^- \varepsilon_{kl}^+ \varepsilon_{k_wk_u}^+  \widehat{\varepsilon}_{k_ul_u}^+ \widehat{\varepsilon}_l\widehat{\varepsilon}_{k_w} \text{ with }   l_u \neq k \neq l , k_w \in I_3.  \label{eq:pII}
\end{align}  Since, again, by rank constraints, we have $k_w \in \{l_u,k,l\}$. From the assumption, we have $k_w \neq l$. Moreover, if $k_w$ is equal to $l_u$ or $k$, we observe that $p_{II}'$ in \eqref{eq:pII} is decomposable. That is, \begin{align*}
    p_{II}' = & \text{ }  p_{l_uk^-;kk_u^+,\widehat{k_ul_u}^+} p_{kl^+;\hat{k},\hat{l}} \text{ if } k_w = k \\
    p_{II}' = & \text{ } p_{k_u l_u^+,\widehat{k_ul_u}^+} p_{k_wk^-;kl^+;\hat{l},\hat{l}_u} \text{ if } k_w = l_u.
\end{align*}  Hence, we must have $  l = k_w \in I_3$, which leads to a contradiction.

In conclusion, if $p_{II}' \in \mathcal{Q}_{B_3}(d)$ in \eqref{eq:poly} is indecomposable, then $1 \leq s_1 = s_2 \leq 3$.
\end{proof}

 All the $S_3$-equivalent roots of \eqref{eq:ind2} are $$(kl_u) (ll) (k_ws_2) \widesim{S_3\text{-span}}  (kl_u)   (k_wl) \widesim{S_3\text{-span}}(k_wl)(kl)(ll_u)(k_wl).$$  The indecomposable monomials of the form $p_{II}' $ are therefore given by \begin{align}
      p_{l_uk^-;kl^+,lk_u^+,\widehat{k_ul_u}^+;\hat{l}^2}= \varepsilon_{l_uk}^- \varepsilon_{kl}^+ \varepsilon_{lk_u}^+  \widehat{\varepsilon}_{k_ul_u}^+ \left(\widehat{\varepsilon}_l\right)^2, \quad  \hat{p}_{II}'  \in \textbf{q}_6     \label{eq:m159}
\end{align} with  $l_u \neq k \neq l ,k_u  \in  I_3$.  Using \eqref{eq:m1}, $p_{II}'$ in \eqref{eq:m159} gives rise to the following forms \begin{align}
\begin{matrix}
      \varepsilon_{32}^- \left(\varepsilon_{12}^+\right)^2 \widehat{\varepsilon}_{23}^+ \left(\widehat{\varepsilon}_1\right)^2, \text{ }\varepsilon_{23}^- \left(\varepsilon_{13}^+\right)^2 \widehat{\varepsilon}_{23}^+ \left(\widehat{\varepsilon}_1\right)^2, \text{ }  \varepsilon_{31}^- \left(\varepsilon_{12}^+\right)^2 \widehat{\varepsilon}_{13}^+ \left(\widehat{\varepsilon}_2\right)^2, \\
 \text{ } \varepsilon_{13}^- \left(\varepsilon_{23}^+\right)^2 \widehat{\varepsilon}_{13}^+ \left(\widehat{\varepsilon}_2\right)^2, \text{ }  \varepsilon_{21}^- \left(\varepsilon_{31}^+\right)^2 \widehat{\varepsilon}_{21}^+ \left(\widehat{\varepsilon}_3\right)^2, \text{ }\varepsilon_{12}^- \left(\varepsilon_{23}^+\right)^2 \widehat{\varepsilon}_{13}^+ \left(\widehat{\varepsilon}_3\right)^2.
\end{matrix} \label{eq:id6010}
\end{align} Together with the $\hat{\cdot}$ action on $p_{II}'$, we can conclude that $  \textbf{q}_6  $ contains $ 12$ indecomposable monomials.

\smallskip
We now list all indecomposable monomials with the forms $p_I$ and $p_{II}$ in \eqref{eq:poly}. Starting with $p  = p_I$. All possible combinations of indices pairs in $R(g_I)$ from the coefficient matrix \eqref{eq:coefficient} are $(al_u)(bk_w)$, $(k_wl_u)$, and $(ab)$. From \eqref{eq:60}, we observe that $R(\varepsilon_{ba}^-\varepsilon_a\widehat{\varepsilon}_b) = 0$. Hence $(al_u)(bk_w)$ and $(ab) $  with $  a \neq b ,k_w \neq l_u \in I_3$ give rise to decomposable monomials. Then using \eqref{eq:m1}, the pair $ (k_wl_u)$ gives us indecomposable monomials as follows: \begin{align*}
   p_I :=  p_{l_ul_w^- ;k_wl_w^+,\widehat{l_wl_u}^+;l_w,\hat{k}_w} = \varepsilon_{l_uk_w}^- \varepsilon_{k_wl_w}^+ \widehat{\varepsilon}_{l_wl_u}^+ \varepsilon_{l_w} \widehat{\varepsilon}_{k_w}
\end{align*}In particular, \begin{align}
   & \begin{matrix}
    p_{23^- ;12^+,\widehat{23}^+;2,\widehat{1}} ,    \quad p_{12^-;12^+,\widehat{23}^+;3,\widehat{2}} ,
   \quad  p_{13^-;12^+,\widehat{13}^+;1,\widehat{2}} , \\
    p_{12^-;12^+,\widehat{13}^+;3,\widehat{1}} ,   \quad p_{13^-;13^+,\widehat{23}^+;2,\hat{3}} ,\quad p_{23^-;13^+,\widehat{23}^+;3,\widehat{1}}
    \end{matrix}  \\
    & \begin{matrix}
    p_{32^- ;\widehat{12}^+,23^+;\widehat{2}, 1} ,  \quad  p_{21^-;\widehat{12}^+, 23^+;\hat{3},2} , \quad  p_{31^-;\widehat{12}^+,13^+;\widehat{1},2} , \\
    p_{21^-;\widehat{12}^+,13^+;\hat{3},1} , \quad p_{31^-;\widehat{13}^+,23^+;\widehat{2},3} , \quad p_{32^-;\widehat{13}^+,23^+;\hat{3},1}
    \end{matrix}
\end{align} which are in $\textbf{q}_5$.  Now, assume that indecomposable polynomials have the form $p_{II}$. For any $  k,l \in I_3$, we then have that
\begin{align}
    \begin{matrix}
        p_{[32^-];12^+;\widehat{1},\hat{3}}, \quad p_{[31^-];12^+;\widehat{2},\hat{3}}, \quad  p_{[23^-];13^+;\widehat{1},\widehat{2}},\quad p_{[21^-];13^+;\widehat{2},\hat{3}}, \quad  p_{[13^-];23^+;\widehat{1},\widehat{2}}, \quad p_{[12^-];23^+;\widehat{1},\hat{3}} \\
        p_{[23^-],\widehat{12}^+;1,3} ,\quad p_{[13^-];\widehat{12}^+;2,3}, \quad    p_{[32^-],\widehat{13}^+;1,2} ,\quad p_{[12^-];\widehat{13}^+;2,3}, \quad    p_{[31^-],\widehat{23}^+;1,2} ,\quad p_{[21^-]:\widehat{23}^+;1,3}.
    \end{matrix}
\end{align}
are indecomposable monomials in $\textbf{q}_4$ and $\textbf{q}_5$.

\smallskip
In conclusion, we have $\mathcal{Q}_{B_3}(d) = \mathbb{C}\langle \textbf{Q}_6 \rangle$. Here, $\textbf{Q}_6 = \bigsqcup_{j=1}^6 \textbf{q}_j$ with \begin{align*}
    \left| \textbf{q}_1\right| = & \text{ }  = 3, \text{ }   \left| \textbf{q}_2\right| = 9, \text{ }  \left| \textbf{q}_3\right|  = 20, \\
    \left| \textbf{q}_4\right|  = & \text{ } 30 , \text{ }  \left| \textbf{q}_5\right|  = 30, \text{ }  \left| \textbf{q}_6\right|  = 12,
\end{align*} where $|\cdot|$ means the number of elements in a set. Hence, $\dim_{FL}\mathcal{Q}_{B_3}(d)= \sum_{j=1}^6|\textbf{q}_j| =104$.  We now deduce all functionally independent generators by checking the Jacobian matrix.  

\begin{proposition}
In the finitely-generating set $\mathbf{Q}_6$, there exist $18$ generators contained in $\mathbf{q}_1\sqcup\mathbf{q}_2\sqcup\mathbf{q}_3$ whose Jacobian has generic rank $18$. In other words, any generator of degree $4, 5$, or $6$ is algebraically dependent on them.
\end{proposition}

\begin{proof}
 Recall that the coordinate functional of $\mathfrak{so}^*(7,\mathbb{C})$ is given by
\begin{align*}
\beta_{B_3} = \{h_1,h_2,h_3\} \,\sqcup\, \{\varepsilon_{ij}^-: i\neq j\} \,\sqcup\, \{\varepsilon_{ij}^+,\widehat{\varepsilon}^+_{ij}:1\leq i<j\leq 3\}\,\sqcup\, \{\varepsilon_a,\widehat{\varepsilon}_a: a=1,2,3\}.
\end{align*}

Consider the following list:
\begin{align}
\nonumber
&\text{Degree 1:} && Q_1=h_1,\quad Q_2=h_2,\quad Q_3=h_3.\\[2pt]
&\text{Degree 2:} && \begin{aligned}[t]
& Q_4=\varepsilon_{12}^-\varepsilon_{21}^-,\quad Q_5=\varepsilon_{13}^-\varepsilon_{31}^-,\quad  Q_6=\varepsilon_{23}^-\varepsilon_{32}^-,\\
& Q_7=\varepsilon_{12}^+\widehat{\varepsilon}_{12}^+,\quad Q_8=\varepsilon_{13}^+\widehat{\varepsilon}_{13}^+,\quad Q_9=\varepsilon_{23}^+\widehat{\varepsilon}_{23}^+,\\
& Q_{10}=\varepsilon_1\widehat{\varepsilon}_1,\quad \hskip  0.1cm Q_{11}=\varepsilon_2\widehat{\varepsilon}_2,\quad \hskip  0.2cm  Q_{12}=\varepsilon_3\widehat{\varepsilon}_3.
\end{aligned}\\[2pt] \label{eq:fdp}
\nonumber
&\text{Degree 3 :} &&
\begin{aligned}[t]
& Q_{13}= \varepsilon_{12}^-\varepsilon_{23}^-\varepsilon_{31}^-, \quad  Q_{14}=\varepsilon_{12}^-\,\widehat{\varepsilon}_1\widehat{\varepsilon}_2,\quad  Q_{15}=\varepsilon_{23}^-\,\widehat{\varepsilon}_2 \varepsilon_3,\\
& Q_{16}=\widehat{\varepsilon}_{12}^+\,\varepsilon_1\varepsilon_2,\qquad Q_{17}=\widehat{\varepsilon}_{13}^+\,\varepsilon_1\varepsilon_3,\quad Q_{18}=\widehat{\varepsilon}_{23}^+\,\varepsilon_2\varepsilon_3.
\end{aligned}
\end{align} These are $18$ monomials of degrees $\leq 3$. We denote the set of functionally independent generators by $\mathcal{F}_{B_3}:=\{Q_1,\dots,Q_{18}\}$.

\medskip
To show the generators in $\mathcal{F}_{B_3}$, it is sufficient to show that the rank of the Jacobian matrix with respect to these generators is exactly $18$. Also, to avoid any singularity, we factor the Jacobian matrix out on a Zariski open set. Let $X:=(\mathbb{C}^\times)^{21}\subset \mathbb{A}^{21} $ be the open subset where all coordinate functions in $\beta_{B_3}$ are nonzero. Write $\mathbb{A}^{21}$ for the affine space over $\mathbb{C}$ with dimension $21$ and coordinate functions $(x_1,\ldots,x_{21})$ with $x\in\beta_{B_3}$. For $i=1,\dots,18$, write \begin{align*}
    Q_i=\prod_{j = 1}^{21} x_j^{n_{i,j}},
\end{align*}  and let $A=(n_{i,j})$ be the $18\times 21$ exponent matrix. For $c=(c_x)_{x\in\beta_{B_3}}\in X$, denote the Jacobian matrix by \begin{align}
    J(c)=\left(\frac{\partial Q_i}{\partial x}\right)_{i,x} (c)\in M_{18\times 21}(\mathbb{C}).
\end{align} A direct computation gives for any $i$ and $x\in\beta_{B_3}$,
\begin{align*}
\frac{\partial Q_i}{\partial x}(c) =\frac{\partial}{\partial x}\Bigl(\prod_{j =1}^{21} x_j^{n_{i,j}}\Bigr)(c)  = n_{i,j} c_j^{\,n_{i,j}-1} \prod_{\ell\neq j} c_\ell^{n_{i,\ell}} = n_{i,j} \frac{\prod_{\ell =1}^{21} c_\ell^{n_{i,\ell}}}{c_j}= n_{i,j}\, \frac{Q_i(c)}{c_j}.
\end{align*} Hence, $J_{i,j}(c)=n_{i,j} Q_i(c) c_j^{-1}$, which decomposes each Jacobian entry into a dependent scale $Q_i(c)$, a column-dependent scale $c_j^{-1}$, and a point-independent coefficient $n_{i,j}$. 
It is therefore natural to isolate the $c$-dependence in two diagonal matrices. Define diagonal matrices \begin{align*}
D(c):=\mathrm{diag}\bigl(Q_1(c),\dots,Q_{18}(c)\bigr),\qquad D_{x^{-1}}(c):=\mathrm{diag}\bigl(c_j^{-1}\bigr)_{1 \leq j \leq 21}.
\end{align*}  With the exponent matrix $A$, we then observe that \begin{align*}
\bigl(D(c) A D_{x^{-1}}(c)\bigr)_{i,j} =\sum_{j =1}^{21}D(c)_{i,i} A_{i,j} \left(D_{x^{-1}}(c)\right)_{j,i} = Q_i(c) n_{i,j} c_j^{-1}.
\end{align*} Thus, $J_{i,j}(c)=\left(D(c) A D_{x^{-1}}(c)\right)_{i,j}$ for all $(i,j)$. Hence, the Jacobian at $c$ has a factorization $J(c)=D(c) A D_{x^{-1}}(c)$, and the rank of $J(c)$ is, in fact, the rank of the exponent matrix $A$.

\medskip
It remains to show $\mathrm{rank}\,A=18$. Select the following $18$ columns (variables) and form the corresponding $18\times 18$ submatrix $A'$ of $A$ based on the coordinates of $\mathfrak{so}^*(7,\mathbb{R})$:
\begin{align*}
&h_1,h_2,h_3 \big|\varepsilon_{21}^-,\varepsilon_{31}^-,\varepsilon_{32}^- \big| \varepsilon_{12}^+,\varepsilon_{13}^+,\varepsilon_{23}^+ \big| \varepsilon_1,\varepsilon_2,\varepsilon_3 \big| \widehat{\varepsilon}_1,\widehat{\varepsilon}_2,\widehat{\varepsilon}_3 \big| \widehat{\varepsilon}_{12}^+,\widehat{\varepsilon}_{13}^+,\widehat{\varepsilon}_{23}^+,
\end{align*} and keep the order of the rows $(Q_1,\dots,Q_{18})$ defined in \eqref{eq:fdp}. With these orders, $A'$ has the $2\times2$ block form
\begin{align*}
A'&= \begin{pmatrix}
I_6 & *\\[2pt]
0 & M
\end{pmatrix},\qquad 
M= \begin{pmatrix}
I_3 & 0   & 0   & I_3\\
0   & I_3 & 0   & 0  \\
0   & U   & I_3 & 0  \\
0   & S   & 0   & I_3
\end{pmatrix},
\end{align*}
where $I_k$ denotes the $k\times k$ identity matrix, and the $6\times 6$ block $I_6$ corresponds to the first six rows $\{Q_1,Q_2,Q_3,Q_4,Q_5,Q_6\}$ and the columns $\{h_1,h_2,h_3,\varepsilon_{21}^-,\varepsilon_{31}^-,\varepsilon_{32}^-\}$. Moreover, with the choices in \eqref{eq:fdp}, the off-diagonal $3\times 3$ blocks $U,S$ are explicitly
\begin{align*}
U=\begin{pmatrix}
0&1&0\\
0&0&1\\
0&0&1
\end{pmatrix},
\qquad
S=\begin{pmatrix}
1&1&0\\
1&0&1\\
0&1&1
\end{pmatrix}.
\end{align*}

Now perform the following elementary column operations on $A'$:
\begin{align*}
&\text{For the long root pairs:}\qquad C_{\widehat{\varepsilon}_{12}^+}\ \leftarrow\ C_{\widehat{\varepsilon}_{12}^+}- C_{\varepsilon_{12}^+},\quad
C_{\widehat{\varepsilon}_{13}^+}\ \leftarrow\ C_{\widehat{\varepsilon}_{13}^+}- C_{\varepsilon_{13}^+},\quad
C_{\widehat{\varepsilon}_{23}^+}\ \leftarrow\ C_{\widehat{\varepsilon}_{23}^+}- C_{\varepsilon_{23}^+},\\[2pt]
&\text{For the short-root pairs:}\qquad C_{\widehat{\varepsilon}_{1}}\ \leftarrow\ C_{\widehat{\varepsilon}_{1}}- C_{\varepsilon_{1}},\qquad 
C_{\widehat{\varepsilon}_{2}}\ \leftarrow\ C_{\widehat{\varepsilon}_{2}}- C_{\varepsilon_{2}},\qquad
C_{\widehat{\varepsilon}_{3}}\ \leftarrow\ C_{\widehat{\varepsilon}_{3}}- C_{\varepsilon_{3}}.
\end{align*}
After these column operations, the resulting $18\times 18$ matrix is block lower triangular, except that the submatrix on the rows $(Q_{13},Q_{14},Q_{15})$ and columns $(\widehat{\varepsilon}_1,\widehat{\varepsilon}_2,\widehat{\varepsilon}_3)$ is an invertible $3\times 3$ block (not necessarily $I_3$). More precisely, if we group the rows and columns as
\begin{align*}
&(Q_1,\dots,Q_6)\,|\, (Q_7,Q_8,Q_9)\,|\, (Q_{10},Q_{11},Q_{12})\,|\, (Q_{13},Q_{14},Q_{15})\,|\, (Q_{16},Q_{17},Q_{18}),\\
&(h_1,h_2,h_3,\varepsilon_{21}^-,\varepsilon_{31}^-,\varepsilon_{32}^-)\,|\,(\varepsilon_{12}^+,\varepsilon_{13}^+,\varepsilon_{23}^+)\,|\,(\varepsilon_1,\varepsilon_2,\varepsilon_3)\,|\,(\widehat{\varepsilon}_1,\widehat{\varepsilon}_2,\widehat{\varepsilon}_3)\,|\,(\widehat{\varepsilon}_{12}^+,\widehat{\varepsilon}_{13}^+,\widehat{\varepsilon}_{23}^+),
\end{align*}
Then, after the column operation, we find that 
\begin{align*}
A' \sim \widetilde{A}'=
\left(
\begin{array}{c|c|c|c|c}
I_6 & 0   & 0   & 0   & 0\\ \hline
0   & I_3 & 0   & 0   & 0\\ \hline
0   & 0   & I_3 & 0   & 0\\ \hline
0   & *   & *   & B   & 0\\ \hline
0   & *   & *   & *   & I_3
\end{array}
\right)= I_6 \oplus I_3 \oplus I_3 \oplus B \oplus I_3,
\end{align*}
where $B\in M_{3\times 3}(\mathbb{Z})$ is the submatrix on rows $(Q_{13},Q_{14},Q_{15})$ and columns $(\widehat{\varepsilon}_1,\widehat{\varepsilon}_2,\widehat{\varepsilon}_3)$, and $B$ is invertible.

Since this $3\times 3$ block $B$ is invertible, we may perform additional elementary column operations restricted to the three columns $(C_{\widehat{\varepsilon}_1},C_{\widehat{\varepsilon}_2},C_{\widehat{\varepsilon}_3})$ such that $B$ also becomes a reduced row echelon form or $I_3$. 
These operations preserve rank and do not affect previously established pivot columns. Thus, the matrix can be decomposed into a block lower triangular form with diagonal blocks as follows:
\begin{align*}
\widetilde{A}' \sim I_6 \oplus I_3 \oplus I_3 \oplus I_3 \oplus I_3.
\end{align*}

The detailed computation of the reduced row action is given below: \begin{enumerate}
\item The rows $Q_7,Q_8,Q_9$ have pivots in the columns $\varepsilon_{12}^+,\varepsilon_{13}^+,\varepsilon_{23}^+$, respectively. Their entries in the columns $\widehat{\varepsilon}_{ij}^+$ have been cleared by the first set of column operations.
\item The rows $Q_{10},Q_{11},Q_{12}$ have pivots in the columns $\varepsilon_1,\varepsilon_2,\varepsilon_3$, respectively. Their entries in the columns $\widehat{\varepsilon}_a$ have been cleared by the second set of column operations.
\item The $3\times 3$ submatrix on rows $Q_{13},Q_{14},Q_{15}$ and columns $\widehat{\varepsilon}_1,\widehat{\varepsilon}_2,\widehat{\varepsilon}_3$ is invertible; after the row reductions within these three rows, we obtain pivots in $\widehat{\varepsilon}_1,\widehat{\varepsilon}_2,\widehat{\varepsilon}_3$, and any remaining nonzero entries lie only in earlier column blocks, which is consistent with lower triangularity.
\item The rows $Q_{16},Q_{17},Q_{18}$ have pivots in the columns $\widehat{\varepsilon}_{12}^+,\widehat{\varepsilon}_{13}^+,\widehat{\varepsilon}_{23}^+$, respectively. Again, any remaining nonzero values are confined to earlier column blocks.
\end{enumerate}  After the preceding row and column operations, we deduce that\begin{align*}
M\sim \begin{pmatrix}
I_3 & 0   & 0   & I_3\\
0   & I_3 & 0   & -I_3\\
0   & 0   & I_3 & I_3\\
0   & 0   & 0   & I_3
\end{pmatrix},\end{align*} whose determinant is $1$. Hence $\det M=1$, and therefore  $\det A'=\det I_6\cdot \det M=1$. Thus, $A'$ is invertible and $\mathrm{rank} \, A=18$. Hence, the Jacobian $J(c)$ has rank $18$ in $X$, so $\mathcal{F}_{B_3}$ is functionally independent in a nonempty Zariski open set.

Finally, we show that there are no further independent generators in degrees $4,5,6$. It is known that, using \eqref{eq:maximum}, the maximal number of functionally independent Cartan commutant generators of type $B_3$ is equal to $\dim \mathfrak{so}(7,\mathbb{C}) -\dim \mathfrak{h} = 18$, implying that no higher-order generators are required. 
\end{proof}

 We now write down the generic formula for the non-trivial commutator relations in a compact form. See \cite{MR4660510,abc} for more detailed information. Without writing each commutator explicitly, let us assume that the explicit generators above in each of the six layer subsets are composed of
\begin{align*}
    \{A_1,A_2,A_3\} \sqcup \left\{B_1,\ldots,B_9\right\} \sqcup   \ldots \sqcup \left\{E_1,\ldots,E_{30} \right\} \sqcup \left\{F_1,\ldots,F_{12} \right\} := \textbf{A} \sqcup \textbf{B} \sqcup \textbf{C} \sqcup \ldots \sqcup \textbf{E} \sqcup \textbf{F}.
\end{align*} Here, the generator of degree one is indicated by the letter $A_i$. Following alphabetical order, elements of degree two are indicated with $B_j$, and so on. The representatives for each subset have already been provided in the construction. Then, to the extent that $\mathcal{Q}_{B_3}(d)$ is closed, the non-trivial commutator relations formally adopt the following structure:
\begin{align}
    \begin{matrix}
        \{\textbf{B},\textbf{B}\} \sim \textbf{C} + \textbf{A} \textbf{B} + \textbf{A}^3 \\
        \{\textbf{B},\textbf{C}\} \sim \textbf{D} + \textbf{B}^2 + \textbf{A} \{\textbf{B},\textbf{B}\} \\
         \{\textbf{B},\textbf{D}\} \sim \textbf{E} + \textbf{BC}   + \textbf{A} \{\textbf{B},\textbf{C}\}    \\
         \{\textbf{B},\textbf{E}\} \sim   \textbf{BD} + \textbf{C}^2 + \{\textbf{B},\textbf{D}\} \\
         \{\textbf{C},\textbf{E}\} \sim  \textbf{BE} + \textbf{CD} + \textbf{B}^2\textbf{C} + \textbf{A} \{\textbf{B},\textbf{E}\} \\
         \{\textbf{D},\textbf{E}\} \sim \textbf{D}^2 + \textbf{CE} + \textbf{B}\textbf{C}^2 + \textbf{B}^4 +   \textbf{A} \{\textbf{C},\textbf{E}\} \\
         \{\textbf{E},\textbf{E}\} \sim \textbf{C}^3 + \textbf{DE} + \textbf{B}^3\textbf{C} +  \textbf{A} \{\textbf{D},\textbf{E}\} \\
          \{\textbf{F},\textbf{E}\} \sim \textbf{E}^2 + \textbf{DF} + \textbf{B}^5 +  \textbf{A} \{\textbf{E},\textbf{E}\} \\
           \{\textbf{F},\textbf{F}\} \sim \textbf{EF} +  \textbf{B}\textbf{C}^3 + \textbf{C}\textbf{D}^2 +     \textbf{A} \{\textbf{E},\textbf{F}\} .
    \end{matrix} \label{eq:65}
\end{align}  In particular, the multiplication of terms is given by \begin{align*}
    \textbf{BC} \sim \sum_{j=1}^9\sum_{k=1}^{20} c_{jk}^{pq} B_jC_k \text{ with }  1 \leq p,q \leq  20.
\end{align*} Progressively expanding the terms in \eqref{eq:65}, we obtain, for example,
\begin{align*}
    \{\textbf{E},\textbf{E}\} \sim \text{ } &  \textbf{C}^3 + \textbf{DE} + \textbf{B}^3\textbf{C} +  \textbf{A} \left(\textbf{D}^2 + \textbf{CE} + \textbf{B}\textbf{C}^2 + \textbf{B}^4\right)+  \textbf{A}^2 \{\textbf{C},\textbf{E}\}   \\
    \sim \text{ } & \textbf{C}^3 + \textbf{DE} + \textbf{B}^3\textbf{C} +  \textbf{A} \left(\textbf{D}^2 + \textbf{CE} + \textbf{B}\textbf{C}^2 + \textbf{B}^4\right)+  \textbf{A}^2 ( \textbf{BE} + \textbf{CD} + \textbf{B}^2\textbf{C}) + \textbf{A}^3 \{\textbf{B},\textbf{E}\} \\
    \sim \text{ } & \textbf{C}^3 + \textbf{DE} + \textbf{B}^3\textbf{C} +  \textbf{A} \left(\textbf{D}^2 + \textbf{CE} + \textbf{B}\textbf{C}^2 + \textbf{B}^4 \right)+  \textbf{A}^2 ( \textbf{BE} + \textbf{CD} + \textbf{B}^2\textbf{C}) + \textbf{A}^3(\textbf{BD} + \textbf{C}^2) \\
    & + \textbf{A}^4\{\textbf{B},\textbf{D}\}  \\
    \sim \text{ } & \textbf{C}^3 + \textbf{DE} + \textbf{B}^3\textbf{C} +  \textbf{A} \left(\textbf{D}^2 + \textbf{CE} + \textbf{B}\textbf{C}^2 + \textbf{B}^4 \right)+  \textbf{A}^2 ( \textbf{BE} + \textbf{CD} + \textbf{B}^2\textbf{C}) + \textbf{A}^3(\textbf{BD} + \textbf{C}^2) \\
     &+ \textbf{A}^4( \textbf{E} + \textbf{BC} )  + \textbf{A}^5 ( \textbf{D} + \textbf{B}^2) + \textbf{A}^6 \textbf{C} + \textbf{A}^7  \textbf{B} + \textbf{A}^9
\end{align*}
Clearly, some of the coefficients listed above will be zero. In any case, it is clear that $d \leq 5$.  Based on an explicit calculation, we obtain, for example,
\begin{align*}
    \left\{p_{II}',\hat{p}_{II}' \right\} = & \text{ }  \underbrace{\left\{\varepsilon_{l_uk}^-,\varepsilon_{kl_u}^-\right\} p_{kl^+,\widehat{kl}^+} p_{lk_u^+,\widehat{lk_u}^+}p_{k_ul_u^+,\widehat{k_ul_u}^+} \left(p_{l,\hat{l}}\right)^2 + \ldots + \left\{\varepsilon_{k_ul_u}^+,\widehat{\varepsilon}_{k_ul_u}^+\right\} p_{l_uk^-,kl_u^-} p_{kl^+,\widehat{kl}^+} p_{lk_u^+,\widehat{lk_u}^+}  \left(p_{l,\hat{l}}\right)^2 }_{\text{$ \in \textbf{A} \textbf{B}^5$}} \\
    & + \ldots
\end{align*}
Here $ \left\{\varepsilon_{l_uk}^-,\varepsilon_{kl_u}^-\right\} \in \{\mathfrak{g}_\beta^*,\mathfrak{g}_{-\beta}^*\} \subset \mathfrak{h}^*$. By \cite[Proposition 2.5]{campoamor2025construction}, \textbf{A} corresponds to the Poisson center. This ensures that the degree of $\mathcal{Q}_{B_3}(d)$ is $5$. That is, $\left(\mathcal{Q}_{B_3}(5),\{\cdot,\cdot\}\right)$ is a quintic algebra.  

\begin{theorem}
    Let $B_3 = \mathfrak{so}(7,\mathbb{C})$. Then the Cartan commutant induced by the reduction chain $\mathfrak{h} \subset B_3$ forms a quintic algebra $\mathcal{Q}_{B_3}(5)$. With the Hamiltonian given by \begin{align*}
    \mathcal{H} = \sum_{i_1,i_2,i_3} \Gamma_{i_1,i_2,i_3} h_1^{i_1} h_2^{i_2} h_3^{i_3} \in S(\mathfrak{h}).
\end{align*}  a superintegrable system $\mathcal{S}$ consists of the integrals of motion defined in \eqref{eq:fdp}.
\end{theorem}

 \subsection{Explicit generators for \texorpdfstring{$\mathcal{Q}_{D_3}(d')$}{Q	extsubscript{D3}(d')}}
 \label{subsec:so6}

 Consider now polynomial (Poisson) algebras arising from the semisimple Lie algebra of type $D_3$, i.e., the Cartan centraliser in $U(D_3)$. The Cartan subalgebra of $\mathfrak{so}(6,\mathbb{C})$ contains $3$ elements $H_1,H_2,H_3$. Take $(\mathfrak{g}^+)^* = \mathrm{span}\left\{\varepsilon_{12}^-,\varepsilon_{13}^-, \varepsilon_{23}^-,\varepsilon_{12}^+,\varepsilon_{13}^+, \varepsilon_{23}^+\right\}$ and $(\mathfrak{g}^-)^* = \mathrm{span}\left\{\varepsilon_{21}^-,\varepsilon_{31}^-, \varepsilon_{32}^-,  \widehat{\varepsilon}_{12}^+, \widehat{\varepsilon}_{13}^+, \widehat{\varepsilon}_{23}^+\right\} $ such that $$\mathfrak{so}^*(6,\mathbb{C}) = \underbrace{\mathrm{span} \{h_1,h_2,h_3\}}_{\text{$= \mathfrak{h}$}} \oplus (\mathfrak{g}^-)^* \oplus (\mathfrak{g}^+)^*.$$ Using the argument in Section \ref{subsec:conson}, we see that all indecomposable monomials have the form in either $g$, $m_1$, or $gm_1 \in \mathcal{SK}_1$ with neither $R(g) = 0$ nor $R(m_1) = 0$. A routine computation shows that there are $23$ indecomposable polynomial solutions in the centraliser $S\left(D_3\right)^\mathfrak{h}$.

We now proceed to construct these elements. By the definition of the polynomial Poisson algebra, we know that $\mathfrak{h} \subset \mathcal{Q}_{D_3}(d')$ forms the first layer. Furthermore, in observation \textbf{(a)}, assume that $p \in \mathcal{Q}_{D_3}(d')$ has the form $g(\boldsymbol{x}_1)$ or $m_1(\boldsymbol{x}_L)$, where $\boldsymbol{x}_1 = \left(\varepsilon_{12}^-,\varepsilon_{13}^-, \varepsilon_{23}^-,\varepsilon_{21}^-,\varepsilon_{31}^-,  \varepsilon_{32}^-\right)$ and $ \boldsymbol{x}_L = \left(\varepsilon_{12}^+,\varepsilon_{13}^+, \varepsilon_{23}^+,\widehat{\varepsilon}_{12}^+, \widehat{\varepsilon}_{13}^+, \widehat{\varepsilon}_{23}^+\right)$. Then $p  \in \textbf{q}_2$ or $\textbf{q}_3$. In particular, we have \begin{align}
  &  p_{12^-,21^-}    = \varepsilon_{12}^-\varepsilon_{21}^-, \quad p_{[13^-],31^-}   = \left[\varepsilon_{13}^-\right]\varepsilon_{31}^-, \quad p_{[23^-],32^-}    = \left[\varepsilon_{23}^-\right]\varepsilon_{32}^-   ,  \label{eq:gena2} \\
& p_{12^+,\widehat{12}^+}   = \varepsilon_{12}^+\widehat{\varepsilon}_{12}^+, \quad p_{13^+,\widehat{13}^+}   = \varepsilon_{13}^+\widehat{\varepsilon}_{13}^+, \quad \quad \text{ } \text{ } p_{23^+,\widehat{23}^+}   = \varepsilon_{23}^+\widehat{\varepsilon}_{23}^+.  \label{eq:so61}
   \end{align} Together with the Cartan elements, the generators of $\mathcal{Q}_{A_2}(2)$ are in \eqref{eq:gena2}.

From observation \textbf{(b)} and the discussion in Subsection \ref{subsec:so5}, taking into account the form of $m_1 $ in \eqref{eq:m1},
   \begin{align}
  & \begin{matrix}
    p_{[12^-];23^+,\widehat{13}^+}    = \left[\varepsilon_{12}^-\right]\varepsilon_{23}^+\widehat{\varepsilon}_{13}^+, \quad  p_{[13^-];23^+,\widehat{12}^+}    = \left[\varepsilon_{13}^-\right]\varepsilon_{23}^+\widehat{\varepsilon}_{12}^+, \quad p_{[23^-];13^+,\widehat{12}^+}   = \left[\varepsilon_{23}^-\right]\varepsilon_{13}^+\widehat{\varepsilon}_{12}^+, \\
   p_{[21^-];13^+,\widehat{23}^+}     = \left[\varepsilon_{21}^-\right]\varepsilon_{13}^+\widehat{\varepsilon}_{23}^+, \quad  p_{[31^-];12^+,\widehat{23}^+}   = \left[\varepsilon_{31}^-\right]\varepsilon_{12}^+\widehat{\varepsilon}_{23}^+,\quad  p_{[32^-];12^+,\widehat{13}^+}    = \left[\varepsilon_{32}^-\right]\varepsilon_{12}^+\widehat{\varepsilon}_{13}^+,
   \end{matrix} \label{eq:so62}
 \end{align}
 are indecomposable monomials from layers $\textbf{q}_3$ and $\textbf{q}_4$. Here, the root vectors of the permutation in $\left[\cdot\right]$ are the same as in \eqref{eq:id5520}. These indecomposable monomials in \eqref{eq:so61} and \eqref{eq:so62} generate  $\mathcal{Q}_{D_3}(d')  =  \mathbb{C}\langle \textbf{Q}_4\rangle    $ with $\textbf{Q}_4 = \textbf{q}_1 \sqcup \dots \sqcup \textbf{q}_4$ and \begin{align*}
      \left| \textbf{q}_1\right|=  3, \text{ }   \left| \textbf{q}_2\right|  = 6, \text{ }   \left| \textbf{q}_3\right|  = 8, \text{ }   \left| \textbf{q}_4\right|  = 6,
 \end{align*}  Hence, $\dim_{FL} \mathcal{Q}_{D_3}(d') = 23. $ Since $A_3 \cong D_3$ are isomorphic as Lie algebras, it is natural to ask about the relations between the polynomial Poisson algebra $\mathcal{Q}_{D_3}(d')  $ and $\mathcal{Q}_{A_3}(3)$. From \cite{campoamor2023algebraic}, we observe  that $\dim_{FL} \mathcal{Q}_{D_3}(d') = \dim_{FL} \mathcal{Q}_{A_3}(3) = 23 $.  The maximal number of functionally independent integrals is $\mathcal{N}(\mathfrak{h}) = \dim  \mathfrak{g} - 3= 12$. In particular, the algebraically dependent relations from the generators above are \begin{align*}
 p_{ij^-,ji^-} p_{ik^-,ki^-} p_{kj^-,jk^-} = & \text{ } p_{[ij]^-,ji^-}p_{[ik^-],ki^-} \\
     p_{ij^-;jk^+,\widehat{ik}^+} p_{jk^-;ik^+,\widehat{ij}^+} = &  \text{ }p_{ik^-;ij^+,\widehat{jk}^+} p_{ik^+,\widehat{ik}^+}  \\
      p_{ik^-,kj^-;jk^+,\widehat{ik}^+} p_{ji^-,ik^-;ik^+,\widehat{ij}^+} = &  \text{ }p_{ik^-;jk^+,\widehat{ij}^+} p_{ik^+,\widehat{ik}^+} p_{ij^-,ji^-}\\
        p_{jk^-;ik^+,\widehat{ij}^+}p_{kj^-;ij^+,\widehat{ik}^+} =  &  \text{ }
 p_{kj^-,jk^-} p_{ij^+,\widehat{ij}^+}  p_{ik^+,\widehat{ik}^+}
 \end{align*} with all $1 \leq i \neq k \neq j \leq 3$. For example, \begin{align*}
     p_{12^-;23^+,\widehat{13}^+} p_{23^-;13^+,\widehat{12}^+}  = &   \text{ }p_{13^-;23^+,\widehat{12}^+}   p_{13^+,\widehat{13}^+} , \\
      p_{12^-;23^+,\widehat{13}^+} p_{31^-;12^+,\widehat{23}^+}  = &   \text{ }p_{32^-;12^+,\widehat{13}^+}   p_{23^+,\widehat{23}^+} , \\
        p_{13^-;23^+,\widehat{12}^+}  p_{21^-;13^+,\widehat{23}^+}  =  & \text{ } p_{23^-;13^+,\widehat{12}^+}   p_{23^+,\widehat{23}^+}, \\
         p_{23^-;13^+,\widehat{12}^+} p_{31^-;12^+,\widehat{23}^+} = & \text{ } p_{21^-;13^+,\widehat{23}^+} p_{12^+,\widehat{12}^+} .
 \end{align*}
 We can consider the following functionally independent elements:
 \begin{align}
        h_1,h_2,h_3,p_{12^-,21^-},p_{13^-,31^-},p_{23^-,32^-},p_{12^-,23^-,31^-}, p_{12^+,\widehat{12}^+},p_{13^+,\widehat{13}^+},p_{23^+,\widehat{23}^+}, p_{12^-;23^+,\widehat{13}^+},p_{23^-;13^+,\widehat{12}^+}    \label{eq:integrals}
 \end{align} forming a set $\mathcal{F}_{D_3}$.

 \medskip
 In analogy to the construction in Subsection \ref{subsec:so5}, we can regroup the generators into sets $\textbf{A} \sqcup \textbf{B} \sqcup \textbf{C} \sqcup \textbf{D}$. The non-trivial brackets in terms of compact notation are thus given by
\begin{align}
     \begin{matrix}
           \{\textbf{B},\textbf{B}\} \sim \textbf{C} + \textbf{A} \textbf{B} + \textbf{A}^3 \\
        \{\textbf{B},\textbf{C}\} \sim \textbf{D} + \textbf{B}^2 + \textbf{A} \{\textbf{B},\textbf{B}\} \\
         \{\textbf{B},\textbf{D}\} \sim \textbf{E} + \textbf{BC}   + \textbf{A} \{\textbf{B},\textbf{C}\}    \\
         \{\textbf{C},\textbf{D}\} \sim   \textbf{BD} + \textbf{C}^2 + \textbf{B}^3 +\{\textbf{B},\textbf{D}\} \\
         \{\textbf{D},\textbf{D}\} \sim  \textbf{CD}   + \textbf{B}^2\textbf{C} + \textbf{A} \{\textbf{C},\textbf{D}\} \\
     \end{matrix} \label{eq:ma67}
 \end{align}
 The last expression in \eqref{eq:ma67} can further be expanded as follows:
 \begin{align*}
   \{\textbf{D},\textbf{D}\} \sim  \textbf{CD}   + \textbf{B}^2\textbf{C} + \textbf{A}(\textbf{C}^2 + \textbf{BD} + \textbf{B}^3) + \textbf{A}^2 \textbf{BC} + \textbf{A}^3(\textbf{D} + \textbf{B}^2) + \textbf{A}^4\textbf{C} + \textbf{A}^5 \textbf{B} + \textbf{A}^7.
 \end{align*}
 Using the explicit generators given above, it can be shown that $\mathcal{Q}_{D_3}(3)$ is a cubic algebra. As an example, some of the explicit expressions of the bilinear operation of $ \{\textbf{B},\textbf{B}\}$ and $\{\textbf{B},\textbf{C}\}$ are given by
 \begin{align*}
     \left\{p_{12^-,21^-},p_{[13^-],31^-}\right\} & = \left(p_{12^-,23^-,31^-} -p_{12^-;23^+,\hat{31}^+}\right)= - \{p_{12^-,21^-},p_{23^-,32^-}\}  , \\
     \left\{p_{12^-,21^-},p_{13^+,\widehat{13}^+}\right\} &= \left(p_{12^-,23^-,31^-} - p_{12^-;23^+,\hat{31}^+}\right)=-\left\{p_{12^-,21^-},p_{13^+,\widehat{13}^+}\right\} \\
      \left\{p_{12^-,21^-},p_{12^-,23^-,31^-}\right\} & =(h_2 -h_1) p_{12^-,23^-,31^-} + p_{12^-,21^-} \left(p_{[13^-],31^-} - p_{23^-,32^-}\right) \\
         \left\{p_{12^-,21^-},p_{21^-;13^+,\hat{32}^+}\right\} & =(h_2 -h_1) p_{21^-;13^+,\hat{32}^+} + p_{12^-,21^-} \left(p_{13^+,\widehat{13}^+}-p_{23^+,\widehat{23}^+}\right) \\
         \left\{p_{12^-,21^-},p_{12^-;23^+,\hat{31}^+}\right\} & = \left(p_{23,12} - p_{13^-,21^-;21^+,\hat{31}}\right) = \left\{p_{12^-,21^-},p_{23^-;12^+,\widehat{13}^+}\right\} ,\\
       \left\{p_{12^-,21^-},p_{13^-;32^+,\hat{21}}\right\} & =  -\left(p_{12^-,13^-;31^+,\widehat{12}^+} + p_{12,23}\right) =\left\{p_{12^-,21^-},p_{23^-;12^+,\widehat{13}^+}\right\} \\
     \left\{p_{12^-,21^-},p_{12^-,13^-;31^+,\widehat{12}^+}\right\} &= (h_2 -h_1)p_{12^-,13^-;31^+,\widehat{12}^+}  - p_{12^-,21^-} \left( \left(p_{31,2}\right)^2 + p_{32^-;21^+,\hat{31}^+}\right)  \end{align*}

Concluding from all of the above results, we deduce the following theorem.

\begin{theorem}
    Let $D_3 = \mathfrak{so}(6,\mathbb{C})$. Then the Cartan commutant induced by the reduction chain $\mathfrak{h} \subset D_3$ forms a cubic algebra $\mathcal{Q}_{D_3}(3)$. With the  Hamiltonian $$\mathcal{H} = \sum_{i_1+i_2+i_3 =2} \Gamma_{i_1,i_2,i_3} h_1^{i_1} h_2^{i_2} h_3^{i_3} \in S(\mathfrak{h}),$$ the integrals of motion in the set $\mathcal{F}_{D_3}$ of \eqref{eq:integrals} form the system $\mathcal{S}$, which is superintegrable by definition, where $\Gamma_{i_1,i_2,i_3}$ are some constants.
\end{theorem}

\smallskip
 It is worth mentioning that, as a consequence of the isomorphism $A_3\simeq D_3$, the polynomial Poisson algebras are, as expected, equivalent. In particular, they are of the same dimension and order. However, although they are not identical in the generators due to the different choice of basis.

\subsection{Explicit generators for \texorpdfstring{$\mathcal{Q}_{C_3}(d'')$}{Q	extsubscript{C3}(d'')}}
\label{subsec:sp5}

 As is well known, the Coxeter diagram for the root systems of $\mathfrak{sp}(6,\mathbb{C})$ and $\mathfrak{so}(7,\mathbb{C})$ is the same, with the difference that short and long roots are interchanged. This allows us to omit observations \textbf{(a)} and \textbf{(b)} from Section \ref{subsubsec:sp}. Suppose that the coordinate ring of $\mathfrak{sp}(6,\mathbb{C})$ is given by $$\beta_{\mathfrak{so}^*(6,\mathbb{C})} =(h_1,h_2,h_3, \varepsilon_{12}^\pm, \varepsilon_{13}^\pm,\varepsilon_{23}^\pm,\widehat{\varepsilon}_{12}^\pm, \widehat{\varepsilon}_{13}^\pm,\widehat{\varepsilon}_{23}^\pm,\varepsilon_1,\varepsilon_2,\varepsilon_3, \widehat{\varepsilon}_1,\widehat{\varepsilon}_2,\widehat{\varepsilon}_3).$$

We first examine the polynomials that contain the short root vectors. Based on observation \textbf{(c)}, the monomials span $\mathcal{SK}_2$ are
\begin{align}
\left.\begin{matrix}
    p_{[12^-]^2;\widehat{1},2}  = \left(\left[  \varepsilon_{12}^-\right]\right)^2\widehat{\varepsilon}_1 \varepsilon_2,\quad  p_{[21^-]^2;1,\widehat{2}}  = \left(\left[  \varepsilon_{21}^-\right]\right)^2\varepsilon_1 \widehat{\varepsilon}_2 \\
    p_{[13^-]^2;\widehat{1},3}  = \left(\left[  \varepsilon_{13}^-\right]\right)^2\widehat{\varepsilon}_1 \varepsilon_3,\quad  p_{[31^-]^2;1,\hat{3}}  = \left(\left[  \varepsilon_{31}^-\right]\right)^2\varepsilon_1 \widehat{\varepsilon}_3 \\
   p_{[23^-]^2;\widehat{2},3}  = \left(\left[  \varepsilon_{23}^-\right]\right)^2\widehat{\varepsilon}_2 \varepsilon_3, \quad  p_{[32^-]^2;2,\hat{3}} = \left(\left[  \varepsilon_{32}^-\right]\right)^2\varepsilon_2 \widehat{\varepsilon}_3
\end{matrix} \right\} \in \textbf{q}_4 \text{ or } \textbf{q}_6. \label{eq:k1k2}
\end{align}
Here, the elements in the set $[\cdot]$ are the same as \eqref{eq:id5520}. For example, as defined in \eqref{eq:id5520}, $[12^-] = \{(12), (13)(32)\} $ contains all the $S_3$-equivalent permutation roots. It is clear that if $\left[  \varepsilon_{ij}^-\right] = \varepsilon_{ij}^-$, then $\deg p = 4$. Otherwise, $\deg p = 6$ for any $p \in \mathcal{K}_1\mathcal{K}_2$ defined in \eqref{eq:k1k2}.

We now look at all the indecomposable monomials that span the subspace $\mathcal{K}_1\mathcal{K}_2$. The procedure to find indecomposable monomials developed in Section \ref{subsec:so5} can also be applied here. It follows that the choice of the monomial $m_1 $ has the same formulation as given in \eqref{eq:classif}, \eqref{eq:m1} and \eqref{eq:classif2}. We first assume that $\alpha_n \notin R\left(m_1 \right)$.  From \eqref{eq:m1}, the indecomposable monomials are  \begin{align}
    \begin{matrix}
      p_{13^+,\widehat{23}^+,13^+,\widehat{23}^+;\widehat{1},2}  =   \left(m_1^{(c)} \right)^2\widehat{\varepsilon}_1 \varepsilon_2,\quad  p_{23^+,\widehat{13}^+,23^+,\widehat{13}^+;1,\hat{3}}  =  \left(\hat{m}_1^{(c)} \right)^2\varepsilon_1 \widehat{\varepsilon}_2 \\
           p_{12^+,\widehat{23}^+,12^+,\widehat{23}^+;\widehat{1},3}  =  \left( m_1^{(a)} \right)^2\widehat{\varepsilon}_1 \varepsilon_3,\quad  p_{23^+,\widehat{12}^+,23^+,\widehat{12}^+;1,\hat{3}}  =  \left(\hat{m}_1^{(a)} \right)^2\varepsilon_1 \widehat{\varepsilon}_3 \\
  p_{12^+,\widehat{13}^+,12^+,\widehat{23}^+;\widehat{2},3} =   \left(m_1^{(b)} \right)^2\widehat{\varepsilon}_2 \varepsilon_3,\quad  p_{13^+,\widehat{12}^+,13^+,\widehat{12}^+;2,\hat{3}}  =  \left(\hat{m}_1^{(b)} \right)^2\varepsilon_2 \widehat{\varepsilon}_3
    \end{matrix}
\end{align} in $\textbf{q}_6$. On the other hand, if $\alpha_n \in R\left(m_1 \right)$, then we can formulate $m_1 $ from \eqref{eq:classif2}, leading to the following indecomposable monomials
\begin{align}
    \left.\begin{matrix}
        p_{(12^+)^2;\widehat{1},\widehat{2}}  = \left(\varepsilon_{12}^+\right)^2 \widehat{\varepsilon}_1\widehat{\varepsilon}_2, \quad  p_{(13^+)^2;\widehat{1},\hat{3}}  = \left(\varepsilon_{13}^+\right)^2 \widehat{\varepsilon}_1\widehat{\varepsilon}_3, \quad  p_{(23^+)^2;\widehat{2},\hat{3}}  = \left(\varepsilon_{23}^+\right)^2 \widehat{\varepsilon}_2\widehat{\varepsilon}_3 \\
    p_{(\widehat{12}^+)^2;1,2}  = \left(\widehat{\varepsilon}_{12}^+\right)^2 \varepsilon_1\varepsilon_2, \quad  p_{(\widehat{13}^+)^2;1,3}  = \left(\widehat{\varepsilon}_{13}^+\right)^2 \varepsilon_1\varepsilon_3, \quad   p_{(\widehat{23}^+)^2;2,3}  = \left(\widehat{\varepsilon}_{23}^+\right)^2 \varepsilon_2\varepsilon_3
\end{matrix}\right\} \in \textbf{q}_4.
\end{align} Moreover, if $m_1 $ has the form of (B2), then \begin{align}
        \begin{matrix}
        p_{13^+,12^+,\widehat{23}^+;\widehat{1} } = \varepsilon_{13}^+m_1^{(a)}  \widehat{\varepsilon}_1, \quad  p_{23^+,12^+,\widehat{13}^+; \widehat{2}}   = \varepsilon_{23}^+ m_1^{(b)}  \widehat{\varepsilon}_2, \quad   p_{ 12^+, 13^+,\widehat{23}^+;\widehat{1}}   = \varepsilon_{12}^+m_1^{(c)} \widehat{\varepsilon}_1, \\
    p_{\widehat{13}^+,23^+,\widehat{12}^+;1}   =\widehat{\varepsilon}_{13}^+\hat{m}_1^{(a)} \varepsilon_1, \quad  p_{ \widehat{23}^+,13^+,\widehat{12}^+;2}   =\widehat{\varepsilon}_{23}^+   \hat{m}_1^{(b)} \varepsilon_2, \quad   p_{\widehat{12}^+,23^+,\widehat{13}^+;1} =\widehat{\varepsilon}_{12}^+ \hat{m}_1^{(c)} \varepsilon_1
    \end{matrix} \label{eq:id79}
\end{align} are indecomposable monomials in $\textbf{q}_4$.

Finally, using observation \textbf{(e)} in Subsection \ref{subsubsec:sp} and the argument from Subsection \ref{subsec:so5}, we obtain the following: \begin{align}
\nonumber
     p_I  = & \text{ } \left[g_I\right] \varepsilon_{kl}^+  \widehat{\varepsilon}_{lj}^+  \varepsilon_a\widehat{\varepsilon}_b \text{ with $ k  \neq j \in I_3$};\\
       p_{II} = & \text{ } \left[g_{II}\right] \varepsilon_{kl}^+  \widehat{\varepsilon}_b  \text{ with $   k\neq  l \in I_3$}  \label{eq:form2}; \\
       \nonumber
      p_{II}'  = & \text{ }\left[g_{II}'\right]\varepsilon_{kl}^+ \varepsilon_{ij}^+\widehat{\varepsilon}_{jm}^+ \widehat{\varepsilon}_b \text{ with $     k,l,j,i \neq m , b \in I_3,$}
\end{align} where $I_3 = \{1,2,3\}$.  We then present explicit generators of the form in \eqref{eq:form2} based on the aforementioned classification. Specifically, if homogeneous generators take the form of $p_I$, then their explicit forms, in $\textbf{q}_6$, are as follows: \begin{align}
   & \begin{matrix}
    p_{32^-,12^- ;12^+,\widehat{23}^+ ;2,\widehat{1}} ,     \quad   p_{ 21^-,23^-;12^+,\widehat{23}^+ ;3,\widehat{2}} ,
    \quad   p_{ 21^-,31^-;12^+,\widehat{13}^+ ;1,\widehat{2}} , \\
    p_{13^-,12^-;12^+,\widehat{13}^+ ;3,\widehat{1}} , \quad    p_{ 31^-,32^-;13^+,\widehat{23}^+ ;2,\hat{3}} , \quad   p_{ 23^-,13^-;13^+,\widehat{23}^+ ;3,\widehat{1}}
    \end{matrix}  \\
    & \begin{matrix}
    p_{ 23^- ,21^-;\widehat{12}^+,23^+ ;\widehat{2}, 1} ,     \quad   p_{ 12^-,32^-;\widehat{12}^+, 23^+ ;\hat{3},2} ,
    \quad   p_{ 13^-,12^-;\widehat{12}^+,13^+  ;\widehat{1},2} , \\
    p_{31^-,21^-;\widehat{12}^+,13^+ ;\hat{3},1} , \quad     p_{23^-,13^-;\widehat{13}^+,23^+ ;\widehat{2},3} , \quad   p_{ 31^-,32^-;\widehat{13}^+,23^+ ;\hat{3},1} .
    \end{matrix}
\end{align}   Here $p_{32^-,12^- ;12^+,\widehat{23}^+ ;2,\widehat{1}} =  \varepsilon_{32}^- \varepsilon_{12}^- m_1^{(a)}   \varepsilon_2\widehat{\varepsilon}_1$. Now, consider that indecomposable polynomials belong to the type $p_{II}$. For any $1 \leq k \neq l \leq 3$, the explicit forms of the monomials in \eqref{eq:form2} are \begin{align}
    \left.\begin{matrix}
        p_{[12^-];12^+;\widehat{1}}, \quad p_{[21^-];12^+;\widehat{2} }, \quad   p_{ 13^-,23^- ;12^+;\hat{3} }, \\  p_{[13^-];13^+;\widehat{1} }, \quad   p_{[31^-];13^+;\hat{3}}, \quad p_{ 21^-,32^- ;13^+;\widehat{2}},\\
        p_{[23^-];23^+;\widehat{2}}, \quad  p_{[32^-];23^+; \hat{3}} ,\quad p_{21^-,31^- ;23^+;\widehat{1}}
    \end{matrix} \right\} \in \textbf{q}_3 \text{ or }\textbf{q}_4, \label{eq:71}
\end{align} where the $\hat{\cdot}$ actions on \eqref{eq:71} are indecomposable monomials. Lastly, assume that the indecomposable homogeneous polynomial is of the form $p_{II}'$ as defined in \eqref{eq:form2}. Given that $\mathrm{rank}\,\mathfrak{g} = 3$, the expressions for the monomials $m_1$ are \begin{align*}
    \left(\varepsilon_{ij}^+\right)^2\widehat{\varepsilon}_{jk}^+ \text{ and } \varepsilon_{ik}^+ \varepsilon_{ij}^+\widehat{\varepsilon}_{jk}^+ \text{ } 1 \leq i \neq j \neq  k \leq 3.
\end{align*}    It turns out that for any $g \in S_3$-span, the monomials with the expression $[g] \varepsilon_{ik}^+ \varepsilon_{ij}^+\widehat{\varepsilon}_{jk}^+\widehat{\varepsilon}_{s_1}$ are decomposable, where $s_1 =i \neq j \neq k \in I_3$. Consequently, we focus solely on $m_1 = \left(\varepsilon_{ij}^+\right)^2\widehat{\varepsilon}_{jk}^+$. The indecomposable monomials are thus given by
\begin{align}
      \begin{matrix}
        p_{ 32^- ;{12^+}^2,\widehat{23}^+;\widehat{1}},  \quad
        p_{ 31^- ;{12^+}^2,\widehat{13}^+;\widehat{2}}  ,  \quad p_{ 23^- ;{13^+}^2,\widehat{23}^+;\widehat{1}}
    \end{matrix} \in \textbf{q}_5 . \label{eq:id71}
\end{align} Furthermore, the monomials in \eqref{eq:id71} under the $\hat{\cdot}$ action are still indecomposable.

In conclusion, $\mathcal{Q}_{C_3}(d'') = \mathbb{C}\langle \textbf{Q}_6 \rangle $ with $\textbf{Q}_6 = \bigsqcup_{j=1}^6 \textbf{q}_j$ \begin{align*}
    \left| \textbf{q}_1\right|  = & \text{ }  3, \text{ }  \left| \textbf{q}_2\right|  = 9, \text{ }  \left| \textbf{q}_3\right|  = 26, \\
    \left| \textbf{q}_4\right| =  & \text{ } 36, \text{ } \left| \textbf{q}_5\right| = 6, \text{ }  \left| \textbf{q}_6\right| = 24.
\end{align*} Hence $\dim_{FL} \mathcal{Q}_{C_3}(d'') = \sum_{j=1}^6|\textbf{q}_j|  = 104$. Similarly to what we did in Subsection \ref{subsec:so5}, computing the Jacobian matrix with respect to all the generators above, we deduce that there are $18$ functionally independent generators given by \begin{align}
\nonumber
  &  Q_1=h_1,\quad Q_2=h_2,\quad Q_3=h_3, \\
  \nonumber
  & Q_4=\varepsilon_{12}^-\varepsilon_{21}^-,\quad Q_5=\varepsilon_{13}^-\varepsilon_{31}^-,\quad  Q_6=\varepsilon_{23}^-\varepsilon_{32}^-,\\
  \nonumber
& Q_7=\varepsilon_{12}^+\widehat{\varepsilon}_{12}^+,\quad Q_8=\varepsilon_{13}^+\widehat{\varepsilon}_{13}^+,\quad Q_9=\varepsilon_{23}^+\widehat{\varepsilon}_{23}^+,\\
& Q_{10}=\varepsilon_1\widehat{\varepsilon}_1,\quad   Q_{11}=\varepsilon_2\widehat{\varepsilon}_2,\quad    Q_{12}=\varepsilon_3\widehat{\varepsilon}_3, \label{eq:integrals2}\\
\nonumber
& Q_{13}= \varepsilon_{12}^-\varepsilon_{23}^-\varepsilon_{31}^-, \quad  Q_{14}=\varepsilon_{12}^-\, \varepsilon_{23}^+\widehat{\varepsilon}_{13}^+,\quad  Q_{15}=\varepsilon_{13}^-\, \varepsilon_{23}^+\widehat{\varepsilon}_{12}^+,\\
\nonumber
&   Q_{16}:= \varepsilon_{12}^-\varepsilon_{12}^+ \varepsilon_1^- ,\qquad
Q_{17}:=\varepsilon_{23}^- \varepsilon_{23}^+ \varepsilon_2^- ,\qquad
Q_{18}:=\varepsilon_{31}^-\varepsilon_{13}^+\varepsilon_3.,
\end{align} which forms a finite set $\mathcal{F}_{C_3}$. Denote the generators by the compact notation $\textbf{A} \sqcup \ldots \sqcup  \textbf{E} \sqcup\textbf{F}$. In particular, the non-trivial Poisson brackets of the highest degree is  \begin{align}
    \{\textbf{F},\textbf{F}\} \sim  \textbf{EF} +  \textbf{B}\textbf{C}^3 + \textbf{C}\textbf{D}^2 +  \textbf{A} \{\textbf{E},\textbf{F}\}.
\end{align} Consider a monomial with a maximum degree. It can be shown that $d'' = 5$. For example, take $p_I = [g_I]m \in \mathcal{Q}_6$, where $p_I$ and $[g_I]$ are defined in \eqref{eq:form2}. By direct calculation, \begin{align*}
    \left\{p_I,\hat{p}_I\right\} = \underbrace{[g_I]\left[\hat{g}_I\right]\left\{\varepsilon_{kl}^+,\widehat{\varepsilon}_{kl}^+\right\} p_{\widehat{lj}^+,lj^+} p_{a,\hat{a}}p_{b,\hat{b}} + \ldots + [g_I]  \left[\hat{g}_I\right]p_{kl^+,\widehat{kl}^+} p_{\widehat{lj}^+,lj^+} p_{a,\hat{a}}\left\{\varepsilon_b,\widehat{\varepsilon}_b\right\}+ \ldots , }_{\text{$ \in \textbf{AB}^5$ } }
\end{align*} implying that $ d'' = 5$.

\begin{theorem}
    Let $C_3 = \mathfrak{sp}(5,\mathbb{C})$. Then the Cartan commutant induced by the reduction chain $\mathfrak{h} \subset C_3$ forms a quintic algebra $\mathcal{Q}_{C_3}(5)$. With the Hamiltonian $$\mathcal{H} = \sum_{i_1+i_2+i_3 =2} \Gamma_{i_1,i_2,i_3} h_1^{i_1} h_2^{i_2} h_3^{i_3} \in S(\mathfrak{h}),$$ the integrals of motion in $\mathcal{F}_{C_3}$ of \eqref{eq:integrals2} form the system $\mathcal{S}$, which is superintegrable by definition, where $\Gamma_{i_1,i_2,i_3}$ are some constants.
\end{theorem}

\smallskip
From the construction in Sections \ref{subsec:so5}, \ref{subsec:so6} and \ref{subsec:sp5}, we, in particular, deduce some embeddings between the different polynomial Poisson algebras. Moreover, $\dim_{FL} \mathcal{Q}_{B_3}(5)= \dim_{FL} \mathcal{Q}_{C_3}(5) =  104$ and
\begin{equation}\label{embe}
 \mathcal{Q}_{A_2} (2) \subset \mathcal{Q}_{A_3} (3) ;\quad
 \mathcal{Q}_{A_2} (2) \subset \mathcal{Q}_{D_3} (3)\subset \mathcal{Q}_{B_3}(5) ;\quad
    \mathcal{Q}_{A_2} (2) \subset \mathcal{Q}_{C_3}(5).
\end{equation}

\section{Conclusions}
\label{sec:conclusion}

In this work, the approach proposed in \cite{campoamor2023algebraic} to determine the centraliser of the Cartan subalgebra in the universal enveloping algebra of a complex semisimple Lie algebra $\mathfrak{g}$ has been extended to the remaining classical series $B_n$, $C_n$ and $D_n$. The method is based on a detailed analysis of the root system, taking into account that for these series, there are two root lengths, for which reason the ansatz must be generalized to properly recover the indecomposable polynomials that generate the commutant. A generic algorithm to determine the number of indecomposable and linearly independent polynomials of a given degree has been provided, as well as the general outline of the classification of the polynomial Poisson algebras associated with each root system, which uses properties of the Weyl group. The detailed proof of the technical details will be given elsewhere \cite{abc}. As an illustration of the method, the rank three Lie algebras $B_3$, $C_3$, and $D_3$ have been treated in detail, with explicit formulas for the indecomposable polynomial and their polynomial Poisson algebras. We also showed that, for the latter case, the isomorphism with $A_3$ implies that the corresponding polynomial Poisson algebras are equivalent, albeit their systems of generators are different, depending on the choice of different bases. In particular, from the explicit presentation of the polynomial Poisson algebras certain embeddings \eqref{embe} can be deduced. In this context, it should be recalled that polynomial Poisson algebras related to the root system of $A_n$ can be realized as higher-rank Racah algebras with a suitable realization \cite{campoamor2023algebraic}. Therefore, the various embeddings of the polynomial Poisson algebras $\mathcal{Q}_{A_{n-1}}(n-1)$, $\mathcal{Q}_{B_n}(d), \mathcal{Q}_{C_n}(d'')$ and $\mathcal{Q}_{D_n}(d')$, suggest that appropriate deformations of higher-rank Racah algebras can provide an alternative interpretation of these commutants, with potential applications in superintegrable systems on manifolds more general than the spheres.
On the other hand, as higher-rank Racah algebras play an important role in recoupling theories, orthogonal polynomials, and superintegrability, also connected to models in the context of Dunkl operators and Dirac equations, it is worth analyzing in detail whether the polynomial Poisson algebras related to $B_n$, $C_n$ and $D_n$ lead to some physically relevant generalization. In order to complete the description, the polynomial Poisson algebras related to the root systems of the exceptional Lie algebras have yet to be determined, with partial results for $G_2$ already having been obtained \cite{MR4710584,C183}. The problem in these cases is primarily of a computational nature, especially for the $E_i$ series, although the ansatz used for the classical series is expected to remain valid for the exceptional root systems. It is hoped that the problem will soon be solved in a satisfactory manner. Work in this direction is currently in progress.

\smallskip
In summary, besides the intrinsic interest of polynomial Poisson algebras in universal enveloping algebras and the construction of superintegrable systems for adequate realizations of the generators by differential operators, the method constitutes a unified approach to reinterpreting symmetry algebras in a purely algebraic setting. It also provides a way to understand quantization via the correspondence between the setting of the symmetric algebra and the universal enveloping algebra of Lie algebras. In this context, the quantum analog, its integrals, and symmetry algebra can be obtained via the symmetrization map. The quantization of superintegrable systems is, in general, complicated. Here, the algebraic setting provides some insight for a large class of superintegrable systems.

\section*{Acknowledgment}

\noindent IM was supported by the Australian Research Council Future Fellowship FT180100099. YZZ was supported by the Australian Research Council Discovery Project DP190101529. RCS acknowledges financial support by the research grants PID2023-148373NB-I00 and PID2024-156578NB-I00 (AEI/ FEDER, UE). The research of D.L.~has been partially funded by MUR - Dipartimento di Eccellenza 2023-2027, codice CUP G43C22004580005 - codice progetto   DECC23$\_$012$\_$DIP and partially supported by INFN-CSN4 (Commissione Scientifica Nazionale 4 - Fisica Teorica), MMNLP project. D.L. is a member of GNFM, INdAM.

\section{Declarations}

\noindent There are no conflicts of interest in the manuscript.

\bibliographystyle{unsrt}
\bibliography{bibliography.bib}

\end{document}